\documentclass[preprint,12pt,authoryear]{elsarticle}

\usepackage{amssymb}
\usepackage{graphicx}
\usepackage{natbib}
\usepackage{amssymb}
\usepackage{amsmath}
\usepackage{color}
\usepackage{rotating}

\usepackage{url}
\font\smallfont=cmsy10 at 10truept
\mathchardef\bigCircle="280D

\font\bigfont=cmsy10 at 14.4truept
\mathchardef\tiMes="2902        %

\font\Bigfont=cmsy10 at 17.28truept
\mathchardef\DiaMond="2A05        %
\mathchardef\cirCle="2A0E
\mathchardef\BigCircle="2A0D

\font\Bbigfont=cmsy10 at 24.88truept
\mathchardef\buLLet="2B0F

\def\bigCirc{\raise 0.3ex\hbox{$\bigCircle$}\nobreak$\,$}

\def\Bullet{\raise-0.35ex\hbox{$\buLLet$}\nobreak$\,$}

\def\triangledown{\raise 0.2em\hbox{$\bigtriangledown$}\nobreak$\,$}

\def\minisquare{\hbox{${\vcenter{
               \hrule height 0.3pt \kern-0.4pt
               \hbox{\vrule width  0.3pt height 3.0pt \kern 2.6pt
               \vrule width  0.3pt height 3.0pt} \kern-0.4pt
               \hrule height 0.3pt}}$}}
\def\ssquare{\raise 0.175ex\hbox{${\vcenter{
               \hrule height 0.5truept       \kern-0.25truept
               \hbox{\vrule width 0.5truept height 3.0truept \kern 2.75truept
                     \vrule width 0.5truept height 3.0truept} \kern-0.25truept
               \hrule height 0.5truept}}$}\nobreak$\,$}
\def\squarex{\raise 0.175ex\hbox{${\vcenter{
               \hrule height 0.8truept       \kern-1.80truept
          \hbox{\vrule width 0.8truept height 8.0truept \kern-1.95truept
                \raise 0.8truept\hbox{$\tiMes$}     \kern-6.70truept
                \vrule width 0.8truept height 8.0truept} \kern-0.80truept
               \hrule height 0.8truept}}$}\nobreak$\,$}

\def\sqbull{\raise0.175ex\hbox{\vrule height 1.4ex width 1.6ex depth 0.2ex}\nobreak$\,$}
\def\smsqbull{\raise0.175ex\hbox{\vrule height 0.8ex width 0.9ex depth 0.2ex}\nobreak$\,$}

\def\Diamondplus{${\vcenter{\vcenter{\DiaMond} \kern-10truept
                            \hbox{\vrule width .4truept}\kern -3truept
                            \hrule height .4truept}}$\nobreak$\,$}
\newcount\ndots

\def\drawline#1#2{\raise 2.5truept\vbox{\hrule width #1truept height #2truept}}
\def\moonspace#1{\hskip #1truept}

\def\shortchain{\drawline{6.0}{0.75}}

\def\shortchainspace{\shortchain\moonspace{2}}

\def\Dashy{\drawline{4.00}{1.00}}     
\def\dashy{\drawline{4.00}{0.75}}     
\def\thindashy{\drawline{4.00}{0.25}}     
\def\dashyspace{\dashy\moonspace{2}}
\def\Dashyspace{\Dashy\moonspace{2}}
\def\thindashyspace{\thindashy\moonspace{2}}
\def\longdashy{\drawline{8.00}{0.75}} 
\def\thinlongdashy{\drawline{8.00}{0.25}} 
\def\longdashyspace{\longdashy\moonspace{2}}
\def\thinlongdashyspace{\thinlongdashy\moonspace{2}}
     
\def\dotty{\drawline{1.00}{0.75}}

\def\dottyspace{\dotty\moonspace{2}}

\def\solid{\drawline{24}{0.75}\nobreak$\,$}

\def\dashbox{\hbox{\dashyspace}}  
\def\Dashbox{\hbox{\Dashyspace}}  
\def\dashed{\hbox {\ndots=0 \loop\ifnum\ndots<3\advance\ndots by 1
        \dashbox\repeat\dashy}\nobreak$\,$}       
\def\Dashed{\hbox {\ndots=0 \loop\ifnum\ndots<3\advance\ndots by 1
        \Dashbox\repeat\Dashy}\nobreak$\,$}       
\def\thindashbox{\hbox{\thindashyspace}}  
\def\thindashed{\hbox {\ndots=0 \loop\ifnum\ndots<3\advance\ndots by 1
        \thindashbox\repeat\thindashy}\nobreak$\,$}       
\def\thindash{\hbox {\ndots=0 \loop\ifnum\ndots<3\advance\ndots by 1
        \thindashbox\repeat\thindashy}\nobreak$\,$}       

\def\longdashbox{\hbox{\longdashyspace}}  
\def\thinlongdashbox{\hbox{\thinlongdashyspace}}  
\def\longdash{\hbox {\ndots=0 \loop\ifnum\ndots<3\advance\ndots by 1
        \longdashbox\repeat\longdashy}\nobreak$\,$}       
\def\thinlongdash{\hbox {\ndots=0 \loop\ifnum\ndots<3\advance\ndots by 1
        \thinlongdashbox\repeat\thinlongdashy}\nobreak$\,$}       

\def\dotdashed{\hbox{\shortchainspace\dottyspace\shortchain}\nobreak$\,$}      

\definecolor{grey}{rgb}{0.7,0.7,0.7}

\journal{International Journal of Heat and Fluid Flow}

\begin{document}

\begin{frontmatter}



\title{Turbulent boundary layers around wing sections up to $Re_{c}=1,000,000$}


\author{R. Vinuesa\footnote{{\it Email address for correspondence:} \url{rvinuesa@mech.kth.se} (R. Vinuesa)}, P. S. Negi, M. Atzori, A. Hanifi, D. S. Henningson \\ and P. Schlatter}

\address{Linn\'e FLOW Centre, KTH Mechanics, SE-100 44 Stockholm, Sweden\\
and Swedish e-Science Research Centre (SeRC), Stockholm, Sweden}

\begin{abstract}
Reynolds-number effects in the adverse-pressure-gradient (APG) turbulent boundary layer (TBL) developing on the suction side of a NACA4412 wing section are assessed in the present work. To this end, we analyze four cases at Reynolds numbers based on freestream velocity and chord length ranging from $Re_{c}=100,000$ to $1,000,000$, all of them with $5^{\circ}$ angle of attack. The results of four well-resolved large-eddy simulations (LESs) are used to characterize the effect of Reynolds number on APG TBLs subjected to \textcolor{black}{ approximately} the same pressure-gradient distribution (defined by the Clauser pressure-gradient parameter $\beta$). Comparisons of the wing profiles with zero-pressure-gradient (ZPG) data at matched friction Reynolds numbers reveal that, \textcolor{black}{ for approximately} the same $\beta$ distribution, the lower-Reynolds-number boundary layers are more sensitive to pressure-gradient effects. This is reflected in the values of the inner-scaled edge velocity $U^{+}_{e}$, the shape factor $H$, \textcolor{black}{ the components of the Reynolds-stress tensor in the outer region and the outer-region production of turbulent kinetic energy.} This conclusion is supported by the larger wall-normal velocities \textcolor{black}{ and outer-scaled fluctuations} observed in the lower-$Re_{c}$ cases. Thus, our results \textcolor{black}{ suggest} that two complementing mechanisms contribute to the development of the outer region in TBLs and the formation of large-scale energetic structures: one mechanism associated with the increase in Reynolds number, and another one connected to the APG. Future extensions of the present work will be aimed at studying the differences in the outer-region energizing mechanisms due to APGs and increasing Reynolds number. 
\end{abstract}


\end{frontmatter}



\section{Introduction}
Turbulent boundary layers (TBLs) subjected to streamwise pressure gradients (PGs) are relevant to a wide range of industrial applications from diffusers to turbines and wings, and pose a number of open questions regarding their structure and underlying dynamics. A number of studies over the years has aimed at shedding light on these open questions through various approaches. In the 1950s, \cite{townsend} employed theoretical analyses of the governing equations and concluded that although the only TBL that can be described through self-similar variables is the so-called sink flow (which corresponds to a strongly-accelerated TBL, see for instance the work by \cite{jones_sink}), certain pressure-gradient conditions exhibit self-similarity in their outer region at high Reynolds numbers \citep{marusic_et_al}. This includes the widely studied zero-pressure-gradient (ZPG) turbulent boundary layer \citep{schlatter_orlu10,sillero_et_al}. These so-called near-equilibrium conditions \citep{dixit_ramesh,bobke_et_al} are obtained when the freestream velocity is described in terms of the streamwise coordinate by a power law, given certain restrictions in the exponent \citep{townsend}. The work by \cite{townsend} was complemented one decade later by \cite{mellor_gibson}, who proposed a theoretical framework to calculate boundary-layer parameters in PG TBLs subjected to a constant pressure-gradient magnitude, represented by a constant value of the Clauser pressure-gradient parameter $\beta= \delta^{*} / \tau_{w} {\rm d}P_{e} / {\rm d} x$ (where $\delta^{*}$ is the displacement thickness, $\tau_{w}$ the wall-shear stress and ${\rm d}P_{e} / {\rm d} x$ is the streamwise pressure gradient). Extensive experimental campaigns, such as the ones by \cite{skare_krogstad} and \cite{harun_et_al}, were later undertaken with the aims of obtaining high-Reynolds-number adverse-pressure-gradient (APG) TBLs with constant $\beta$ (in the case of the former), and further understanding the energizing mechanisms present in favorable-pressure-gradient (FPG) and APG TBLs (in the latter study). The challenges of performing high-fidelity simulations of PG TBLs, especially when it comes to a proper definition of boundary conditions, were addressed by \cite{spalart_watmuff}. However, and as stated by \cite{monty_et_al}, the large number of parameters influencing the structure of PG TBLs raises serious difficulties when comparing databases from different experimental or numerical databases. The present work is focused on the analysis of APG effects in a specific case, namely the TBL that forms on the suction side of wings. As the boundary layer develops, it encounters a progressively larger resistance through the increased pressure in the streamwise direction. This APG decelerates the boundary layer and increases its thickness while reducing the wall-shear stress. As a result of the larger boundary-layer thickness the wake parameter in the mean velocity profile increases \citep{perry_et_al,vinuesa_aiaa}, and more energetic turbulent structures develop in the outer region \citep{maciel_et_al}. The recent work by \cite{bobke_et_al} highlights the importance of the flow development in the establishment of an APG TBL, and in particular the streamwise evolution of the Clauser pressure-gradient parameter $\beta$. In their study, \cite{bobke_et_al} compared different APG TBLs subjected to various $\beta(x)$ distributions, including several flat-plate cases and one APG developing on the suction side of a wing section \citep{hosseini_et_al}. Their main conclusion states that the effect of APGs is more prominent in the cases where the boundary layer has been subjected to a stronger pressure gradient for a longer streamwise distance, a conclusion that demonstrates the relevance of accounting for the $\beta(x)$ distribution when assessing pressure-gradient effects on TBLs \citep{diagnostic_ftac}. Along these lines, the numerical studies by \textcolor{black}{ \cite{kitsios_et_al}}, \cite{lee} and \cite{bobke_et_al} aim at characterizing the effect of APGs on TBLs in cases with a constant pressure-gradient magnitude, {\it i.e.}, in flat-plate boundary layers exhibiting long regions with constant values of $\beta$.  

The aim of the present work is to assess the effect of the Reynolds number ($Re$) on four APG TBLs subjected to \textcolor{black}{ approximately} the same $\beta(x)$ distribution. In particular, we consider the turbulent flow around a NACA4412 wing section at four Reynolds numbers based on freestream velocity $U_{\infty}$ and chord length $c$, ranging from $Re_{c}=U_{\infty } c / \nu=100,000$ to $1,000,000$. As discussed by \cite{pinkerton}, the NACA4412 wing section is characterized by exhibiting a pressure-gradient distribution essentially independent of $Re$ at moderate angles of attack, a fact that makes this particular airfoil a suitable candidate to study Reynolds-number effects on TBLs given a particular pressure-gradient history. Due to this, but also because the NACA4412 airfoil presents benign stalling properties, and also sufficient thickness from a structural point of view, a number of experimental studies have considered the NACA4412 profile. Most notably, the flying hot-wire measurements by \cite{coles_wadcock} and the later laser Dopler velocimetry (LDV) measurements by \cite{wadcock} and \cite{hastings_williams} provided data of the turbulent boundary layers developing at specific conditions. The NACA4412 wing section has become a general benchmark case, and therefore also a number of numerical studies have been conducted with the aim of characterizing turbulent flows developing around this airfoil, including the large-eddy simulations (LESs) by \cite{jansen} or our earlier direct numerical simulations (DNSs) at $Re_{c}=400,000$ \citep{hosseini_et_al,wing_ftac}.  Similarly, simpler symmetric profiles such as the NACA0012 airfoil have been studied through DNS by \textcolor{black}{ \cite{hoarau_et_al}} and \cite{rodriguez_et_al}. Other recent LESs include the studies on symmetric airfoils by \textcolor{black}{ \cite{kitsios_wing}}, \cite{wolf_et_al} and \cite{sato_et_al}. To the authors' knowledge this is the first numerical study where the pressure-gradient effects on the turbulent boundary layers are characterized systematically for the same airfoil at several Reynolds numbers, including the comparably high $Re_{c}$ of 1 million.

The present manuscript is organized as follows: the numerical setup is described in $\S$\ref{setup_section}, the results are presented for the boundary layer developing on the suction side of the wing in $\S$\ref{suction_section}, and finally in $\S$\ref{conclusions_section} a summary of the main conclusions of this study is presented.


\section{Computational setup} \label{setup_section}

Well-resolved large-eddy simulations of the flow around a NACA4412 wing section at various Reynolds numbers were carried out using the spectral-element code Nek5000 \citep{fischer_et_al}, developed at the Argonne National Laboratory. In the spectral-element method (SEM) the computational domain is decomposed into elements, where the velocity and pressure fields are expressed in terms of high-order Lagrange interpolants of Legendre polynomials, at the Gauss--Lobatto--Legendre (GLL) quadrature points. In the present work we used the $\mathbb{P}_{N}-\mathbb{P}_{N-2}$ formulation, which implies that the velocity and pressure fields are expressed in terms of polynomials of order $N$ and $N-2$, respectively. The time discretization is based on an explicit third-order extrapolation for the nonlinear terms, and an implicit third-order backward differentiation for the viscous ones. The code is written in Fortran 77 and C and the message-passing-interface (MPI) is used for parallelism. 

\begin{figure}
\centering
\includegraphics[width=0.49\textwidth]{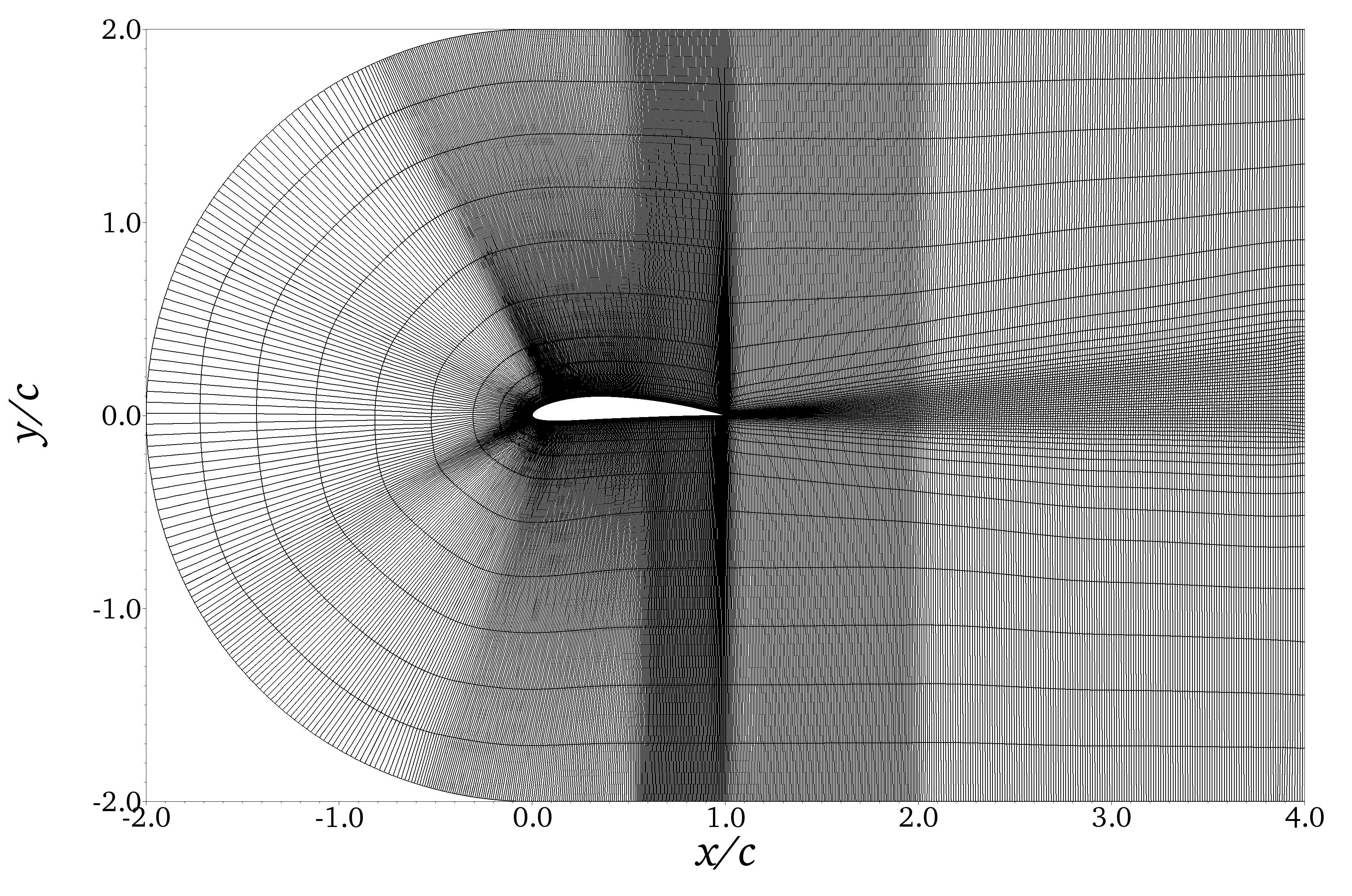}
\includegraphics[width=0.49\textwidth]{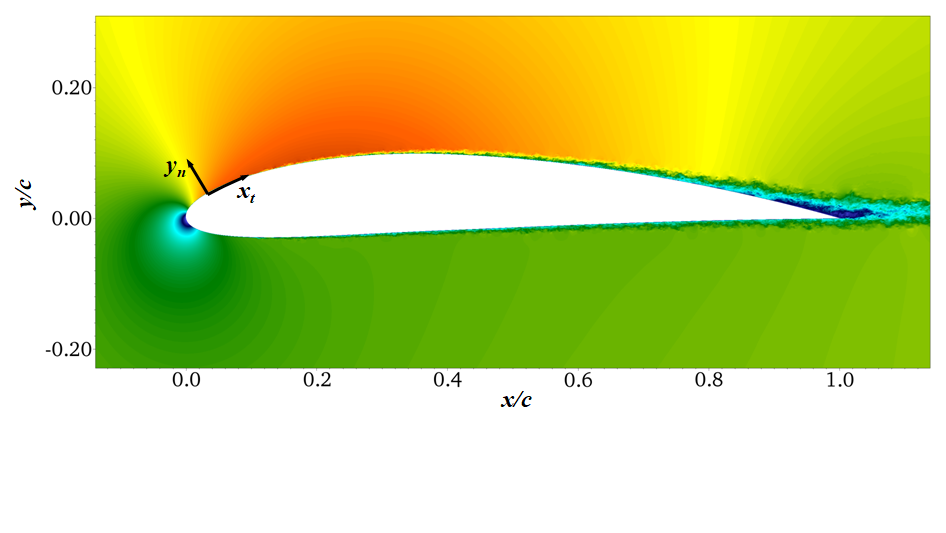}
\caption{(Left) Two-dimensional slice of the computational domain showing the spectral-element distribution, but not the individual GLL points. (Right) Instantaneous two-dimensional visualization of the horizontal velocity for the $Re_{c}=1,000,000$ case. Dark blue represents a horizontal velocity of $-0.1$ and dark red a value of $2$. The global frame of reference, as well as the local one expressed in the directions tangential and normal to the wing surface, are also shown. \textcolor{black}{ The $z$ direction, normal to the $xy$ plane, is common to the global and the local frames of reference.}}
\label{mesh_frame_reference}
\end{figure}

A total of four Reynolds numbers, namely $Re_{c}=100,000$, $200,000$, $400,000$ and $1,000,000$, is considered in the present study. The various cases, as well as their respective domain sizes and color codes for the rest of the manuscript, are summarized in Table \ref{wing_cases}. A two-dimensional slice of the computational domain employed to simulate the $Re_{c}=1,000,000$ case is shown in Figure \ref{mesh_frame_reference} (left), where $x$, $y$ and $z$ denote the horizontal, vertical and spanwise directions, respectively. In all the cases under consideration the domain is periodic in the spanwise direction, and a width of $L_{z}=0.2c$ was considered in the $Re_{c}=1,000,000$ wing. In this case, a total of $4.5$ million spectral elements was used to discretize the domain with a polynomial order $N=7$, which amounts to a total of 2.28 billion grid points. As in the DNS by \cite{hosseini_et_al}, a Dirichlet boundary condition extracted from an initial RANS (Reynolds-Averaged Navier--Stokes) simulation was imposed on all the boundaries except the outflow. In the present study, the boundary condition by \cite{dong_et_al} was employed at the outflow. The initial RANS simulation was carried out with the $k-\omega$ SST (shear-stress transport) model \citep{menter} implemented in the commercial software ANSYS Fluent. An angle of attack of $5^{\circ}$ was considered in all the Reynolds-number cases. An instantaneous two-dimensional visualization of the horizontal velocity around the wing section at $Re_{c}=1,000,000$ is shown in Figure \ref{mesh_frame_reference} (right), where both the stagnation point and the development of the TBLs on the suction and pressure sides can be observed. A local frame of reference expressed in the directions tangential and normal to the wing surface ({\it i.e.} given by the coordinates $x_{t}$ and $y_{n}$) is also shown in this figure.


\begin{table}
\scriptsize
\caption{Summary of wing cases under study and numerical details of the simulations. The normalized averaging eddy-turnover times ${\rm ETT}_{a}^{*}$ are obtained at $x_{ss}/c=0.7$.}
\label{wing_cases}
\centering
\begin{tabular}{c c c c c c c c c c c}
\hline\noalign{\smallskip}
$Re_{c}$ & $L_{x}/c$ & $L_{y}/c$ & $L_{z}/c$& $\Delta x_{t}^{+}$ & $\Delta y_{n,w}^{+}$ & $\Delta z^{+}$&Wake & Grid points & ${\rm ETT}_{a}^{*}$ & Color \\
\noalign{\smallskip}\hline \noalign{\smallskip}
$100,000$ & 5 & 4 & 0.1 & $18$ & $0.64$ & $9$ & $\Delta x / \eta <9$ & $48.4 \times 10^{6}$ & $16.2$ &  {\color{blue}\solid} \\
$200,000$ &  6 & 4 & 0.2 & $18$ & $0.64$ & $9$ & $\Delta x / \eta <9$ & $336 \times 10^{6}$ & $14.7$ &  {\color{red}\solid} \\
$400,000$ &  \textcolor{black}{ 6} & \textcolor{black}{ 4} & 0.1 & $18$ & $0.64$ & $9$ & $\Delta x / \eta <9$ & \textcolor{black}{ $463 \times 10^{6}$} & \textcolor{black}{ $10.3$} &  {\color{green}\solid} \\
$1,000,000$ & 6 & 4 & 0.2 & $27$ & $0.96$ & $13.5$ & $\Delta x / \eta <13.5$ & $2.28 \times 10^{9}$ & $10.5$ &  {\color{cyan}\solid} \\
\hline\noalign{\smallskip}
\end{tabular}
\end{table}

\begin{figure}
\centering
\includegraphics[width=0.49\textwidth]{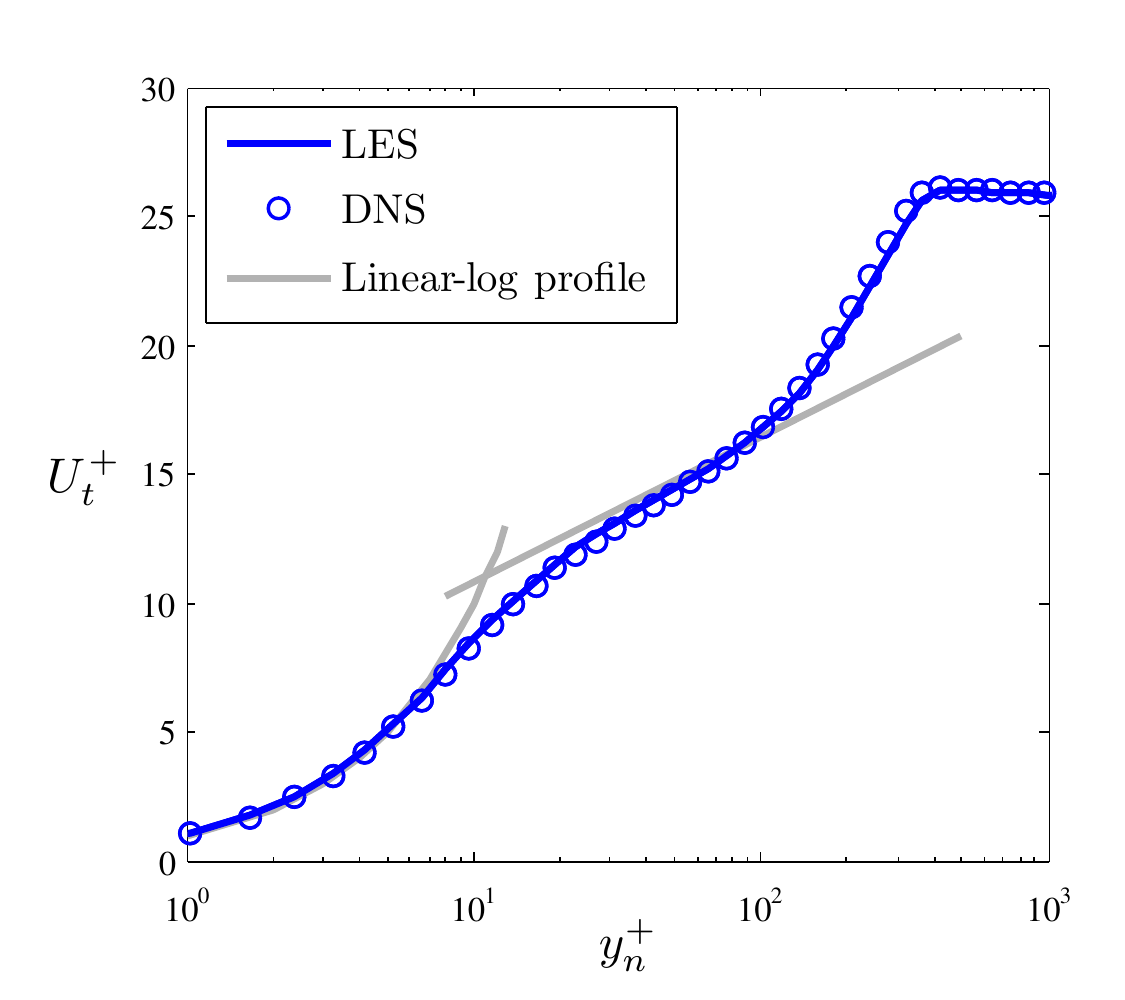}
\includegraphics[width=0.49\textwidth]{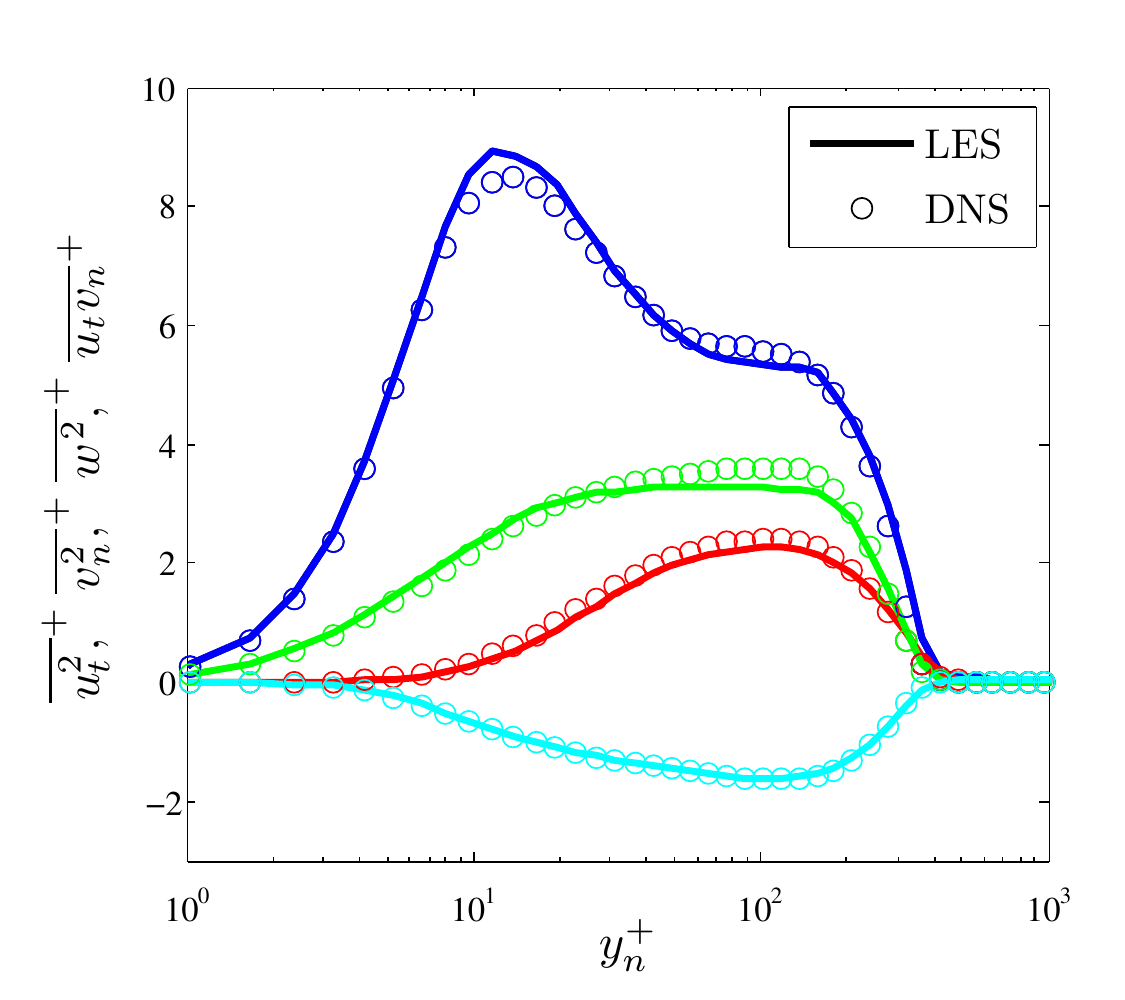}
\caption{Comparison of $Re_{c}=400,000$ DNS by \cite{hosseini_et_al} and LES with the resolution employed in the $Re_{c}=1,000,000$ case. (Left) Inner-scaled mean velocity profile and (right) components of the Reynolds-stress tensor at $x_{ss}/c=0.7$.}
\label{validation_N7}
\end{figure}

The LES approach is based on a relaxation-term (RT) filter, which provides an additional dissipative force in order to account for the contribution of the smallest, unresolved, turbulent scales \citep{schlatter_et_al_2004}. A validation of the method in turbulent channel flows and the flow around a NACA4412 wing section at $Re_{c}=400,000$ is given by~\cite{negi_et_al}. \textcolor{black}{ To give a quantification of the contribution of the subgrid-scale (SGS) model, around $90\%$ of the total dissipation of turbulent kinetic energy (TKE) is due to viscosity and thus resolved in the LES, whereas the remaining $10\%$ originates from the SGS model (evaluated at $x_{ss}/c=0.7$). The SGS model contribution is approximately the same as that in the ZPG TBL study by \cite{eitel_amor_et_al} and the work on APG TBLs by \cite{bobke_et_al}, using a similar RT-based LES approach.} \textcolor{black}{ The same resolution as that employed by \cite{negi_et_al} was considered in the present simulations from $Re_{c}=100,000$ to $400,000$ as shown in Table~\ref{wing_cases}.} In the $Re_{c}=1,000,000$ case, the mesh resolution around the wing follows these guidelines: $\Delta x^{+}_{t} < 27$, $\Delta y^{+}_{n,w} < 0.96$ (which is the distance between the wall and the first grid point in the wall-normal direction) and $\Delta z^{+} < 13.5$.  \textcolor{black}{ In this work} the superscript `+' denotes scaling in terms of the friction velocity $u_{\tau}=\sqrt{\tau_{w} / \rho}$ (with $\rho$ being the fluid density) \textcolor{black}{ and the viscous length $\ell^{*}=\nu / u_{\tau}$ (where $\nu$ is the fluid kinematic viscosity).} Regarding the wake region, we defined the criterion $\Delta x / \eta < 13.5$, where $\eta=\left ( \nu^{3} / \varepsilon \right )^{1/4}$ is the Kolmogorov scale \textcolor{black}{ ($\varepsilon$ the local isotropic dissipation)}. Note that the overall strategy to build the wing mesh is analogous to the one described by \cite{hosseini_et_al}. It can be observed in Table \ref{wing_cases} that \textcolor{black}{ for} $Re_{c}=1,000,000$ we employed a coarser resolution than in the three other LES cases due to its high computational cost. In order to assess the accuracy of this coarser resolution, we performed an LES \textcolor{black}{ at $Re_{c}=400,000$ with the same domain size as that in \cite{hosseini_et_al}, and compared the results with the ones from DNS \citep{wing_ftac}.} In Figure \ref{validation_N7} we compare the inner-scaled mean velocity profile and selected components of the Reynolds-stress tensor at $x_{ss}/c=0.7$ (note that $ss$ denotes suction side), expressed in the tangential and wall-normal frame of reference, from both simulations. This figure shows an excellent agreement in the mean velocity profile, as well as in the Reynolds-shear stress and in the outer region of the tangential velocity fluctuations. The effect of the reduced resolution is manifested in a small attenuation of the wall-normal and spanwise fluctuations, and in a slight over-prediction of the near-wall peak of $\overline{u^{2}_{t}}^{+}$. However, since the results presented below will be focused on the mean flow, the tangential velocity fluctuations and the Reynolds-shear stress, we consider that the present resolution is adequate for the scope of this study. The averaging times (after discarding initial transients) considered to compute turbulence statistics are reported in Table \ref{wing_cases} for all the wing cases in terms of normalized eddy-turnover times. The eddy-turnover time ${\rm ETT}=\delta_{99} / u_{\tau}$ \textcolor{black}{ (here $\delta_{99}$ is the $99\%$ boundary-layer thickness)} is the \textcolor{black}{ characteristic time scale} of the large scales, and as discussed by \cite{vinuesa_meccanica} it is possible to define a normalized ${\rm ETT}^{*}$, in which the length of the domain in the (homogeneous) periodic directions is taken into account. The complete process to compute statistics, which involves averaging in time and in the periodic spanwise direction, as well as tensor rotation to express all the terms in the tangential and wall-normal frame of reference, is described by \cite{wing_ftac}. The full-resolution runs were performed on the Cray XC40 system ``Beskow'' at the \textcolor{black}{ PDC Center for High-Performance Computing,} and on the IBM System ``MareNostrum'' at the \textcolor{black}{ BSC in Barcelona (Spain).} Depending on the size of the problem, we used 2048, 4096 or 8192 cores. 

\begin{figure}
\centering
\includegraphics[width=0.49\textwidth]{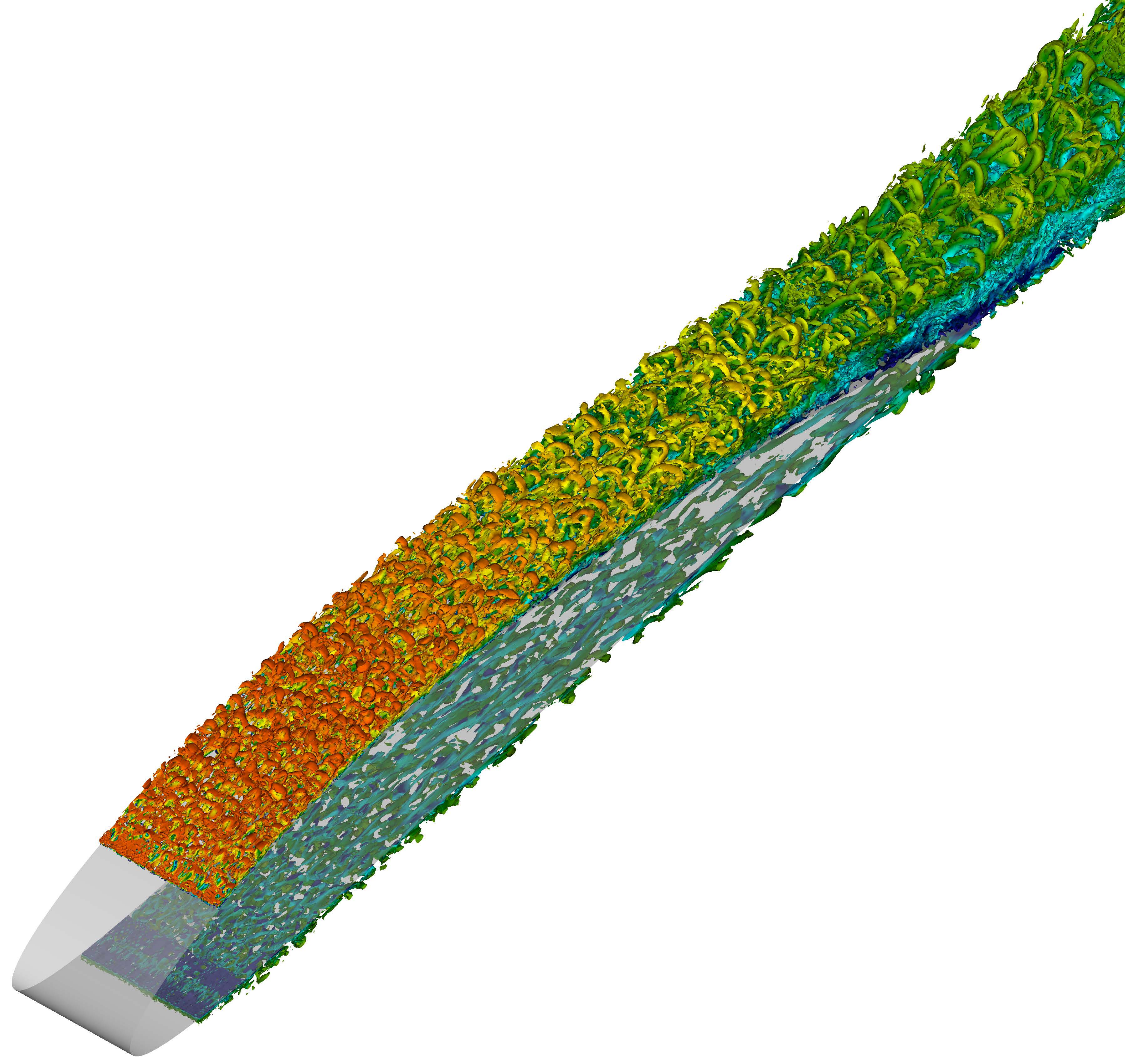}
\includegraphics[width=0.49\textwidth]{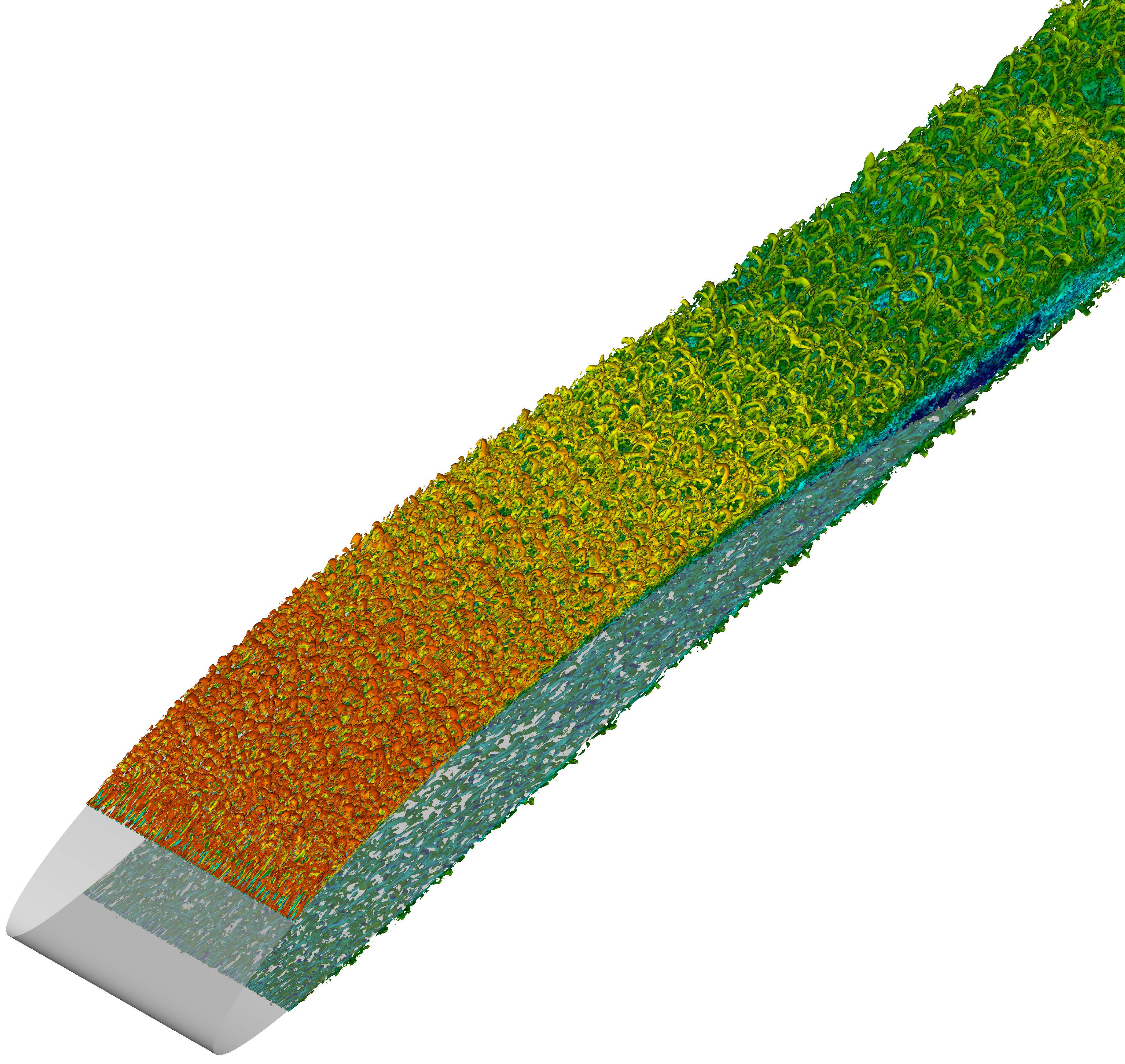}
\includegraphics[width=0.49\textwidth]{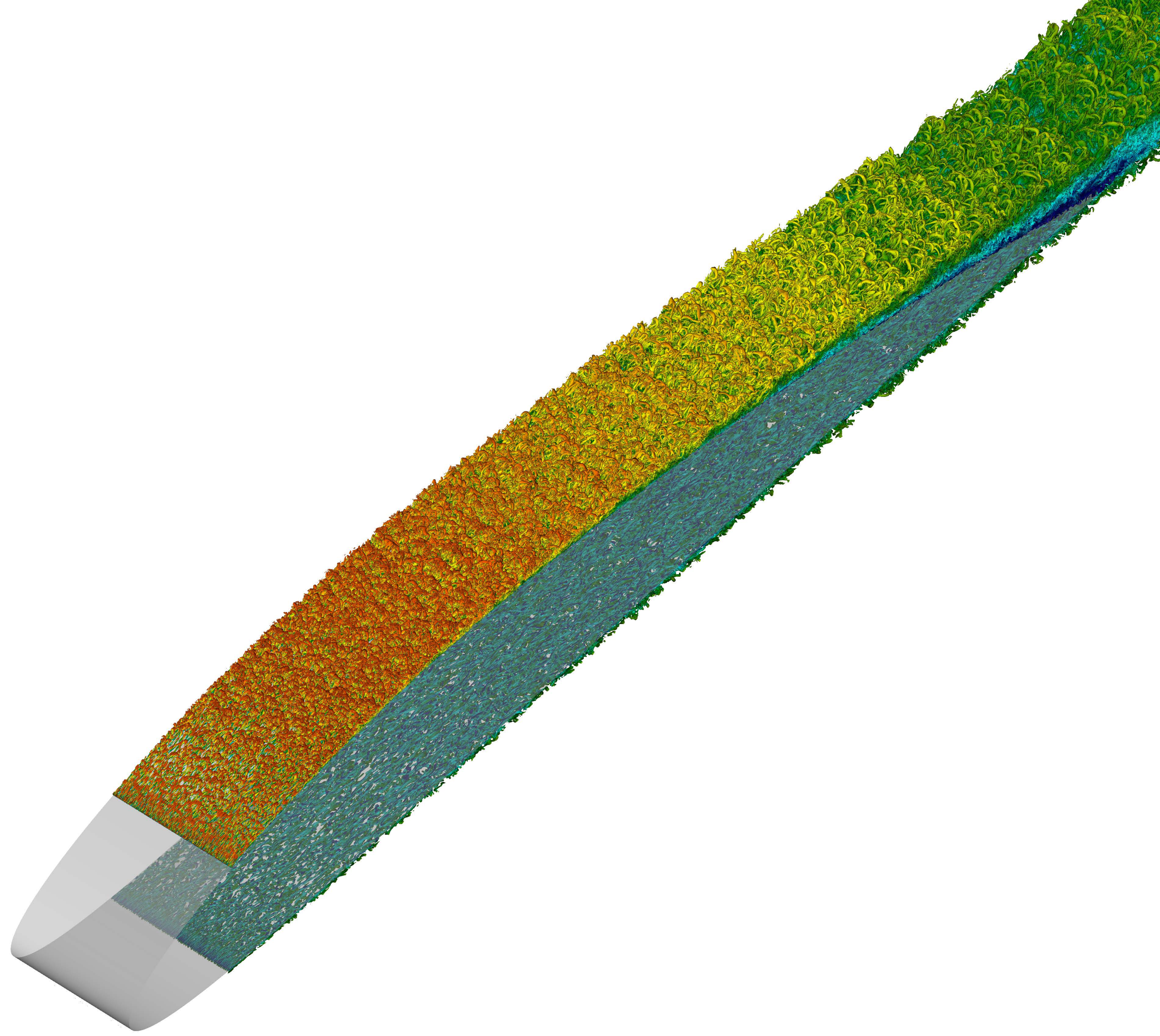}
\includegraphics[width=0.49\textwidth]{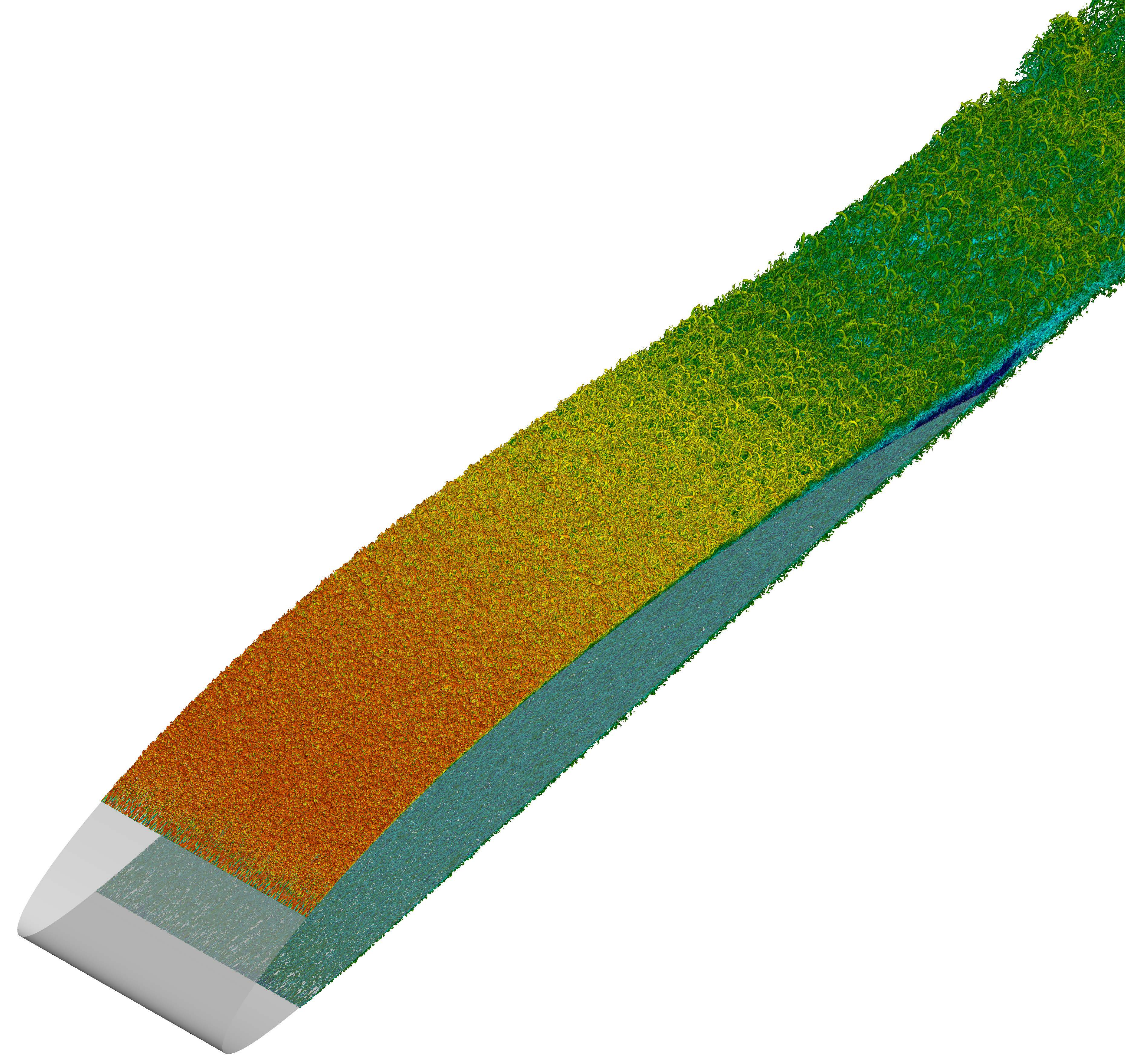}
\caption{Instantaneous visualizations showing coherent vortices identified using the $\lambda_{2}$ method \citep{jeong_hussain} in the four wing simulations under consideration. An inner-scaled isosurface of $\lambda_{2}^{+}=-10^{-4}$ (based on the value of $u_{\tau}$ at $x_{ss}/c=0.4$) is shown in all the cases. The structures are colored by the magnitude of the streamwise velocity, where dark blue denotes $-0.1$ and dark red $2$. The following cases are shown: (top-left) $Re_{c}=100,000$, (top-right) $Re_{c}=200,000$, (bottom-left) $Re_{c}=400,000$ and (bottom-right) $Re_{c}=1,000,000$.}
\label{flow_field}
\end{figure}

In Figure~\ref{flow_field} we show instantaneous visualizations of coherent vortices identified with the $\lambda_{2}$ method \citep{jeong_hussain} in the four wing cases under study. \textcolor{black}{ The validation of the numerical setup discussed above and in the work by~\cite{negi_et_al}, together with the smooth representation of the vortical structures inside the elements and across element boundaries observed in Figure~\ref{flow_field}, suggest that the present LES approach is adequate to simulate the flow.} The boundary layers on the suction and pressure sides were tripped in all the cases using the volume-force method described by \cite{schlatter_orlu12}, at $x/c=0.1$. This figure also illustrates how the scale separation increases with Reynolds number, for a flow case where the geometry is fixed. Whereas the very low $Re_{c}=100,000$ case exhibits predominance of hairpin-like structures \citep{theodorsen,adrian,schlatter_hairpin} \textcolor{black}{ over a large portion of the chord length,} these become progressively less common as $Re$ increases, being essentially absent at $Re_{c}=1,000,000$, except for the region close to the tripping \textcolor{black}{ \citep{eitel_amor_et_al_2}}. It is interesting to note how the vortical structures are affected by the pressure-gradient distributions on both sides: on the suction side the progressively stronger APG leads to a thicker boundary layer with \textcolor{black}{ large structures}, whereas on the pressure side the slight FPG produces a thinner boundary layer with weaker turbulent motions. In particular, the low $Re$ in the $Re_{c}=100,000$ case, combined with the slight FPG, lead to partial relaminarization of the boundary layer on the pressure side. 

The lift and drag coefficients in airfoils are defined as $C_{l}=l/ \left (1/2 \rho U_{\infty}^{2} c \right )$ and $C_{d}=d/ \left (1/2 \rho U_{\infty}^{2} c \right )$, where $l$ and $d$ are the lift and drag forces per unit span, respectively. In Table~\ref{Cl_Cd} we show the values of $C_{l}$ and $C_{d}$ obtained through integration of the viscous and pressure forces around the wing section, in the four cases under consideration. \textcolor{black}{ In this table we also introduce the case names, from W1 to W10, which will be used in the subsequent discussions.} These values show an increase in $C_{l}$ and a decrease in $C_{d}$ with Reynolds number, where changes become progressively smaller as the $Re_{c}=1,000,000$ case is approached. As a consequence, the aerodynamic efficiency (defined as the ratio $C_{l}/C_{d}$) also increases with Reynolds number, and the efficiency at $Re_{c}=1,000,000$ (63.7) exceeds by a factor of 1.85 the one at $Re_{c}=100,000$ (34.4). This conclusion shows that the NACA4412 wing section is designed to operate at higher Reynolds numbers than the ones under consideration in the present work. \textcolor{black}{ The lift and drag coefficients reported in Table~\ref{Cl_Cd} for the W4 case differ slightly from the ones reported by \cite{hosseini_et_al} in their DNS at the same $Re_{c}$. This is due to the fact that the present LES was performed with an extent of the domain in the vertical direction of $L_{y}/c=4$, whereas in the DNS this height was $L_{y}/c=2$. Note that the LES validation by \cite{negi_et_al} was performed with $L_{y}/c=2$, and the agreement with the DNS was excellent. In this study we considered $L_{y}/c=4$ for all the cases because according to \cite{vinuesa_negi_lic} this height eliminates the minor effect of the boundary condition on the wing statistics present in the case with $L_{y}/c=2$.} The impact of the pressure-gradient distribution on the features of the TBL on the suction side of the wing will be assessed in the next section.
\begin{table}
\scriptsize
\caption{Lift and drag coefficients, and aerodynamic efficiency, of the four wing cases under study.}
\label{Cl_Cd}
\centering
\begin{tabular}{c c c c c}
\hline\noalign{\smallskip}
\textcolor{black}{ Case name} & $Re_{c}$ & $C_{l}$ & $C_{d}$ & $C_{l}/C_{d}$  \\
\noalign{\smallskip}\hline \noalign{\smallskip}
\textcolor{black}{ W1} & $100,000$ & $0.8315$ & $0.0242$ & $34.4$ \\
\textcolor{black}{ W2} & $200,000$ & $0.8854$ & $0.0185$ & $47.9$ \\
\textcolor{black}{ W4} & $400,000$ & \textcolor{black}{ $0.9041$} & \textcolor{black}{ $0.0162$} & \textcolor{black}{ $55.8$} \\
\textcolor{black}{ W10} & $1,000,000$ & $0.9112$ & $0.0143$ & $63.7$ \\
\hline\noalign{\smallskip}
\end{tabular}
\end{table}


\section{Results and discussion} \label{suction_section}
  
\begin{figure}
\centering
\includegraphics[width=0.65\textwidth]{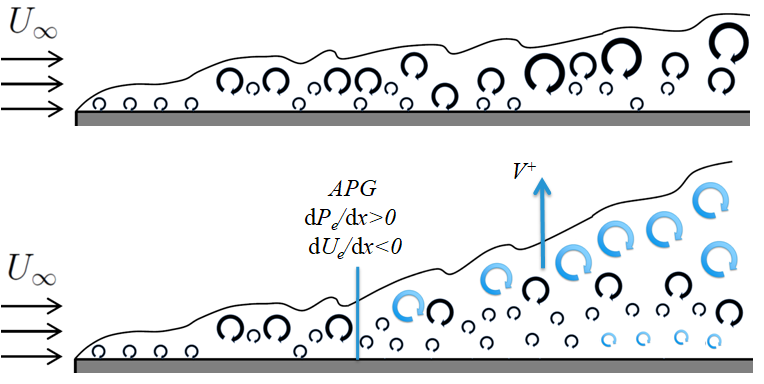}
\includegraphics[width=0.34\textwidth]{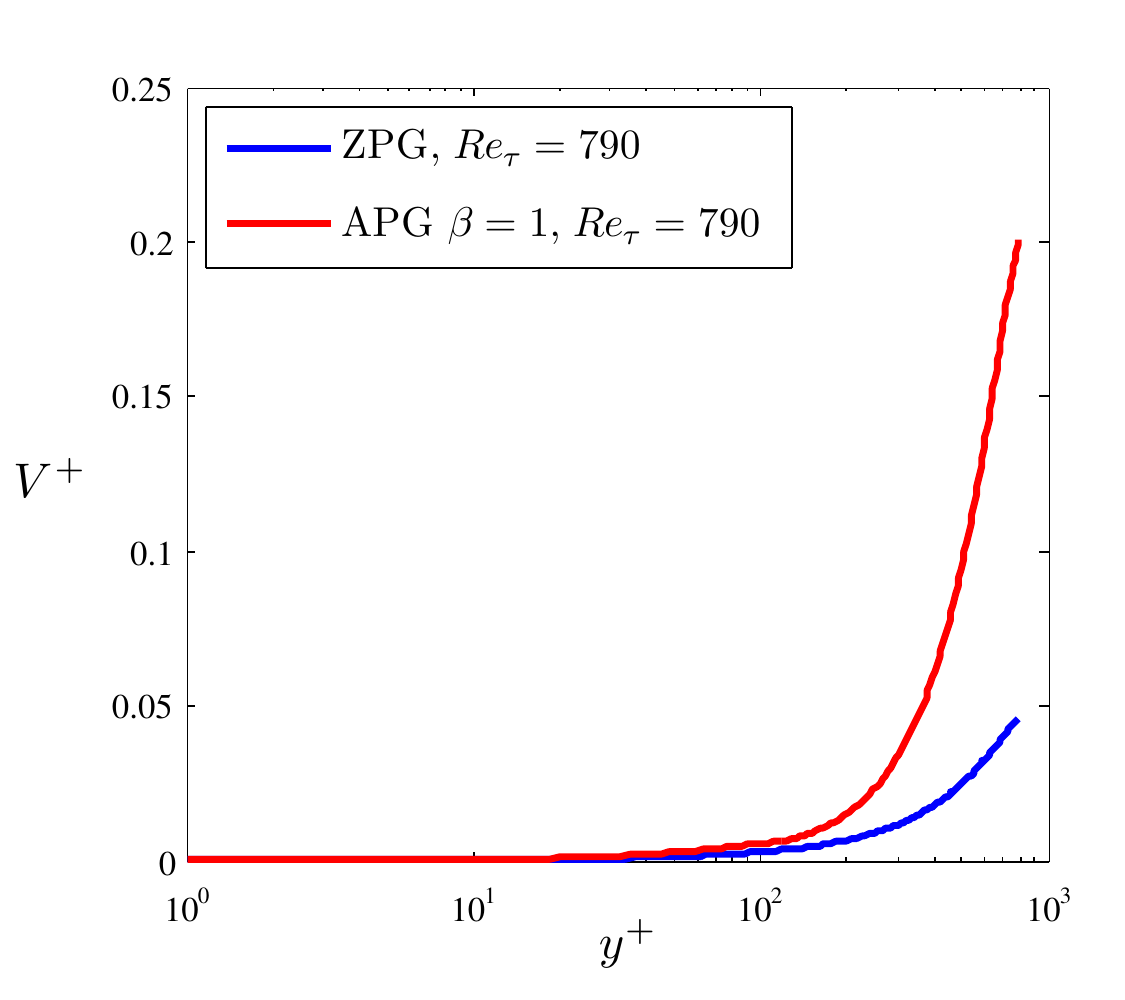}
\caption{(Left) Sketch showing the streamwise development of (top) a ZPG and (bottom) an APG TBL. In black we represent turbulent structures characteristic of ZPGs, and in blue the ones affected by the wall-normal convection in APG TBLs. (Right) Comparison of inner-scaled wall-normal velocity profile at $Re_{\tau}=790$ from a ZPG \citep{schlatter_orlu10} and an APG TBL \citep{bobke_et_al}. Profiles truncated at the boundary-layer edge.}
\label{sketch_Vp}
\end{figure}

As discussed in the Introduction, the aim of the current study is to investigate the Reynolds-number effects on APG TBLs subjected to \textcolor{black}{ approximately} the same $\beta(x)$ distribution. In particular, we aim to assess such effects on the turbulent boundary layer developing on the suction side of a NACA4412 wing section with $5^{\circ}$ angle of attack. To this end, we compare the results from the various numerical databases summarized in Table \ref{wing_cases}, with $Re_{c}$ ranging from $100,000$ to $1,000,000$. In Figure \ref{sketch_Vp} (left) we show a sketch comparing the streamwise development of a ZPG and an APG TBL. In the ZPG case, the increase in local Reynolds number in the streamwise direction produces the development of the outer region in the boundary layer, with progressively more energetic large-scale motions \citep{eitel_amor_et_al,sillero_et_al}. This figure also shows that when a streamwise APG is imposed, the wall-normal convection increases significantly, leading to a more pronounced growth of the boundary layer, and consequently to a much more prominent outer region. The APG \textcolor{black}{ also produces energetic large-scale structures} in the outer region \citep{harun_et_al,wing_ftac}, and as discussed by \cite{maciel_et_al} these large structures are taller, shorter in the streamwise direction, and they form a larger angle with respect to the wall than their ZPG counterparts due to the effect of the wall-normal convection. As shown in Figure \ref{sketch_Vp} (right), the inner-scaled wall-normal velocity is significantly higher in an APG TBL than in a ZPG, where the former is around 4 times larger than the latter in the boundary-layer edge. In order to avoid the effect of flow history, in Figure \ref{sketch_Vp} (right) we show $V^{+}$ profiles for a ZPG TBL ({\it i.e.} a TBL with a constant value of $\beta=0$) from the database by \cite{schlatter_orlu10}, and for an APG with a constant value of $\beta(x)=1$ \citep{bobke_et_al}, both at the same friction Reynolds number $Re_{\tau}=790$. Note that $Re_{\tau}=\delta_{99} u_{\tau} / \nu$ is defined in terms of the $99\%$ boundary-layer thickness $\delta_{99}$,  which was determined following the method described by \cite{vinuesa_diagnostic} for pressure-gradient TBLs. \textcolor{black}{ In the rest of this section, we analyze the boundary-layer development and the profiles of mean velocity and Reynolds stresses.}

\subsection{Boundary-layer development}
\begin{figure}
\centering
\includegraphics[width=0.65\textwidth]{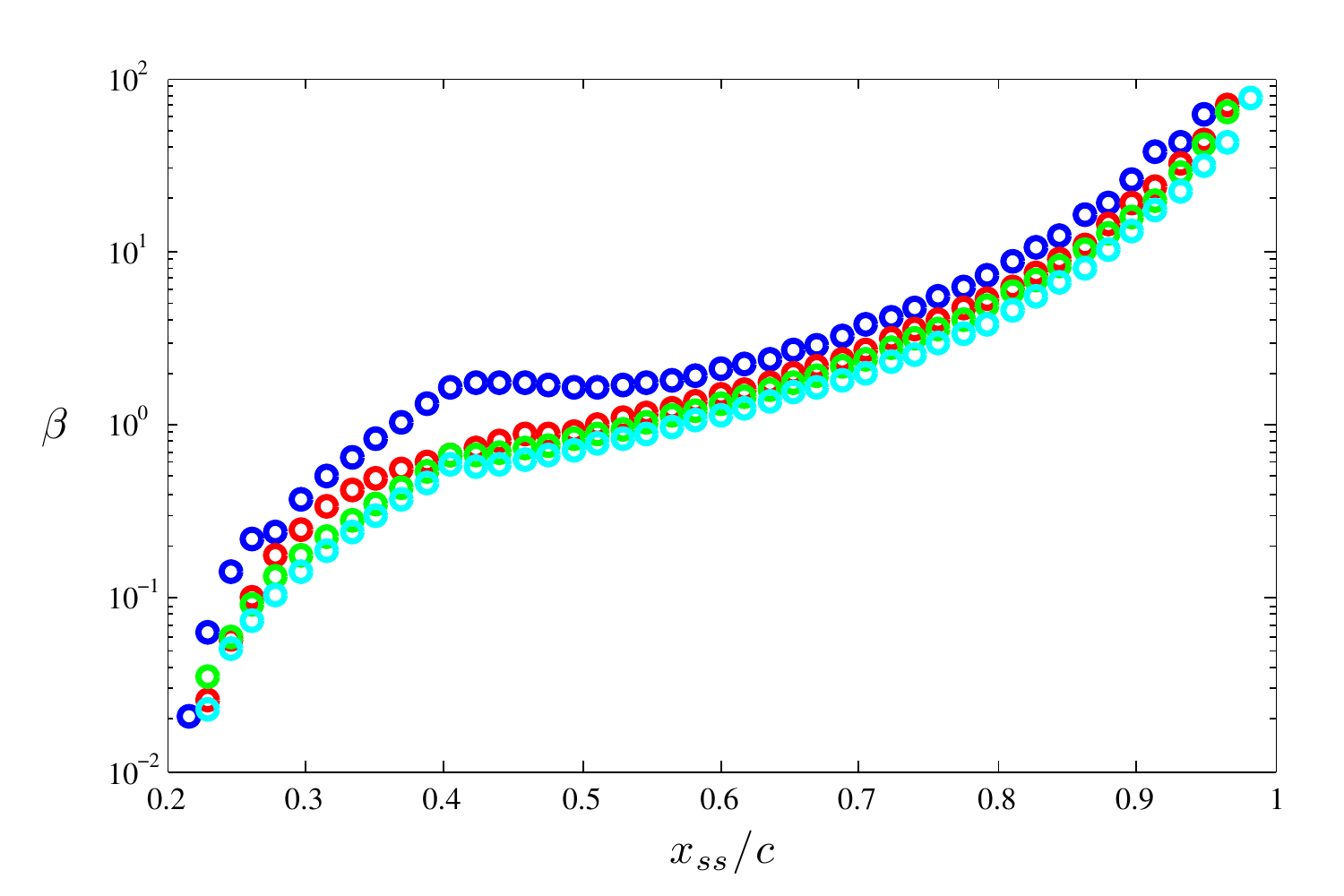}
\includegraphics[width=0.65\textwidth]{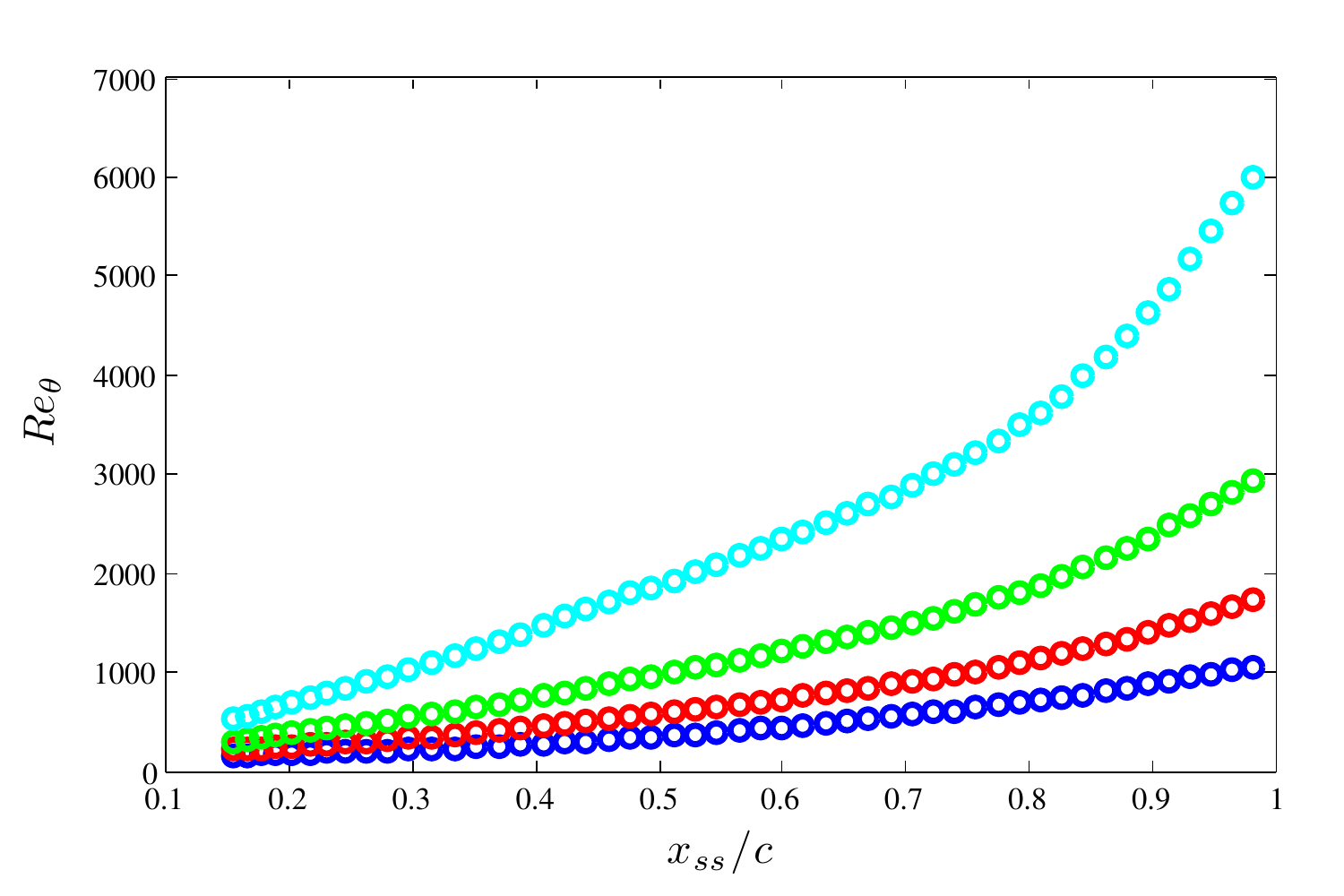}
\includegraphics[width=0.65\textwidth]{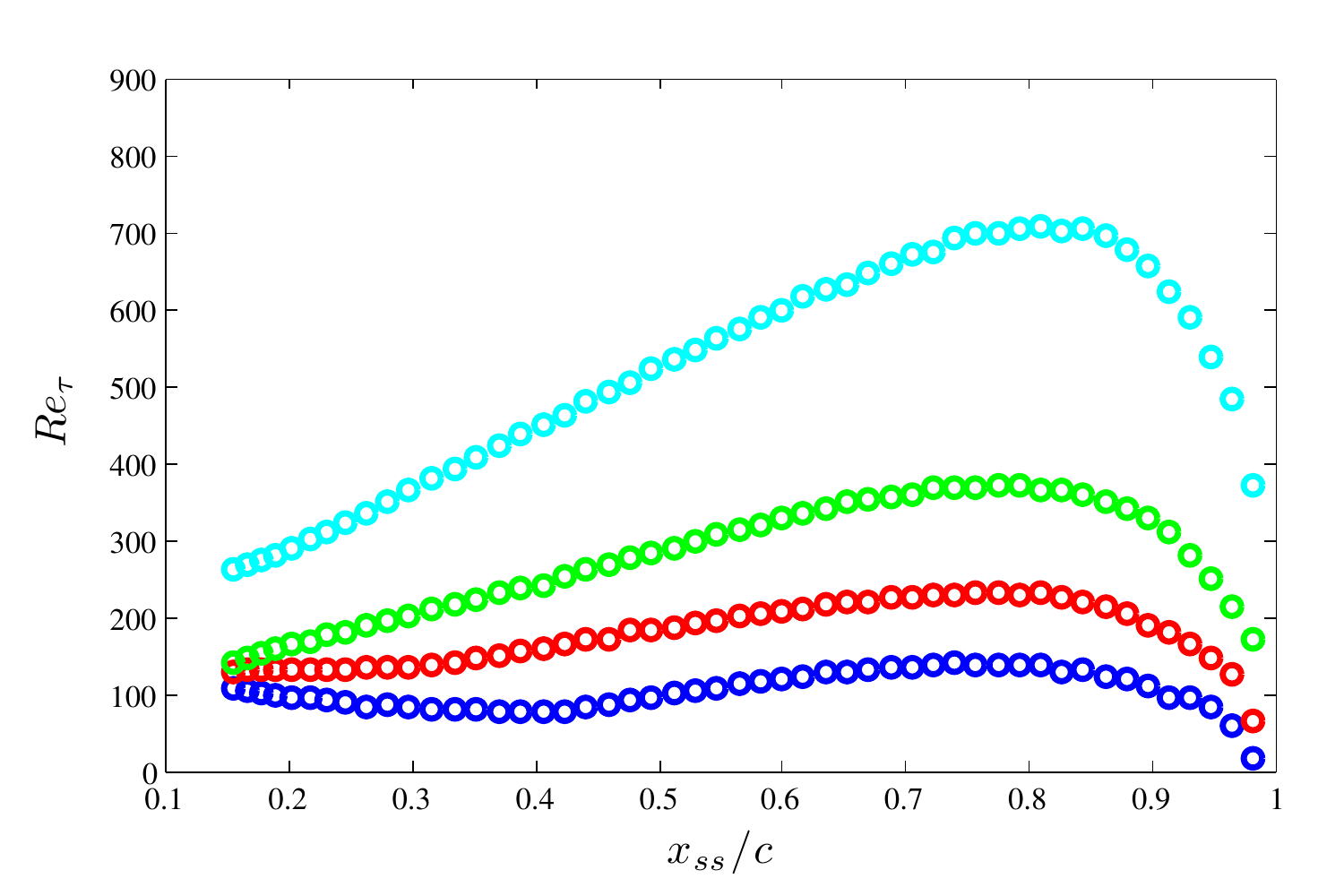}
\caption{Streamwise evolution of (top) the Clauser pressure-gradient parameter $\beta$, (middle) the Reynolds number based on momentum thickness $Re_{\theta}$ and (bottom) the friction Reynolds number $Re_{\tau}$ on the suction side of the wing. The colors correspond to the cases summarized in \textcolor{black}{ Table~\ref{wing_cases}.}}
\label{beta_Reth_Ret}
\end{figure}

In Figure \ref{beta_Reth_Ret} (top) we show the streamwise evolution of the Clauser pressure-gradient parameter $\beta$ for the TBLs on the suction side of the four wing cases under study. As expected \citep{pinkerton}, the three higher-$Re$ boundary layers are subjected to \textcolor{black}{ approximately the same} $\beta(x)$ distributions, \textcolor{black}{ with some relative differences  (on the order of $10\%$)} only arising beyond $x_{ss}/c > 0.9$. The \textcolor{black}{ W1} case shows slightly higher $\beta$ values throughout the wing chord, especially at $x_{ss}/c=0.4$, a fact that is connected to the low Reynolds number of this case. Note that all the boundary layers are subjected to conditions close to zero pressure gradient up to $x_{ss}/c \simeq 0.3$, the point after which the value of $\beta$ increases beyond 0.1. In the next section we will study the velocity profiles at $x_{ss}/c=0.4$ and 0.7, where the local pressure-gradient magnitude is moderate ($\beta \simeq 0.6$) and strong ($\beta \simeq 2$), respectively. Although the value of $\beta$ increases throughout the whole suction side of the wing, an inflection point is observed at $x_{ss}/c=0.4$, which is the point of maximum thickness of the NACA4412 airfoil. Beyond this point, the rate of change of $\beta$ increases significantly with \textcolor{black}{ $x_{ss}$}, a fact that is explained by the progressive reduction in airfoil thickness, which produces a larger increase in streamwise adverse pressure gradient.

In Figure \ref{beta_Reth_Ret} (middle) and (bottom) we show the streamwise evolution of the Reynolds number based on momentum thickness $Re_{\theta}$, and the friction Reynolds number $Re_{\tau}$. \textcolor{black}{ Note that $Re_{\theta}$ is calculated in terms of the edge velocity $U_{e}$ and the momentum thickness $\theta$. As mentioned above, the method proposed by \cite{vinuesa_diagnostic}, based on the diagnostic scaling, is used to calculate the $99\%$ boundary-layer thickness $\delta_{99}$ in the present APG TBLs. Then, the momentum thickness is obtained through integration up to $y_{n}=\delta_{99}$ for each profile under consideration.} The $Re_{\theta}$ trends show a monotonic increase for all four boundary layers, due to the fact that both Reynolds number and APG promote the increase of the boundary-layer thickness. In particular, the thickening experienced by the TBLs due to the APG significantly increases $Re_{\theta}$ in all the cases, with maximum values of $Re_{\theta}=1,050$, $1,730$, \textcolor{black}{ 2,930} and $6,000$ (from low to high $Re_{c}$), all of them observed close to the trailing edge. Regarding the friction Reynolds number, \textcolor{black}{ in all the} boundary-layer cases the maximum is located at $x_{ss} / c\simeq 0.8$, and not at the trailing edge as for $Re_{\theta}$. This is due to the fact that, although the APG increases the boundary-layer thickness, it also decreases the wall-shear stress; thus, the very strong APGs beyond $x_{ss} /c \simeq 0.8$ (where $\beta \simeq 4.5$ in the three higher-$Re$ cases) produce a larger reduction in $u_{\tau}$ than the increase in $\delta_{99}$. The maximum $Re_{\tau}$ values are $139$, $231$, \textcolor{black}{ 366} and $707$ for the various wing cases. The decreasing trend for the \textcolor{black}{ W1} case up to $x_{ss} \simeq 0.4$ is associated to the very low Reynolds number of this \textcolor{black}{ case}, and will be further discussed below. 



\begin{figure}
\centering
\includegraphics[width=0.65\textwidth]{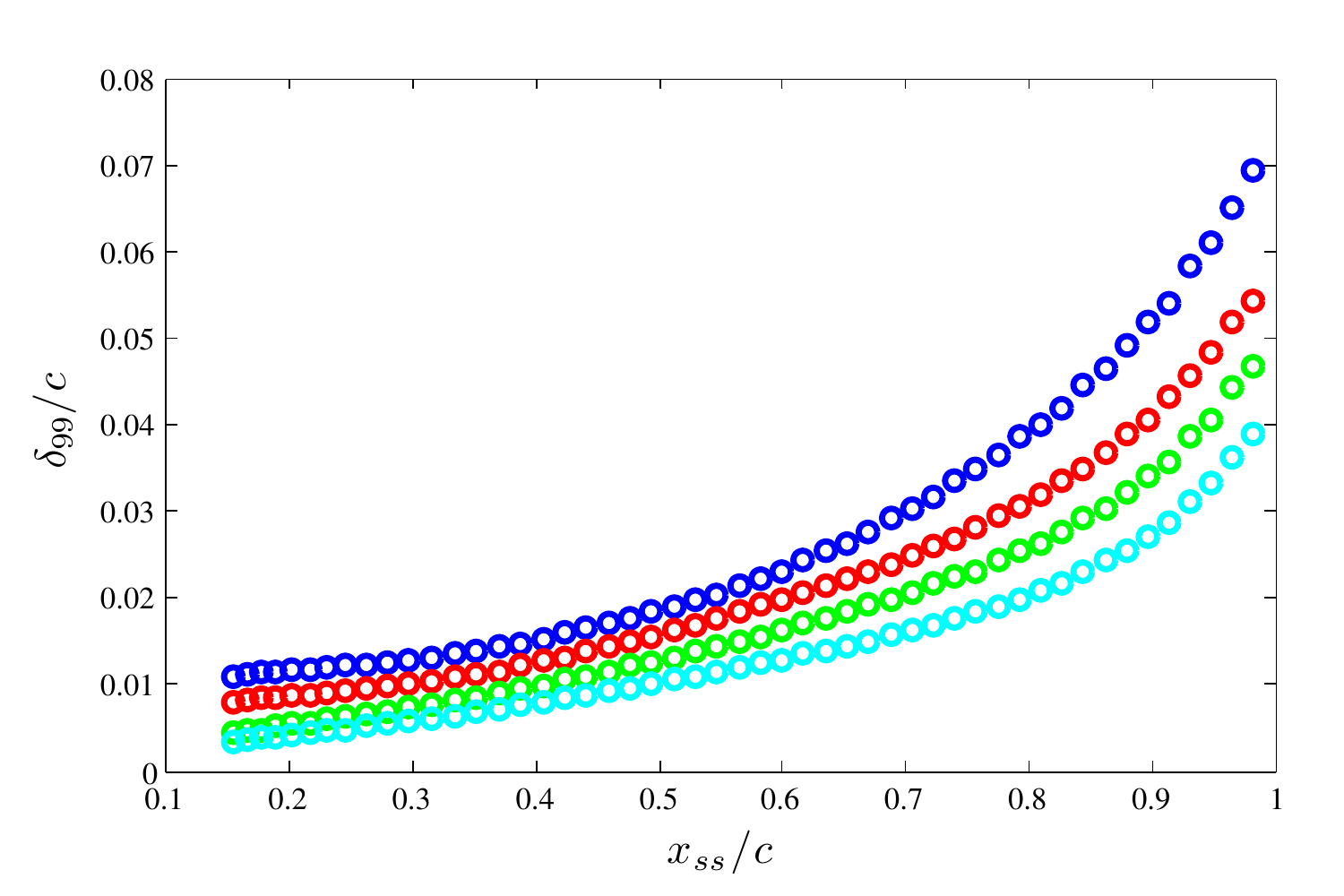}
\caption{\textcolor{black}{ Streamwise evolution of the $99\%$ boundary-layer thickness on the wing suction side. The colors correspond to the cases summarized in Table~\ref{wing_cases}.}}
\label{delta99_fig}
\end{figure}

\textcolor{black}{ The $99\%$ boundary-layer thicknesses from the four wings are shown, as a function of $x_{ss}$, in Figure~\ref{delta99_fig}. It can be clearly observed that, as expected, the boundary-layer thickness decreases with Reynolds number. All the $\delta_{99}$ values increase with $x_{ss}$ as a result of the wall-normal convection induced by the APG, and the most pronounced growth rate is observed for $x_{ss}/c > 0.8$. The method used to calculate $\delta_{99}$ was thoroughly validated in ZPG TBLs by \cite{vinuesa_diagnostic}, who compared it with the results obtained from composite profiles \citep{nickels} and the technique based on the intermittency factor by \cite{li_schlatter}. The study by \cite{vinuesa_diagnostic} concluded that the present method, based on the diagnostic scaling, provides the most robust determination of the boundary-layer edge in PG TBLs.}

\begin{figure}
\centering
\includegraphics[width=0.65\textwidth]{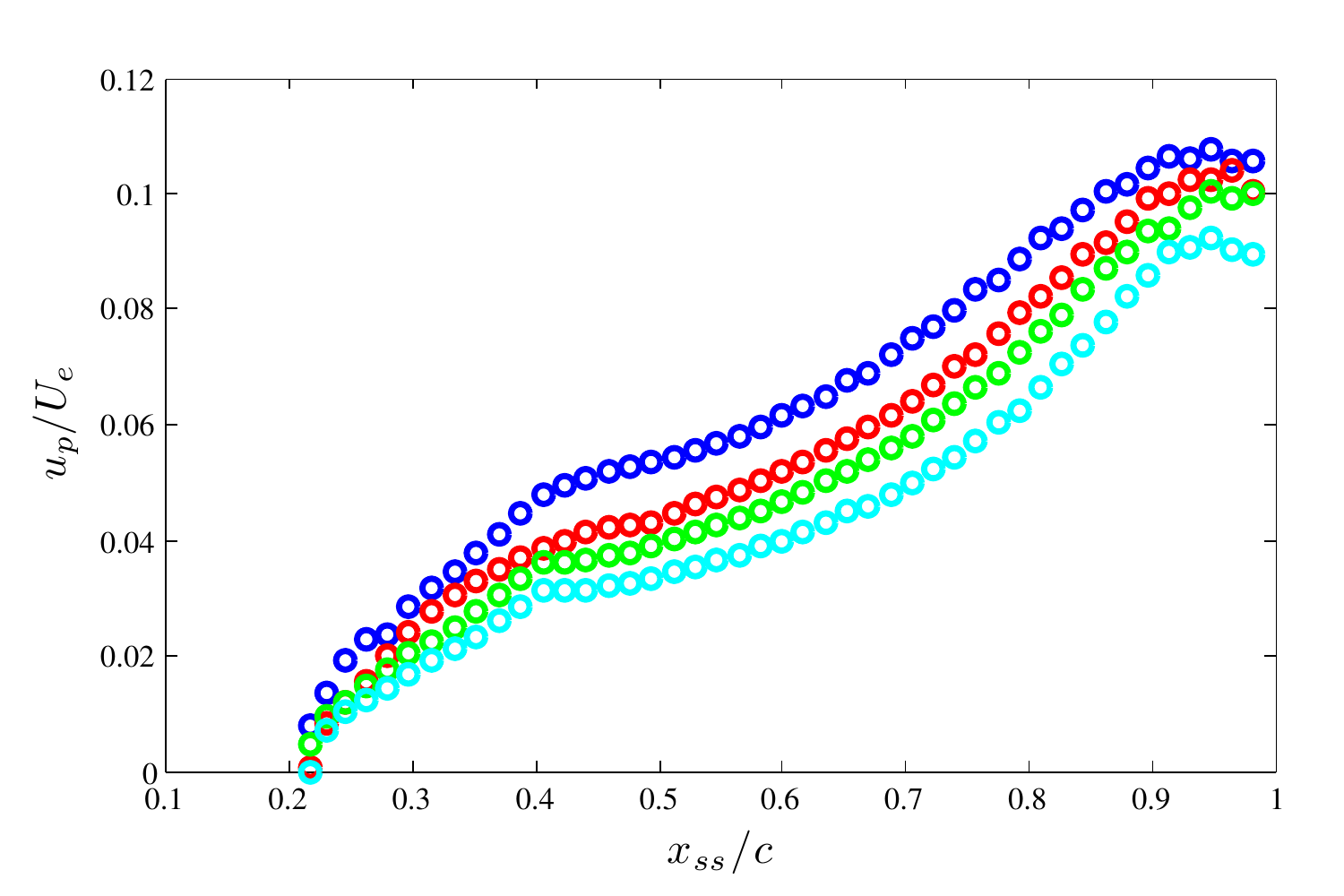}
\caption{\textcolor{black}{ Streamwise evolution of the pressure velocity scaled with the edge velocity on the suction side of the wing. The colors correspond to the cases in Table~\ref{wing_cases}.}}
\label{up_Ue_fig}
\end{figure}

\textcolor{black}{ The pressure-gradient distributions based on $\beta$ shown in Figure~\ref{beta_Reth_Ret}~(top) indicate that the APG increases significantly towards the wing trailing edge. Since the local wall-shear stress reduces as the trailing edge is approached (and $\beta$ is inversely proportional to $u_{\tau}$), it is interesting to analyze the APG magnitude in outer scaling. \cite{mellor_gibson} defined the pressure velocity $u_{p}=\sqrt{\delta^{*} / \rho {\rm d}P_{e} / {\rm d} x}$, which is related to the Clauser pressure-gradient parameter as $\beta^{2}=u_{p}/u_{\tau}$. Similarly, it is possible to define an outer-scaled APG parameter as $u_{p}/U_{e}$, as in the study by \cite{kitsios_et_al}, who also considered a strongly-decelerated APG with $\beta=39$. The ratio $u_{p}/U_{e}$ is shown in Figure~\ref{up_Ue_fig} for the four wing cases under consideration. It can be observed that although the outer-scaled APG increases in all the cases throughout the wing suction side, beyond $x_{ss}/c \simeq 0.95$ the curves reach approximately constant values, with a subtle decrease in the highest-$Re$ case. This indicates that the very small local wall-shear stress is responsible for the significant increase in $\beta$ very close to the trailing edge. Another relevant observation from the behavior of $u_{p}/U_{e}$ is the fact that the outer-scaled APG magnitude does not collapse for the various wing cases, as it was the case for $Re_{c} \geq 200,000$ when using $\beta$. The use of $\beta$ to characterize the pressure-gradient magnitude and to identify near-equilibrium TBLs was introduced by \cite{clauser1,clauser2}, together with \cite{mellor_gibson}, who developed a theoretical framework for such TBLs. Although in the very strong APG conditions beyond $x_{ss}/c \simeq 0.95$ the $\beta$ parameter may not be adequate to characterize the effect of the pressure gradient on the TBL (as in the work by \cite{kitsios_et_al}), the APG on the flow upstream can be assessed in terms of $\beta$. The fact that the lower-$Re$ wings are subjected to a larger $u_{p}/U_{e}$ ratio is connected to their stronger wall-normal velocity, as discussed below. Therefore, in the following we will analyze the various APG TBLs in terms of the $\beta(x)$ distributions, focusing on the region with $x_{ss}/c < 0.95$.}

\begin{figure}
\centering
\includegraphics[width=0.65\textwidth]{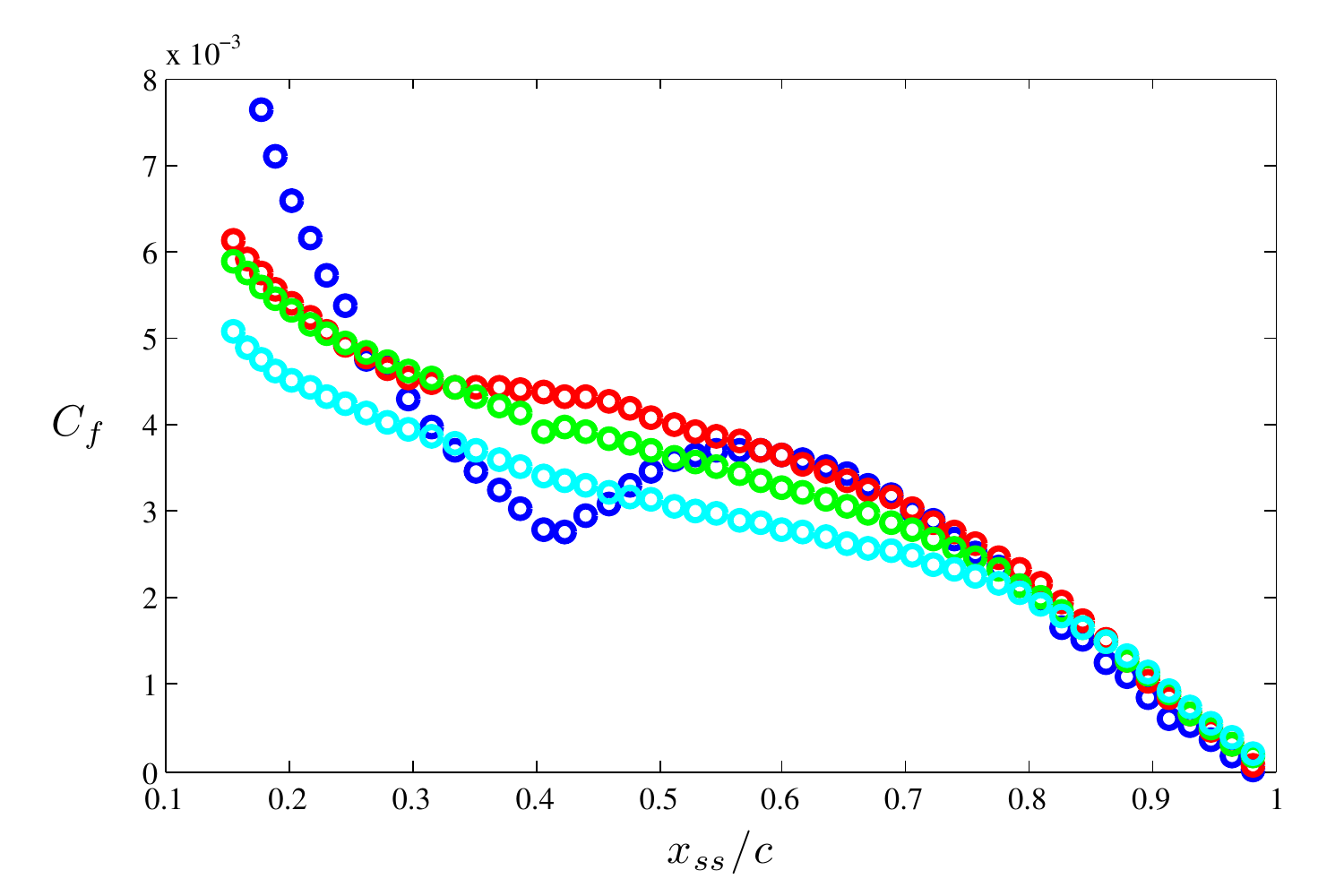}
\includegraphics[width=0.65\textwidth]{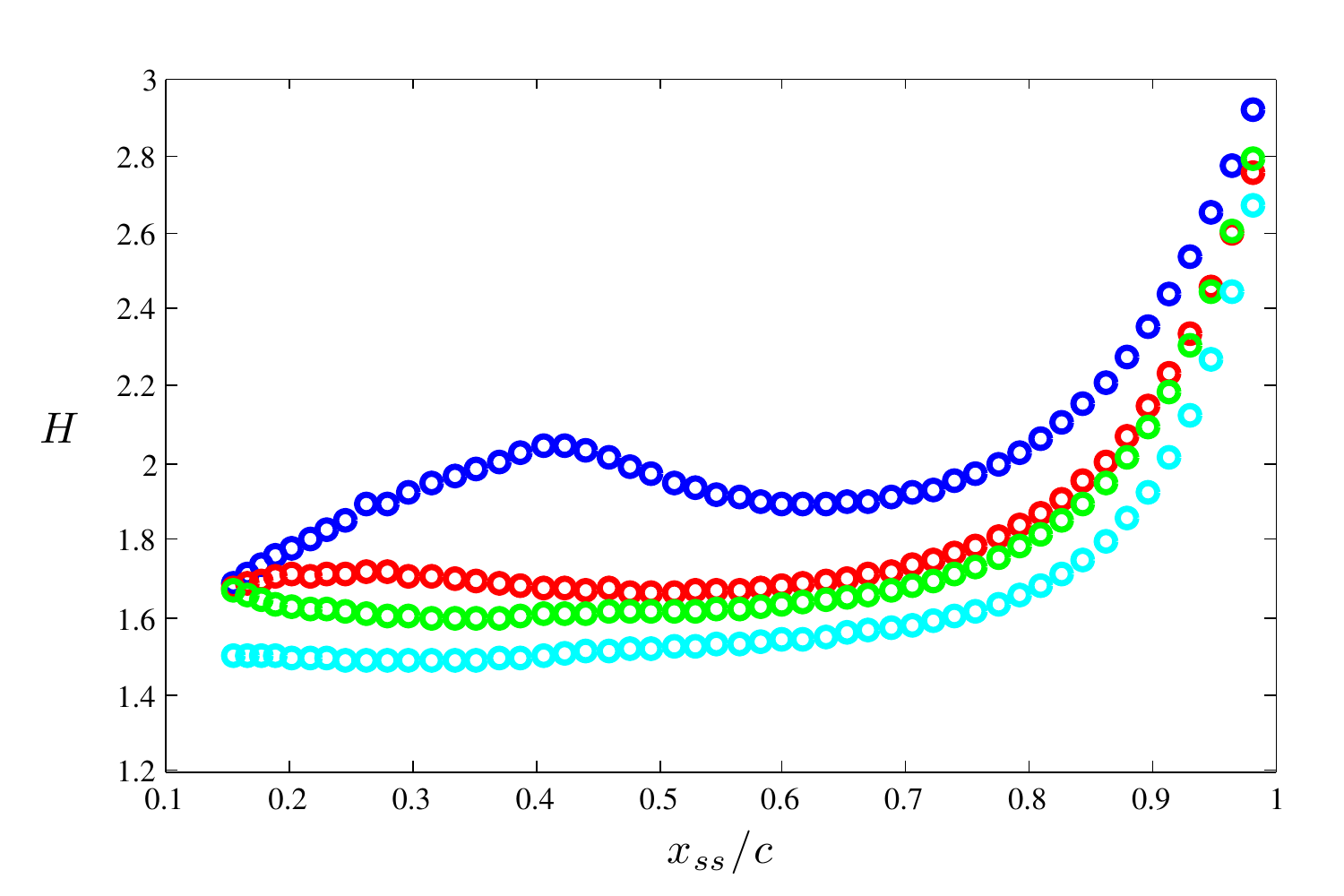}
\caption{Streamwise evolution of (top) the skin-friction coefficient $C_{f}$ and (bottom) the shape factor $H$ on the suction side of the wing. The colors correspond to the cases summarized in Table \ref{wing_cases}.}
\label{Cf_H}
\end{figure}

The skin-friction coefficient $C_{f}= 2 \left (u_{\tau} / U_{e} \right )^{2}$ (where $U_{e}$ is the velocity at the boundary-layer edge) and the shape factor $H=\delta^{*} / \theta$ are shown, as a function of the streamwise position on the suction side of the wing, in Figure \ref{Cf_H}. It can be observed that the \textcolor{black}{ W2} curve is slightly above the one of the \textcolor{black}{ W4 case,} which in turn also exhibits larger values than that of the \textcolor{black}{ W10 case.} Interestingly, the differences between these cases are significantly reduced beyond $x_{ss}/c \simeq 0.9$. Since these three boundary layers are subjected to \textcolor{black}{ approximately} the same $\beta(x)$ distribution, it can be argued that the differences between the various curves are due to Reynolds-number effects, a fact that is consistent with what is observed in ZPG TBLs since $C_{f}$ decreases with $Re$. \textcolor{black}{ The} effect of Reynolds number becomes essentially negligible beyond $x_{ss}/c \simeq 0.9$, where the very strong APG conditions (with a value of $\beta \simeq 14$ at $x_{ss}/c=0.9$) define the state of the boundary layer. Regarding the shape factor, note that APG and Reynolds number have opposite effects on a TBL: whereas the former increases $H$ (due to the thickening of the boundary layer), the latter decreases the shape factor. This can also be observed in Figure \ref{Cf_H} (bottom), where the $H$ curve from the \textcolor{black}{ W10 case} is below the one from the \textcolor{black}{ W4} throughout the whole suction side of the wing. The \textcolor{black}{ W2} curve is above the \textcolor{black}{ W4} one. Note that, since the three boundary layers are subjected to essentially the same pressure-gradient effects, the lower values of $H$ are produced by the higher Reynolds number, again consistent with what is observed in ZPGs.

\begin{figure}
\centering
\includegraphics[width=0.49\textwidth]{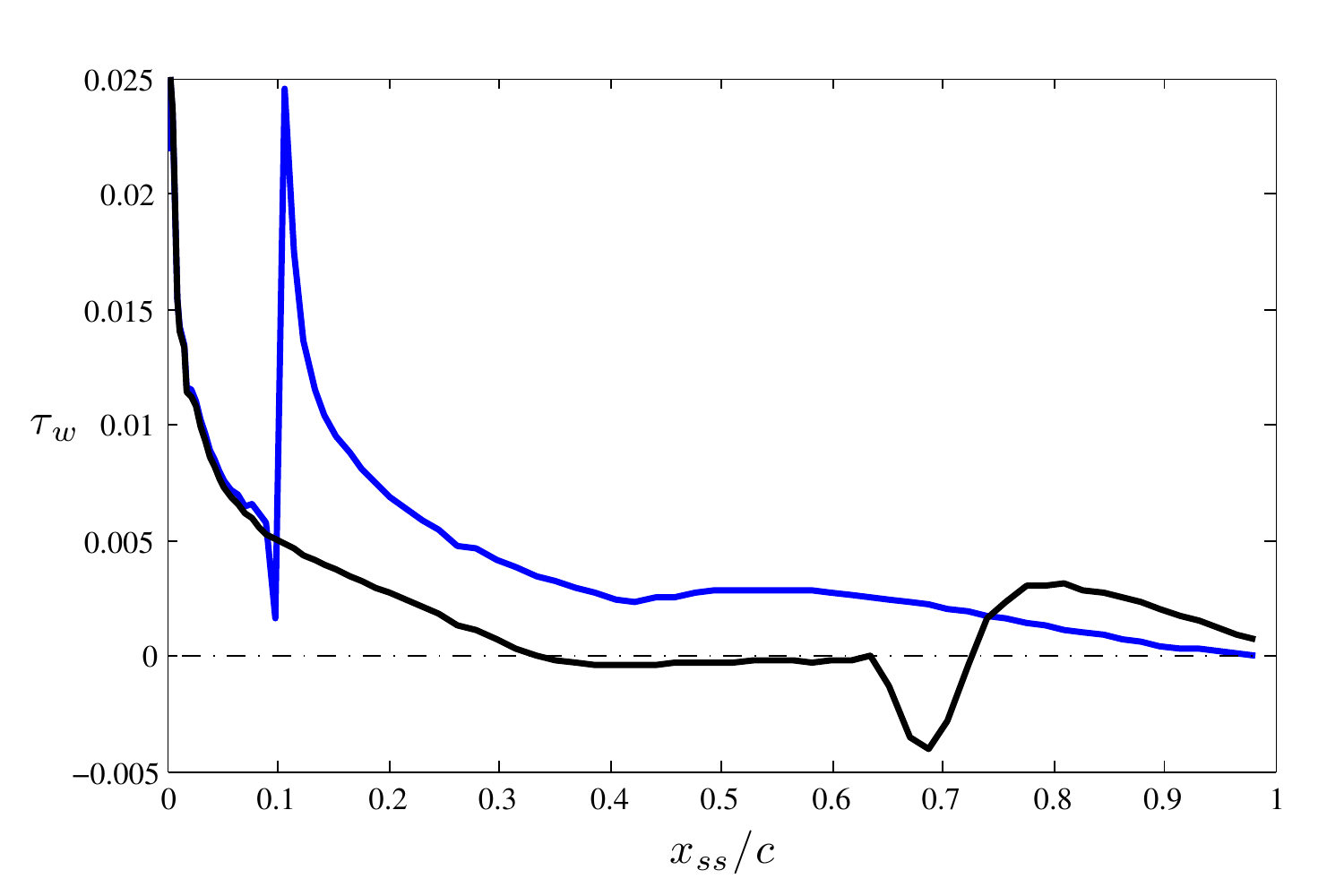}
\includegraphics[width=0.49\textwidth]{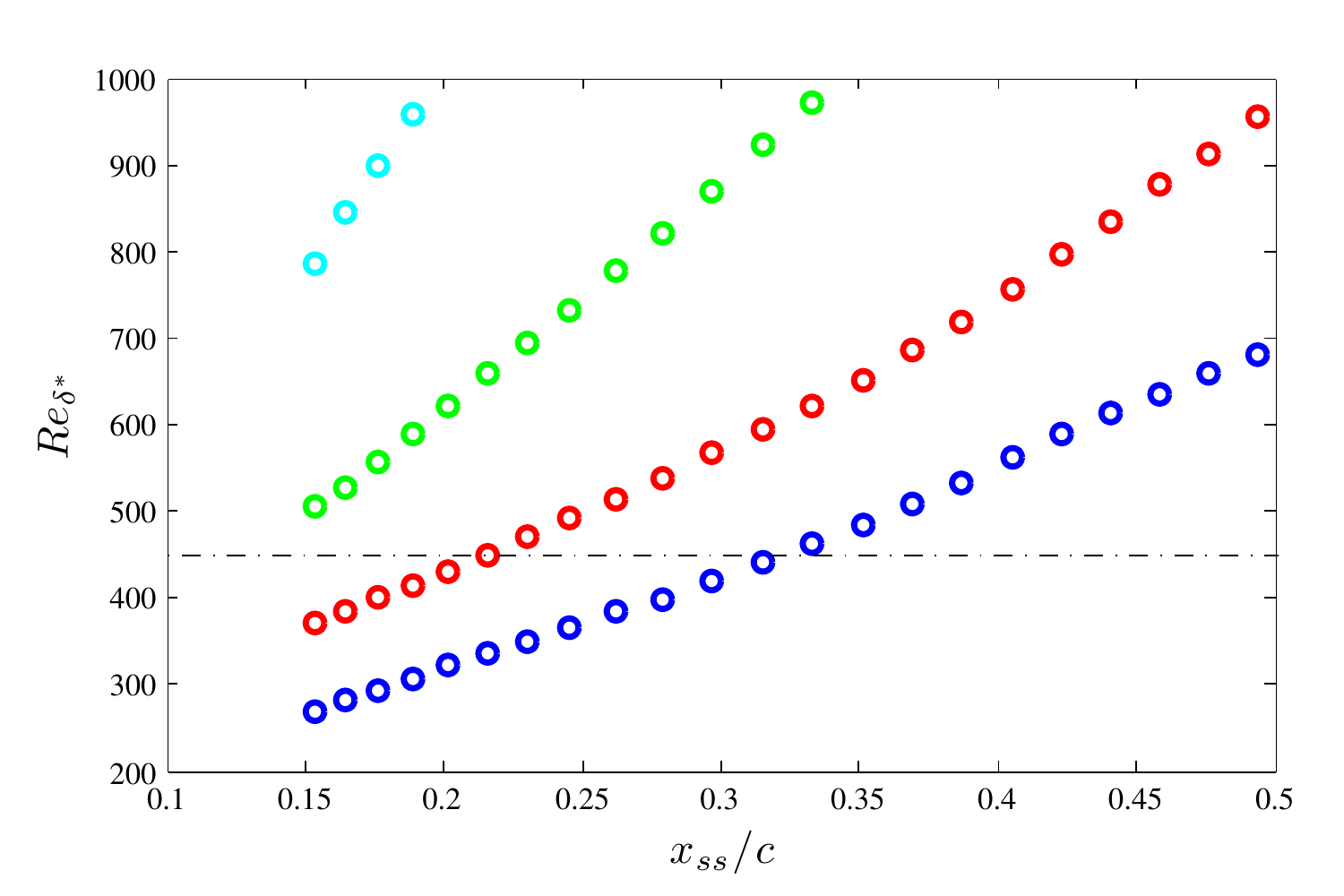}
\caption{(Left) Streamwise evolution of the wall-shear stress on the suction side of the wing at $Re_{c}=100,000$, where {\color{blue}\solid} corresponds to the tripped case from Table \ref{wing_cases} and {\color{black}\solid} to a case without tripping. The zero-wall-shear-stress level is denoted by {\color{black}\dotdashed}. (Right) Streamwise evolution of the Reynolds number based on displacement thickness on the suction side of the wing, where colors correspond to the cases summarized in Table \ref{wing_cases}. Here {\color{black}\dotdashed} denotes the value $Re_{\delta^{*}}=450$.}
\label{Cf_100k}
\end{figure}

Figure \ref{Cf_H} also shows that the \textcolor{black}{ W1} case exhibits a different trend in $C_{f}$ and $H$, compared to the ones observed at higher Reynolds numbers. In particular, a steep decrease in the skin-friction coefficient is present up to $x_{ss}/c \simeq 0.4$, followed by an increase up to $x_{ss}/c \simeq 0.6$, the point after which the $C_{f}$ experiences an evolution close to the higher-$Re$ cases. Regarding the shape factor, a more pronounced growth is present up to $x_{ss}/c \simeq 0.4$, followed by a slight decrease, and by another region of increasing $H$ beyond $x_{ss}/c \simeq 0.6$. Despite the different behavior in both quantities, which resembles that of a laminar boundary layer experiencing transition up to $x_{ss}/c \simeq 0.4$, note that the flow exhibits coherent vortical structures characteristic of TBLs in this region, as observed in the visualization from Figure \ref{flow_field}. In Figure \ref{Cf_100k} (left) we show the streamwise evolution of the \textcolor{black}{ average} wall-shear stress $\tau_{w}$ from the \textcolor{black}{ W1} case, compared to that of an additional simulation performed using the same setup, with one difference: in the latter, the volume-force tripping was disabled. This figure shows that the $\tau_{w}$ curves from both cases are identical up to $x_{ss}/c=0.1$, {\it i.e.} the location where the volume-force tripping is active in one of them. After the peak in $\tau_{w}$, the tripped case shows values of wall-shear stress higher than those in the case without tripping, indicating that this portion of the flow is indeed not laminar. \textcolor{black}{ The case without tripping exhibits negative values of the average wall-shear stress beyond $x/c \simeq 0.3$, a fact that indicates that the mean boundary-layer profile separates at this location. In general, although the average wall-shear stress is larger than zero there may be instances of negative instantaneous wall shear, a fact that would indicate instantaneous separation. This has been studied by \cite{vinuesa_backflow}, who documented up to $30\%$ of reverse flow in their DNS of the APG TBL on a wing section, and by \cite{bross_kahler}, who investigated backflow events in APG TBLs experimentally.} Natural transition is observed at $x_{ss}/c \simeq 0.65$, and beyond $x_{ss}/c \simeq 0.7$ the boundary layer reattaches, exhibiting larger \textcolor{black}{ average} wall-shear stress values than the tripped case. It can then be argued that the very low Reynolds number in the tripped configuration does not exhibit sufficient scale separation for a well-behaved turbulent state to develop, and only after $x_{ss}/c \simeq 0.4$, where the strongest APG effects are present, the boundary layer starts to converge towards the trend observed for the cases at higher $Re$. In Figure \ref{Cf_100k} we also show the evolution of the Reynolds number based on displacement thickness $Re_{\delta^{*}}$, where the typical value of $Re_{\delta^{*}}=450$ employed as inflow condition in our flat-plate boundary-layer simulations \citep{eitel_amor_et_al,bobke_et_al}, before the volume-force tripping, is highlighted. This figure indicates that the tripped \textcolor{black}{ W1} case only reaches $Re_{\delta^{*}}=450$ at $x_{ss}/c \simeq 0.35$. As discussed below, the very low Reynolds number of this case shows that, albeit turbulent, this boundary layer is not well-behaved, therefore justifying the different behavior compared to the higher-$Re$ wings.

\begin{figure}
\centering
\includegraphics[width=0.49\textwidth]{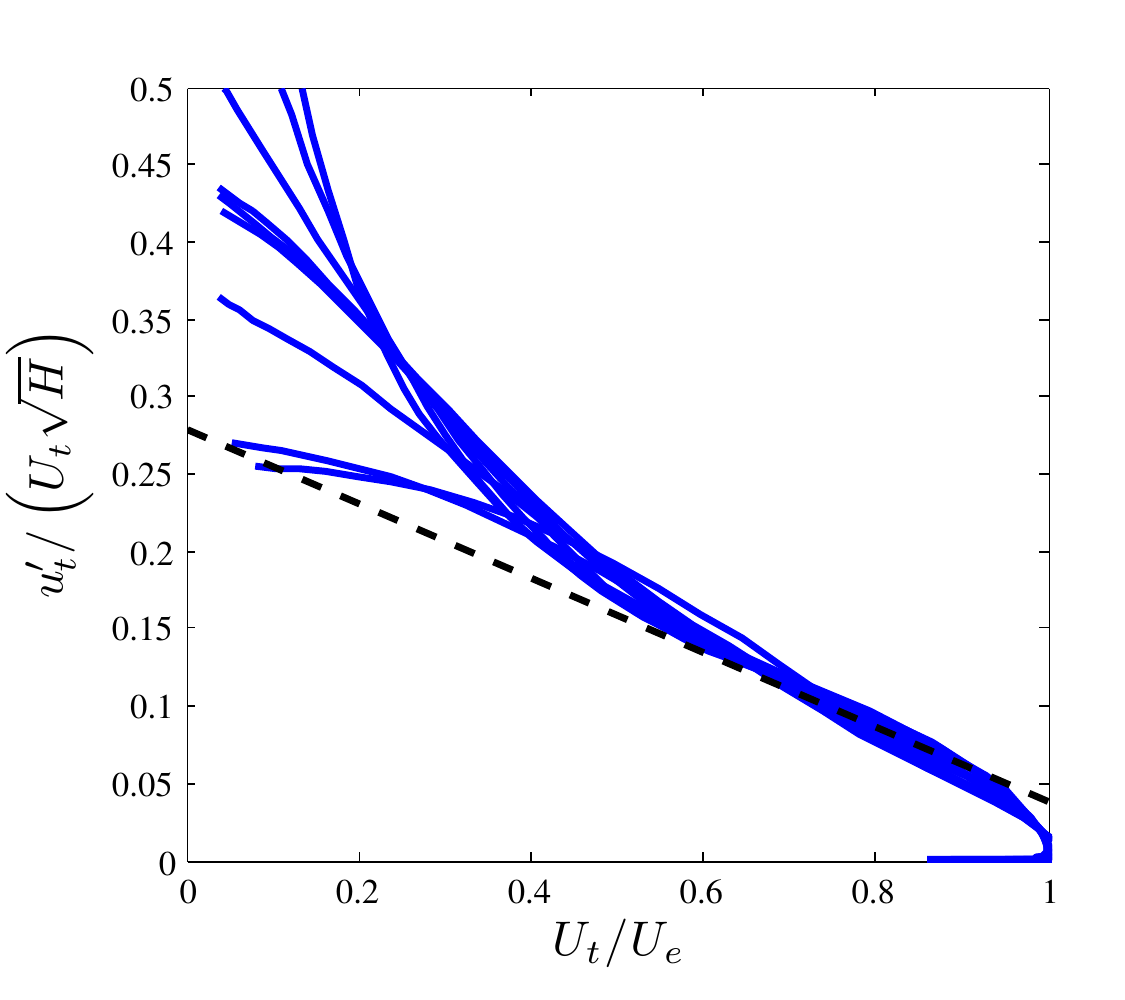}
\includegraphics[width=0.49\textwidth]{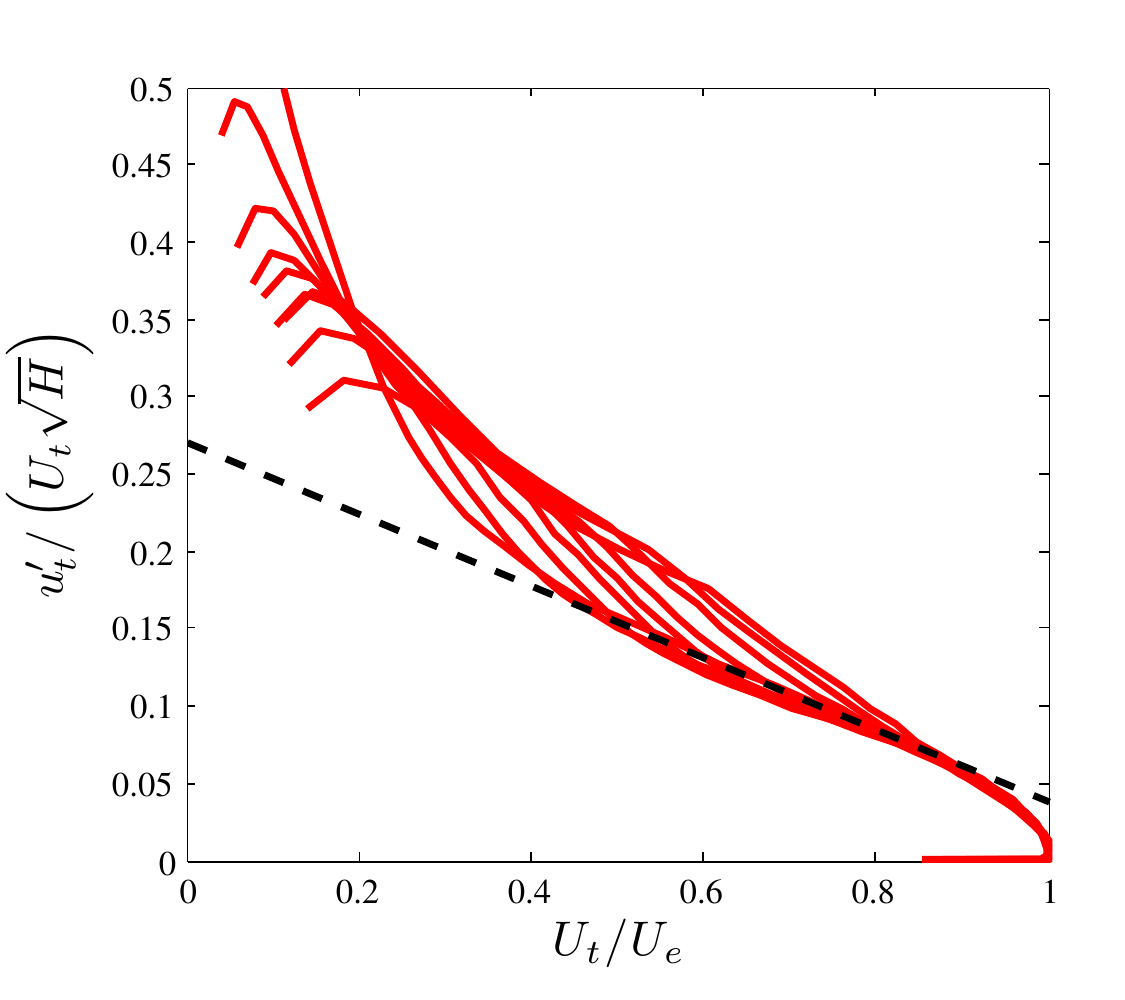}
\includegraphics[width=0.49\textwidth]{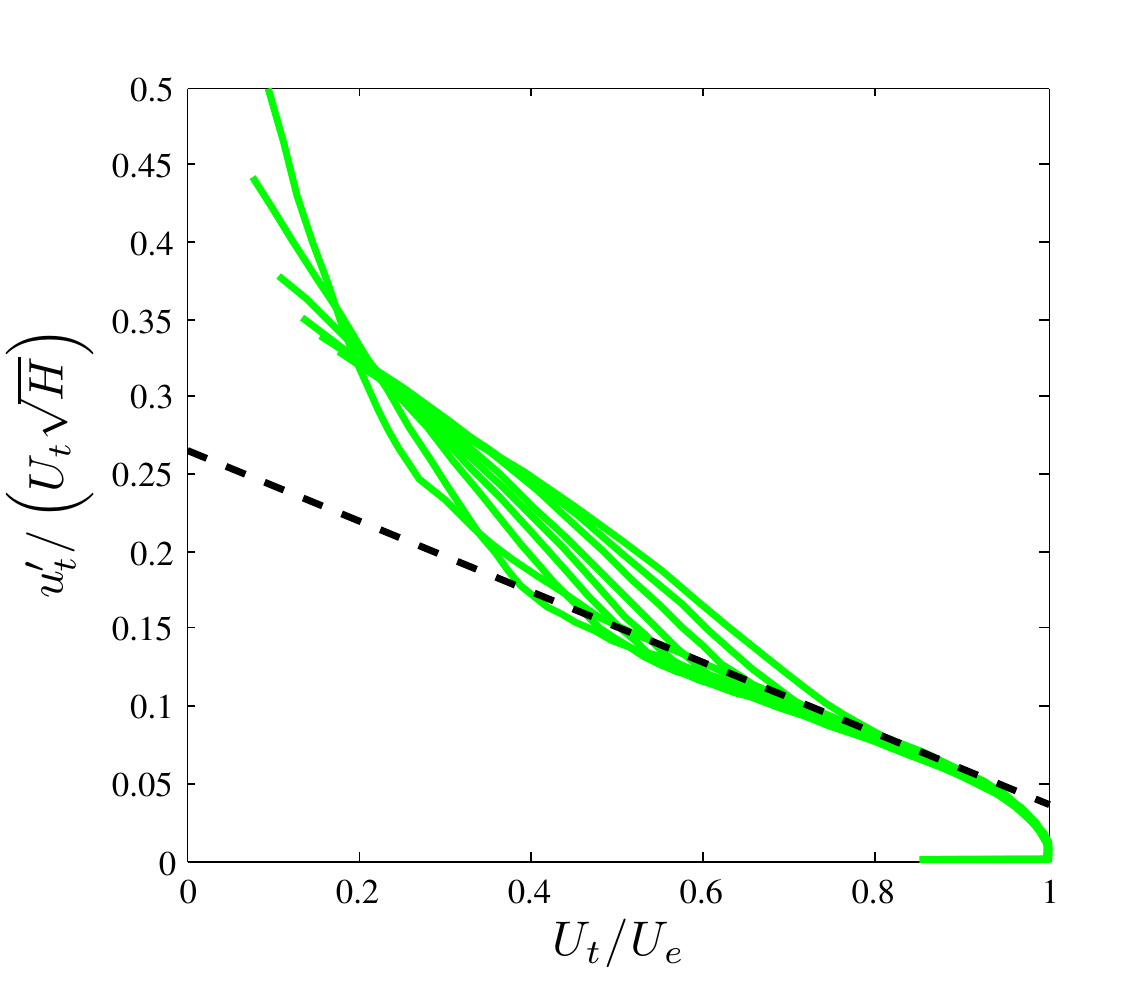}
\includegraphics[width=0.49\textwidth]{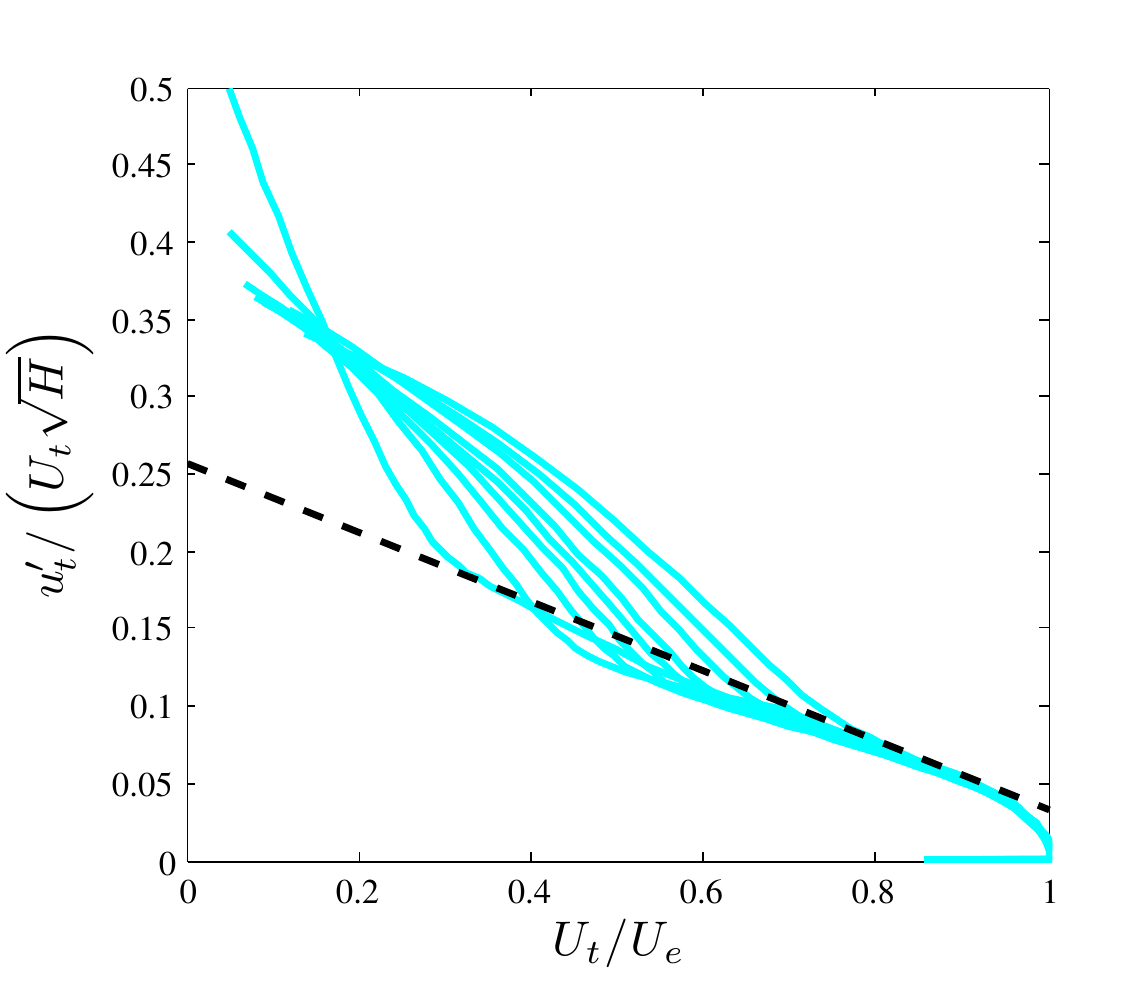}
\caption{Diagnostic-plot scaling modified with the shape factor $H$, applied to several profiles over the whole suction side of the wing in the four $Re_{c}$ cases under study. The colors correspond to the cases summarized in Table \ref{wing_cases}, and {\color{black}\dashed} represents equation (\ref{fit_diag}), where $\alpha_{H}$ and $\beta_{H}$ are obtained from the correlations by \cite{diagnostic_ftac}, using the largest $Re_{\theta}$ value in each case.}
\label{diag_fig}
\end{figure}

In order to further explore the differences between the \textcolor{black}{ W1} case and the others, we apply \textcolor{black}{ one of the methods by} \cite{diagnostic_ftac} to identify well-behaved TBLs, \textcolor{black}{ {\it i.e.}, with correct setups and sufficient scale separation}. \cite{drozdz_et_al} proposed a modified version of the so-called diagnostic-plot scaling for PG TBLs, in which the root-mean-squared profile of the streamwise velocity fluctuations, when scaled with the local man velocity and the shape factor as $u^{\prime} / \left (U \sqrt{H} \right )$, exhibits excellent collapse in the outer region when represented as a function of $U/U_{e}$. \textcolor{black}{ After analyzing a total of 12 high-quality databases available in the literature (10 numerical and 2 experimental), \cite{diagnostic_ftac} also observed the excellent collapse of the profiles in the modified diagnostic scaling documented by \cite{drozdz_et_al}. In particular, they reported that the region defined by $0.8 \leq U/U_{e} \leq 0.9$ exhibits linear behavior, and that this part of the profile could be expressed as:}
\begin{equation}\label{fit_diag}
\frac{u^{\prime}}{U \sqrt{H}}=\alpha_{H}-\beta_{H} \frac{U}{U_{e}},
\end{equation}
where $\alpha_{H}$ and $\beta_{H}$ are $Re$-dependent coefficients. \textcolor{black}{ The linear region between $U/U_{e}=0.8$ and 0.9 was observed in all the cases over a wide range of Reynolds-number and pressure-gradient conditions. Although \cite{diagnostic_ftac} also reported that using the two experimental datasets (up to $Re_{\theta}=18,700$ and $56,100$) the region of linear behavior was observed up to lower $U/U_{e}$ values, the rest of databases (with $Re_{\theta}$ ranges similar to the ones under study here) exhibited the limits reported above.} In Figure \ref{diag_fig} we show, for each wing case, several profiles along the suction side scaled using the modified diagnostic plot by \cite{drozdz_et_al}. We also show, for each case, the curve described by equation (\ref{fit_diag}), where the $\alpha_{H}$ and $\beta_{H}$ values were obtained from the correlations by \cite{diagnostic_ftac}, using the highest $Re_{\theta}$ on the suction side for each case. This figure shows that the three cases with $Re_{c} \geq 200,000$ exhibit excellent collapse in the outer region, and the part of the profile between $U/U_{e}=0.8$ and 0.9 is perfectly described by equation (\ref{fit_diag}). On the other hand, this collapse is not observed in the \textcolor{black}{ W1} case, a fact that also supports the claim that this boundary layer is not well-behaved. \textcolor{black}{ This is due to the combination of a very low Reynolds number (as illustrated in Figure~\ref{Cf_100k}~(right) with $Re_{\delta^{*}}$ and the related discussion) with very strong streamwise pressure-gradient conditions.} 

\begin{figure}
\centering
\includegraphics[width=0.45\textwidth]{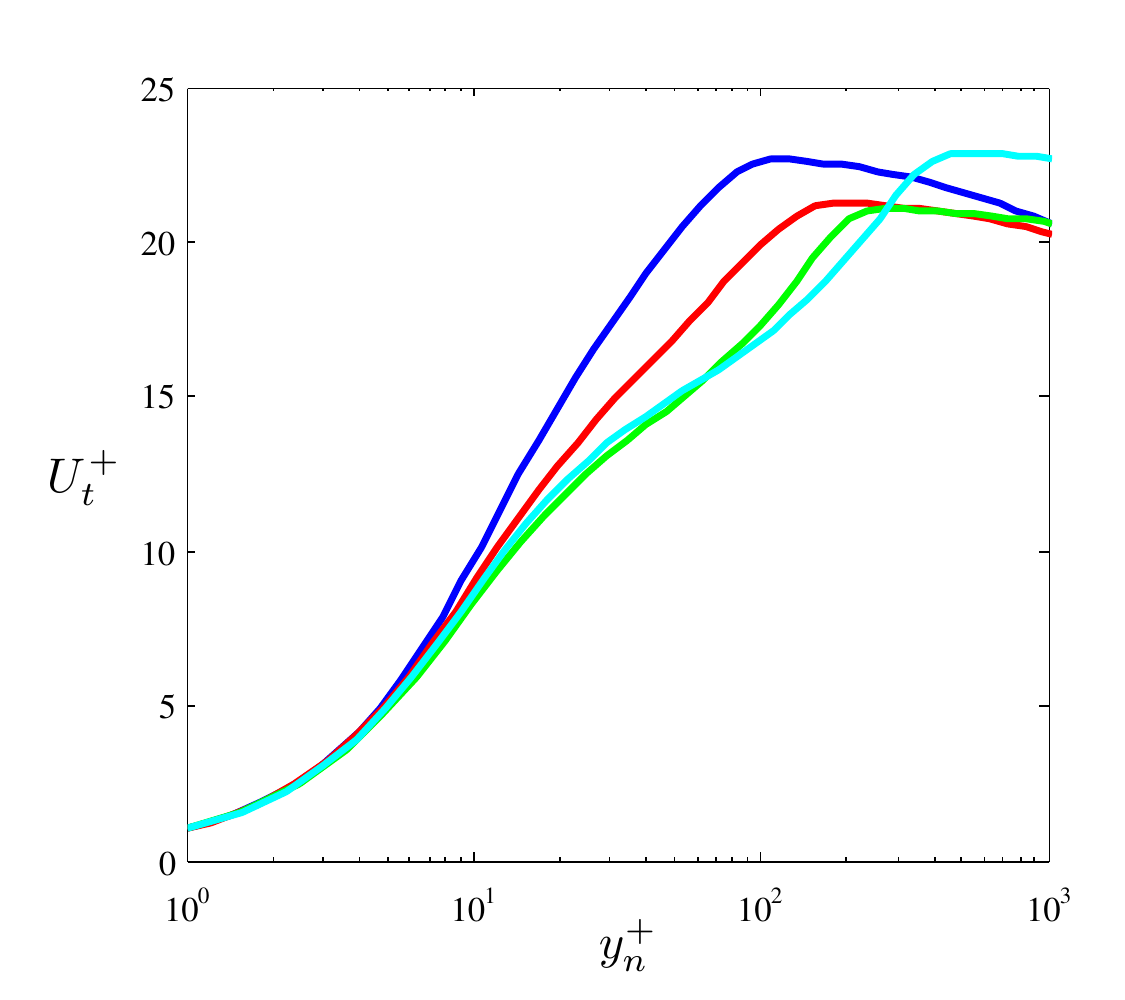}
\includegraphics[width=0.45\textwidth]{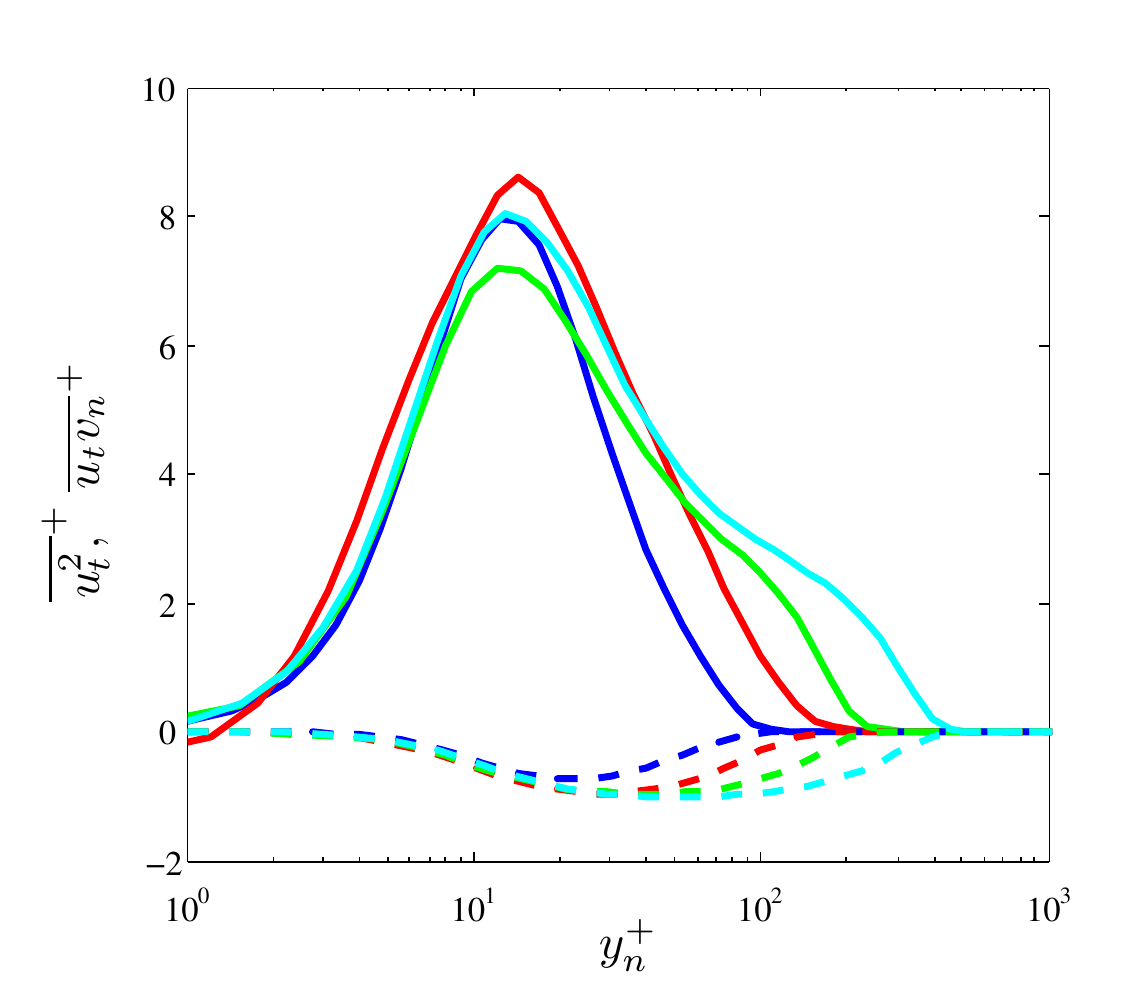}
\caption{\textcolor{black}{ (Left) Inner-scaled mean velocity profiles and (right) selected components of the Reynolds-stress tensor for the four wings at $x_{ss}/c=0.3$. The dashed lines denote the Reynolds-shear stress profiles, and colors as in Table~\ref{wing_cases}.}}
\label{data_03}
\end{figure}

\textcolor{black}{ The flow case under study is moderately complex due to the rapidly growing APG. In Figure~\ref{data_03} we show the inner-scaled mean velocity profile and selected components of the Reynolds-stress tensor at $x_{ss}/c=0.3$. This figure also indicates that the \textcolor{black}{ W1} case, with a local $Re_{\tau}=81$, combines low-$Re$ effects with an increasing impact of the pressure gradient (here $\beta=0.5$). The mean velocity profile confirms the analysis based on the diagnostic scaling, and the fact that this TBL is not well-behaved. The \textcolor{black}{ W4} and \textcolor{black}{ W10} cases exhibit the characteristic turbulence statistics of TBLs subjected to mild APGs (with a local $\beta \simeq 0.2$ in both cases), as discussed in further detail in $\S$\ref{statistics}. Regarding the \textcolor{black}{ W2} case, the local friction Reynolds number is low ($Re_{\tau}=139$), and the slightly higher near-wall peak of $\overline{u^{2}_{t}}^{+}$ indicates that there may be marginally-turbulent flow effects present \citep{marin_et_al}. However, these effects are absent farther downstream (see $\S$\ref{statistics}), a fact that shows (together with the diagnostic-scaling analysis) that this TBL is well-behaved. }


\subsection{Inner-scaled mean velocity and Reynolds-stress profiles} \label{statistics}

\begin{figure}
\centering
\includegraphics[width=0.45\textwidth]{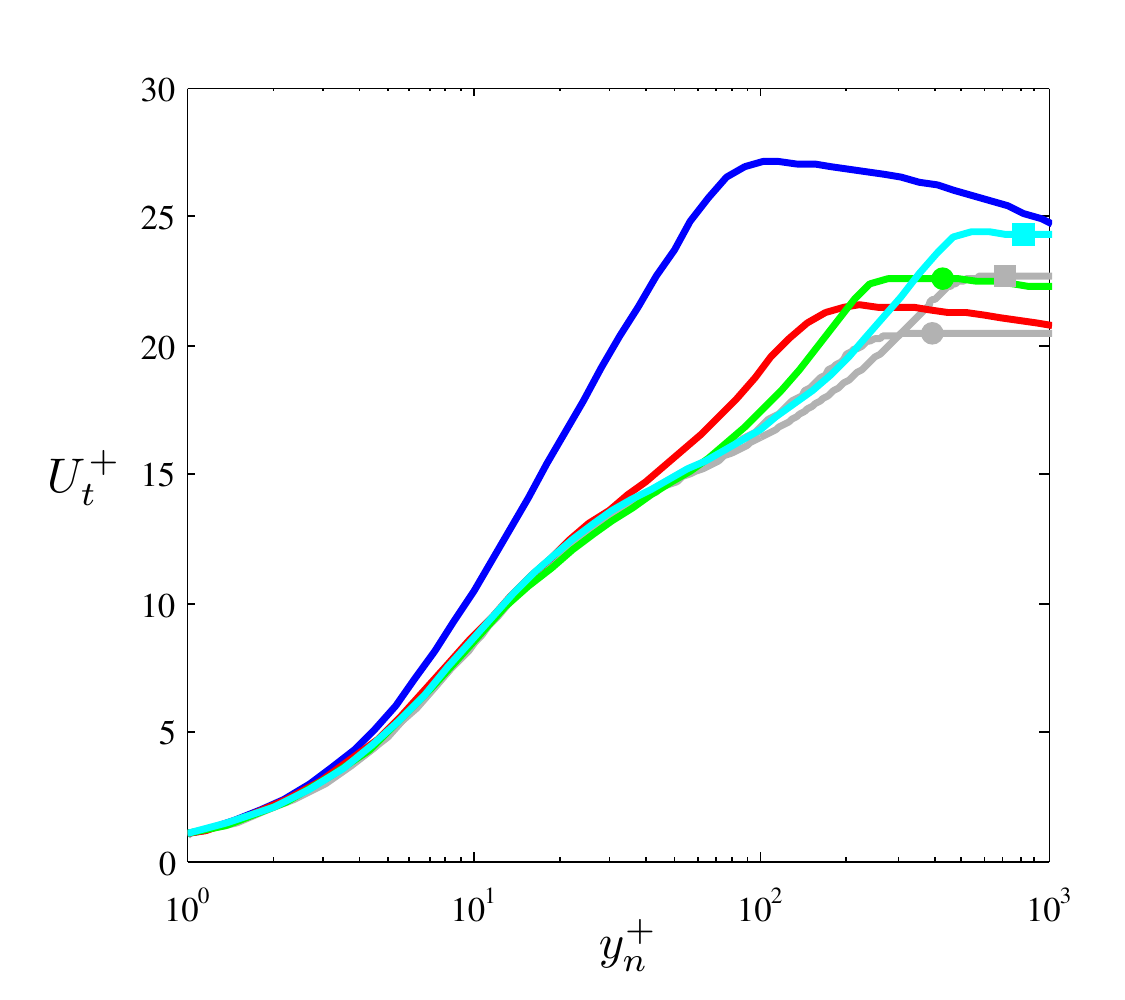}
\includegraphics[width=0.45\textwidth]{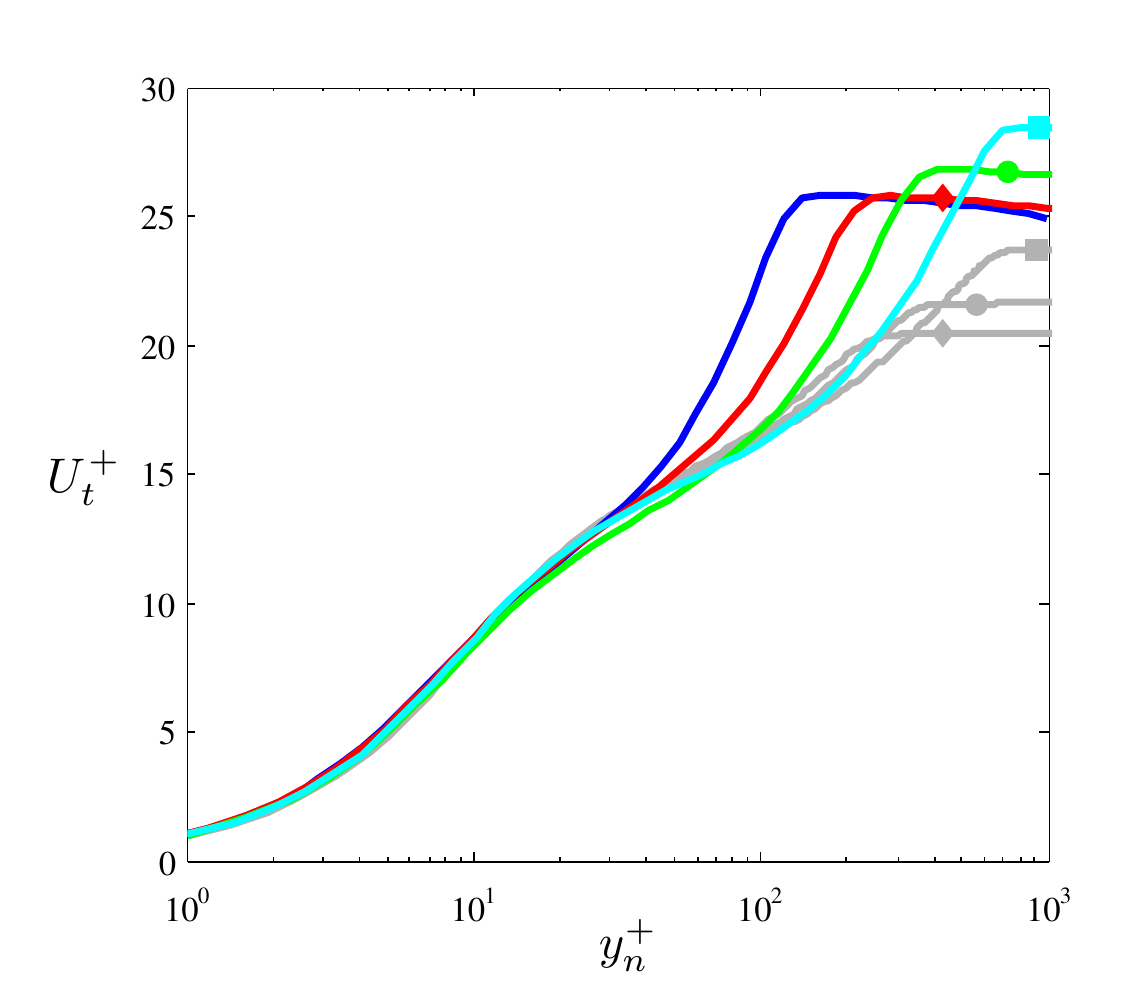}
\includegraphics[width=0.45\textwidth]{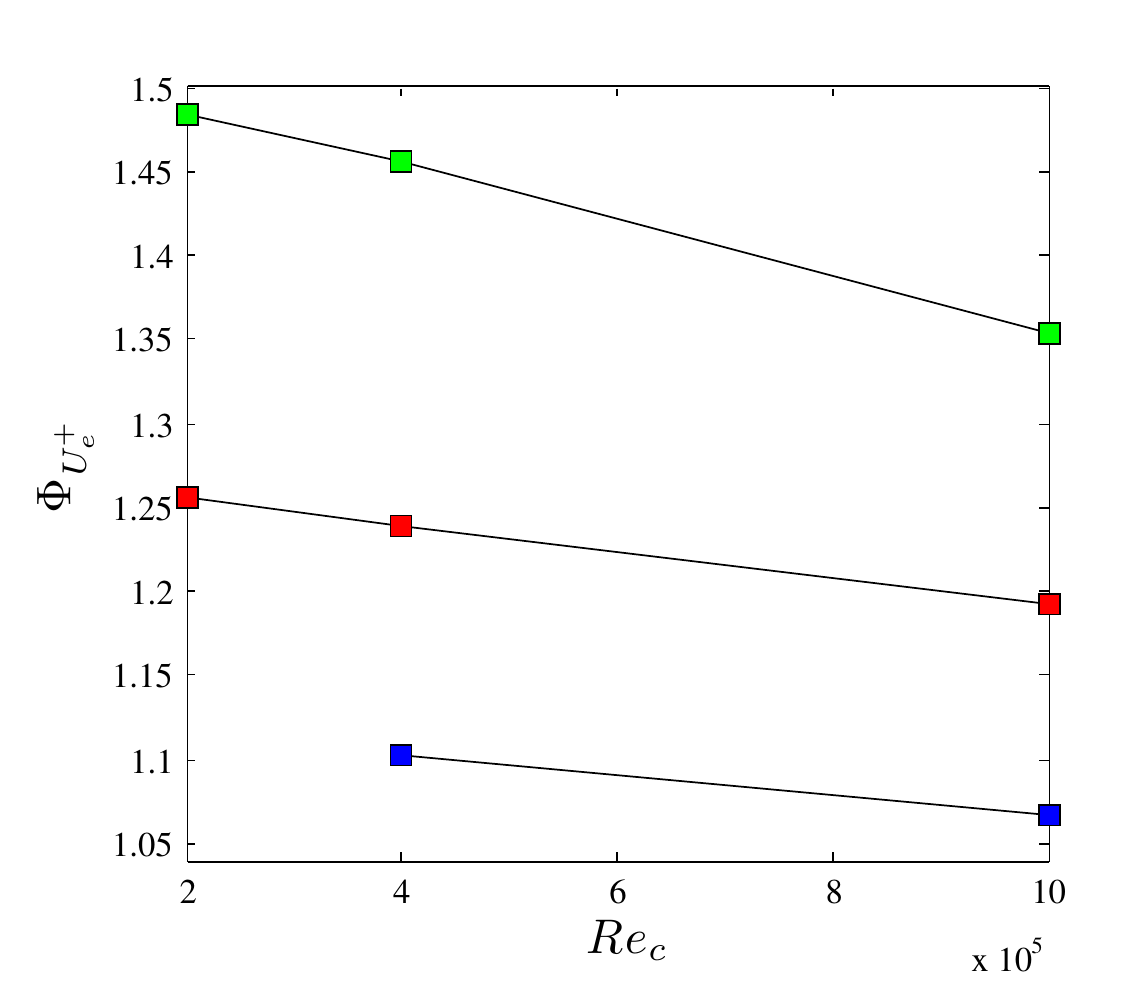}
\includegraphics[width=0.45\textwidth]{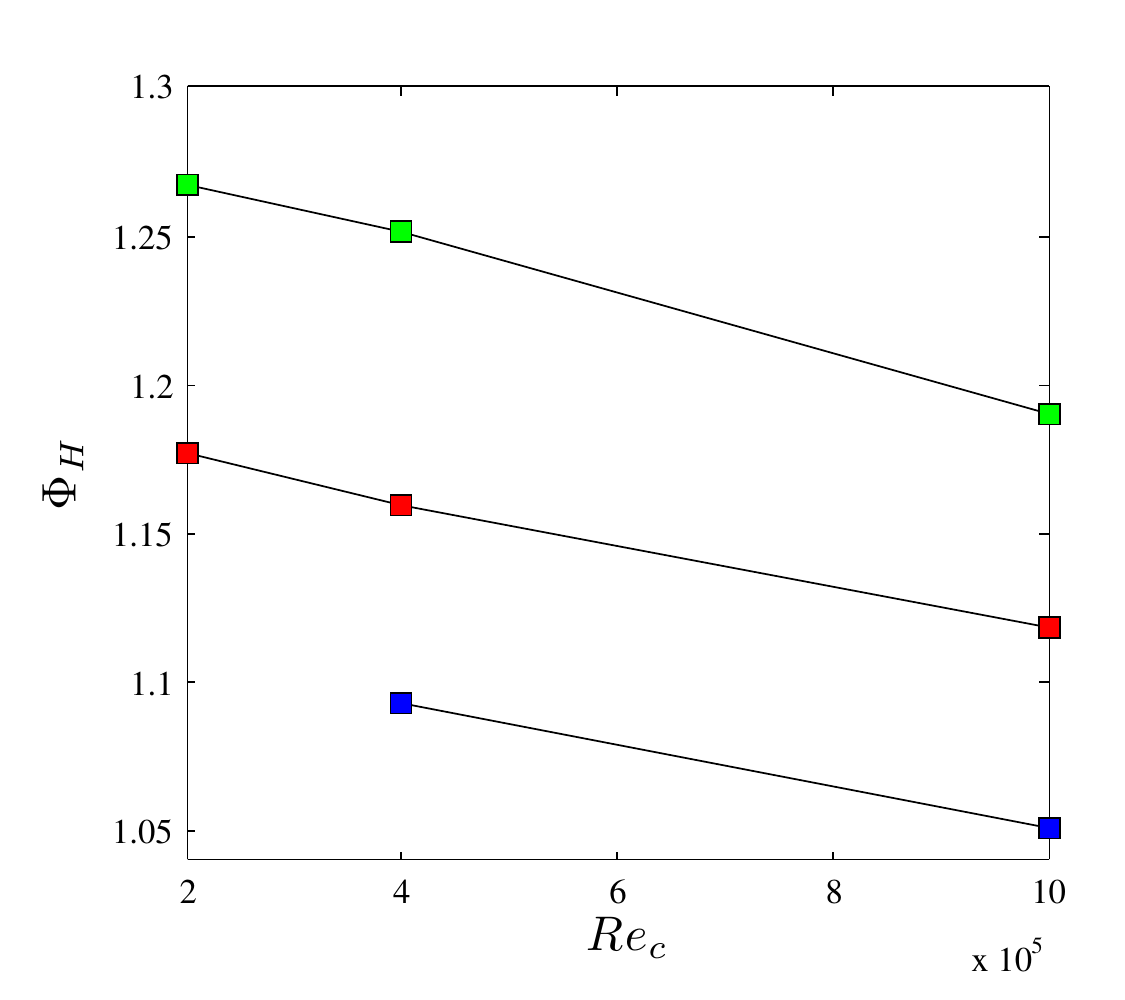}
\caption{Inner-scaled tangential mean velocity profiles at (top-left) $x_{ss}/c=0.4$ and (top-right) $x_{ss}/c=0.7$, for the four wing cases under study, compared with the DNS results of ZPG TBL by \cite{schlatter_orlu10} at approximately matching $Re_{\tau}$ values. Colors from wing cases as in Table \ref{wing_cases}, and {\color{grey}\solid} denotes ZPG TBL data. \textcolor{black}{ The matched $U^{+}_{t}$ profiles for W10 and ZPG10 are denoted by $\left ( \blacksquare \right )$, for W4 and ZPG4 by $\left ( \bullet \right )$ and for W2 and ZPG2 by $\left ( \blacklozenge \right )$.} Ratio of (bottom-left) $U^{+}_{e}$ and (bottom-right) $H$, between wing and ZPG at approximately matching $Re_{\tau}$. Here (\textcolor{blue}{$\blacksquare$}), (\textcolor{red}{$\blacksquare$}) and (\textcolor{green}{$\blacksquare$}) denote ratios at $x_{ss}/c=0.4$, $0.7$ \textcolor{black}{ and $0.8$}, respectively.}
\label{Up_vs_yp}
\end{figure}

Figure \ref{Up_vs_yp} (top) shows the inner-scaled mean velocity profiles at $x_{ss} /c =0.4$ and $0.7$ for the four wing cases, where $U^{+}_{t}$ is the inner-scaled mean velocity in the direction tangential to the wing surface, whereas $y^{+}_{n}$ is the inner-scaled wall-normal coordinate. In Tables \ref{profilesx04} and \ref{profilesx07} we show the boundary-layer parameters at those two streamwise locations, together with the ones from ZPG TBL profiles \citep{schlatter_orlu10} at approximately matching $Re_{\tau}$ values. Note that the lowest $Re_{\tau}$ in the DNS database by \cite{schlatter_orlu10} is $252$, which implies that at $x_{ss}/c=0.4$ the two lower-$Re$ wings do not have a matching ZPG profile, whereas at $x_{ss}/c=0.7$ only the \textcolor{black}{ W1} case is left without a matching ZPG case. These tables also reflect that, except the lowest-$Re$ case, all the wing profiles are subjected approximately to the same values of $\beta \simeq 0.6$ and $2$ at $x_{ss}/c=0.4$, and $0.7$. The mean-flow comparisons between wing and APG are performed at approximately matching values of $Re_{\tau}$, with the aim of assessing the effect of the APG with respect to the baseline ZPG case. Although this comparison can be done by matching several quantities (such as $Re_{\delta^{*}}$ or $Re_{\theta}$), in the present work we fixed $Re_{\tau}$ as in the studies by \cite{monty_et_al}, \cite{harun_et_al} or \cite{bobke_et_al}. Note that by fixing $Re_{\tau}$ we compare two boundary layers which essentially exhibit the same range of spatial scales, but subjected to different pressure-gradient conditions. The first noticeable conclusion is the more prominent wakes present in the APG TBLs compared with the corresponding ZPG TBLs at the same $Re_{\tau}$, which is due to the lower skin-friction coefficient. In addition to this, the shape factor is larger in APGs than in the corresponding ZPG boundary layers (as noted in Tables \ref{profilesx04} and \ref{profilesx07}), a fact that is connected to the boundary-layer thickening, also induced by the APG. Note that the \textcolor{black}{ W1} profile at $x_{ss}/c=0.4$ is not well-behaved, which is manifested in the mean velocity profile as well. A first step towards characterizing the effect of $Re$ in the TBLs subjected to this particular $\beta(x)$ distribution (see Figure \ref{beta_Reth_Ret} (top)) is to observe the evolution of $U^{+}_{e}$ and $H$ over the given $Re_{\tau}$ range in the ZPG and APG cases. In Figure \ref{Up_vs_yp} \textcolor{black}{ (bottom)} we show the ratios $\Phi_{U_{e}^{+}}$ and $\Phi_{H}$ as a function of $Re_{c}$ \textcolor{black}{ at the two previous streamwise locations, together with $x_{ss}/c=0.8$. This additional position was chosen because the friction Reynolds number shows its maximum value at approximately this location for all the wing cases.} These ratios are defined as the value of $U^{+}_{e}$ (or $H$) from the TBL on the wing at a certain $Re_{c}$ and $x_{ss}$, divided by the same quantity in a ZPG TBL with \textcolor{black}{ approximately} the same $Re_{\tau}$. As noted above, both $U^{+}_{e}$ and $H$ are larger in APGs, thus the ratios $\Phi_{U_{e}^{+}}$ and $\Phi_{H}$ have values larger than 1, ranging from 1.07 and \textcolor{black}{ 1.48} in the case of $\Phi_{U_{e}^{+}}$, and from 1.05 to \textcolor{black}{ 1.27} for $\Phi_{H}$ depending on the case under consideration. \textcolor{black}{ The difference with respect to the ZPG becomes larger as one moves downstream due to the increased deceleration experienced by the TBL.} Interestingly, the two indicators exhibit a decreasing trend with $Re_{c}$ at \textcolor{black}{ the three} streamwise positions, a fact that indicates that the values of $U^{+}_{e}$ and $H$ are more severely affected by the pressure gradient at lower Reynolds numbers, when all the TBLs were subjected to \textcolor{black}{ approximately} the same $\beta(x)$ distribution. Additional support for this claim can be found in the mean velocity profiles at $y^{+}_{n} \simeq 25$, where the ZPG cases and the wing at $Re_{c}=1,000,000$ exhibit almost identical values of the inner-scaled velocity $U^{+}_{t}$, but the lower-$Re$ wings show values below these in the two streamwise positions. Lower velocities in the buffer layer with respect to the ZPG are associated with strong effects of the APG, as documented for instance by \cite{spalart_watmuff} or \cite{bobke_et_al}. This is another indication of the fact that the effect of the APG is more pronounced at lower $Re$.
\begin{table}
\scriptsize
\caption{Boundary-layer parameters at $x_{ss}/c=0.4$ for the various wing cases under study. \textcolor{black}{ ZPG4 and ZPG10} denote the DNS ZPG TBL cases \citep{schlatter_orlu10} approximately matching the $Re_{\tau}$ values of the wing profiles at $Re_{c}=400,000$ and $1,000,000$, respectively.}
\label{profilesx04}
\centering
\begin{tabular}{c c c c c c c}
\hline\noalign{\smallskip}
Parameter & W1 & W2 & W4 & W10 & ZPG4 &ZPG10 \\
\noalign{\smallskip}\hline \noalign{\smallskip}
$\beta$ & $1.6$ & $0.67$ & \textcolor{black}{ 0.67} & $0.58$ & $\simeq 0$ & $\simeq 0$ \\
$Re_{\tau}$ & $76$ & $160$ & \textcolor{black}{ 241} & $449$ & $252$ & $492$ \\
$Re_{\theta}$ & $276$ & $452$ & \textcolor{black}{ 759} & $1,465$ & $678$ & $1,421$ \\
$C_{f}$ & $2.8 \times 10^{-3}$ & $4.4 \times 10^{-3}$ & \textcolor{black}{ $3.9 \times 10^{-3}$} & $3.4 \times 10^{-3}$ & $4.8 \times 10^{-3}$ & $3.9 \times 10^{-3}$ \\
$H$ & $2.04$ & $1.68$ & \textcolor{black}{ 1.61} & $1.50$ & $1.47$ & $1.43$ \\
\hline\noalign{\smallskip}
\end{tabular}
\end{table}

\begin{table}
\scriptsize
\caption{Boundary-layer parameters at $x_{ss}/c=0.7$ for the various wing cases under study.  \textcolor{black}{ ZPG2, ZPG4 and ZPG10} denote the DNS ZPG TBL cases \citep{schlatter_orlu10} approximately matching the $Re_{\tau}$ values of the wing profiles at $Re_{c}=200,000$, $400,000$ and $1,000,000$, respectively.}
\label{profilesx07}
\centering
\begin{tabular}{c c c c c c c c}
\hline\noalign{\smallskip}
Parameter & W1 & W2 & W4 & W10 & ZPG2 & ZPG4 &ZPG10 \\
\noalign{\smallskip}\hline \noalign{\smallskip}
$\beta$ & $3.7$ & $2.6$ & \textcolor{black}{ 2.4} & $2.0$ & $\simeq 0$ & $\simeq 0$ & $\simeq 0$ \\
$Re_{\tau}$ & $136$ & $228$ & \textcolor{black}{ 360} & $671$ & $252$ & $359$ & $671$ \\
$Re_{\theta}$ & $566$ & $901$ & \textcolor{black}{ 1,506} & $2,877$ & $678$ & $1,007$ & $2,001$ \\
$C_{f}$ & $3.0 \times 10^{-3}$ & $3.0 \times 10^{-3}$ & \textcolor{black}{ $2.8 \times 10^{-3}$} & $2.5 \times 10^{-3}$ & $4.8 \times 10^{-3}$ & $4.3 \times 10^{-3}$ & $3.5 \times 10^{-3}$ \\
$H$ & $1.92$ & $1.73$ & \textcolor{black}{ 1.68} & $1.58$ & $1.47$ & $1.45$ & $1.41$ \\
\hline\noalign{\smallskip}
\end{tabular}
\end{table}

\begin{figure}
\centering
\includegraphics[width=0.43\textwidth]{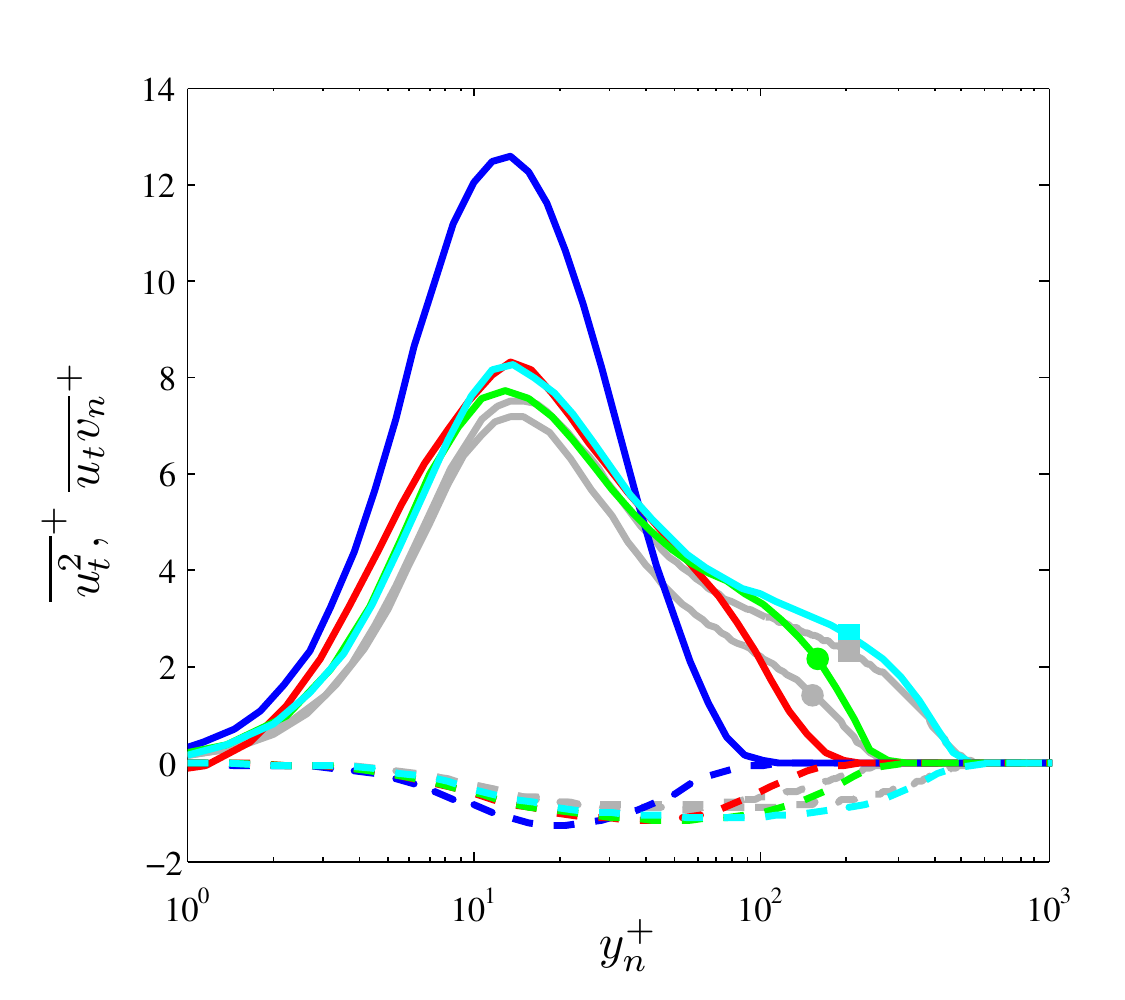}
\includegraphics[width=0.43\textwidth]{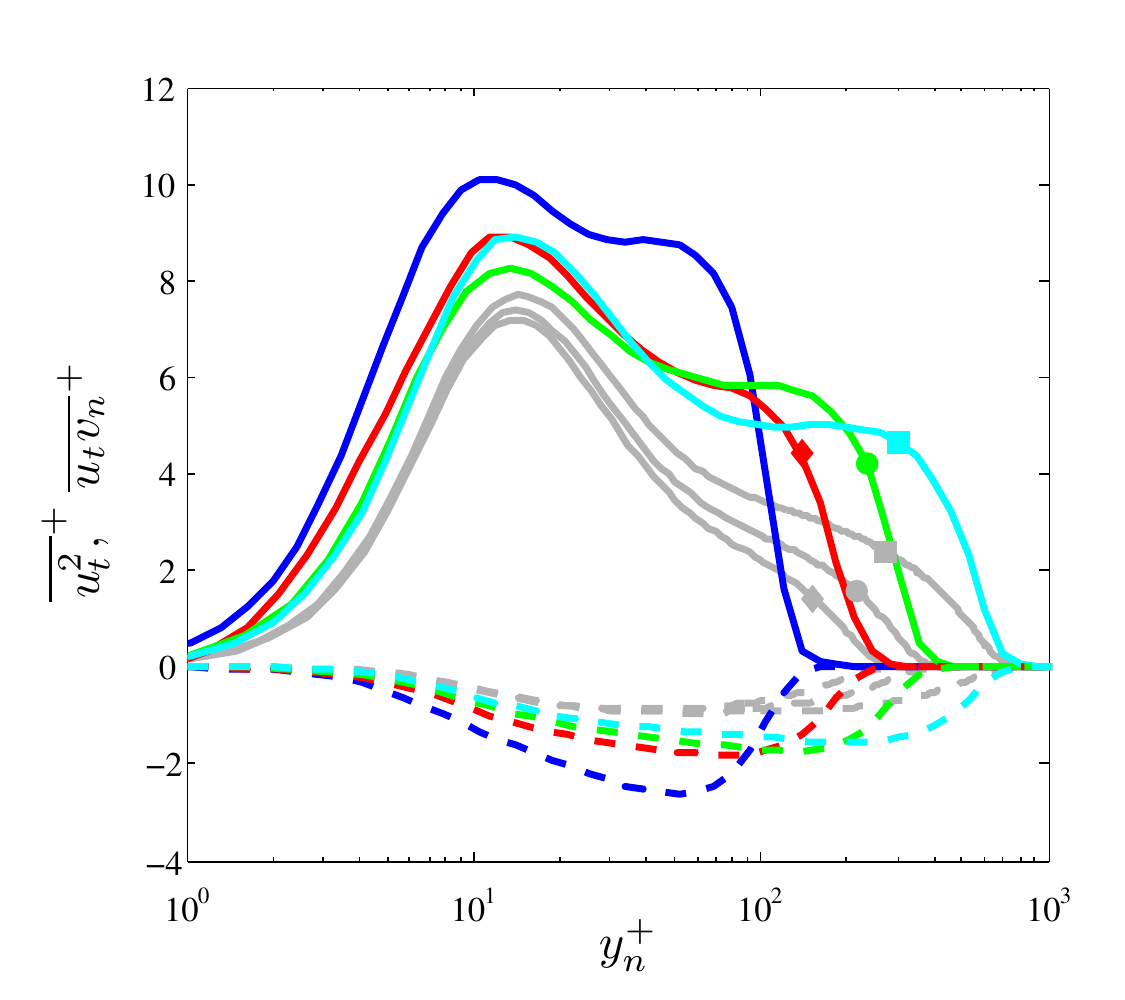}
\includegraphics[width=0.43\textwidth]{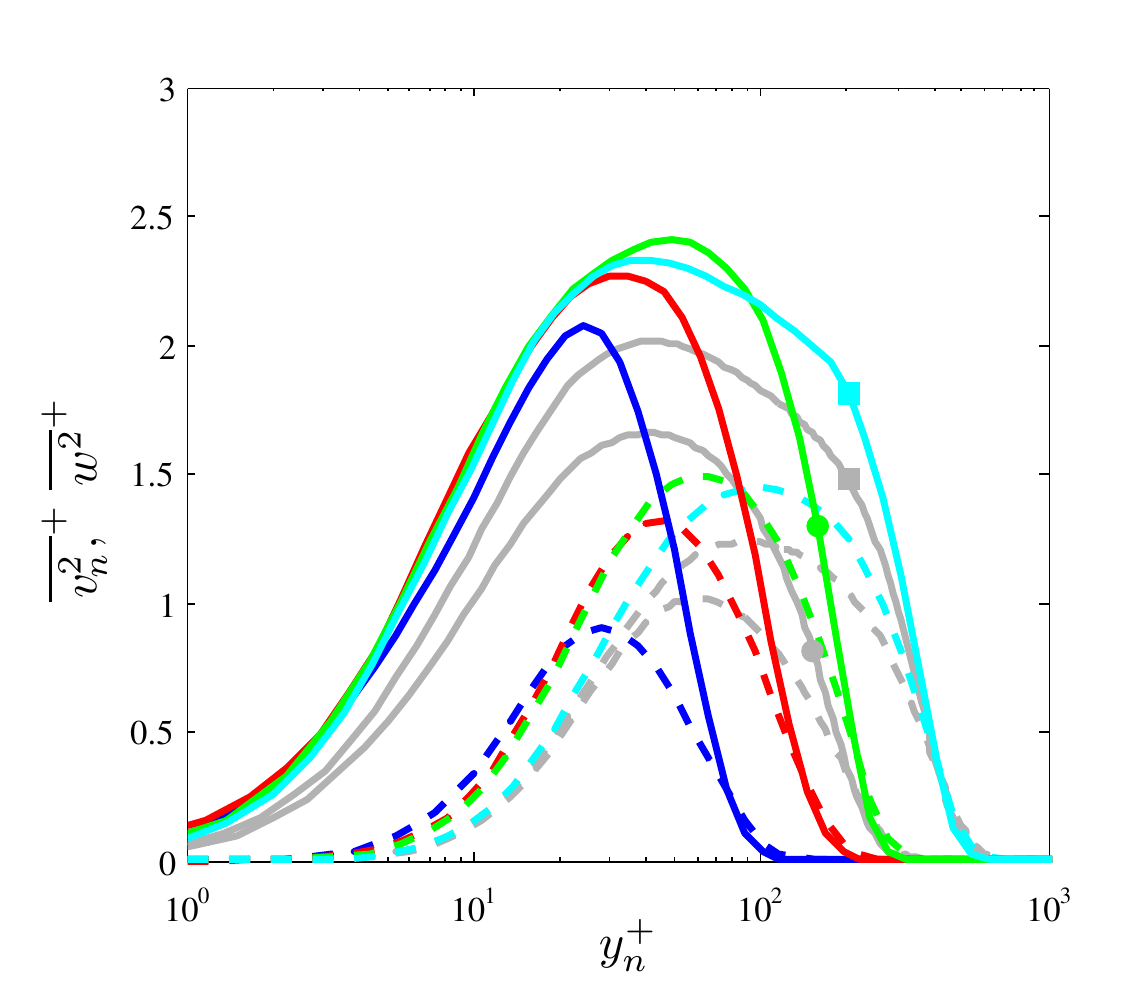}
\includegraphics[width=0.43\textwidth]{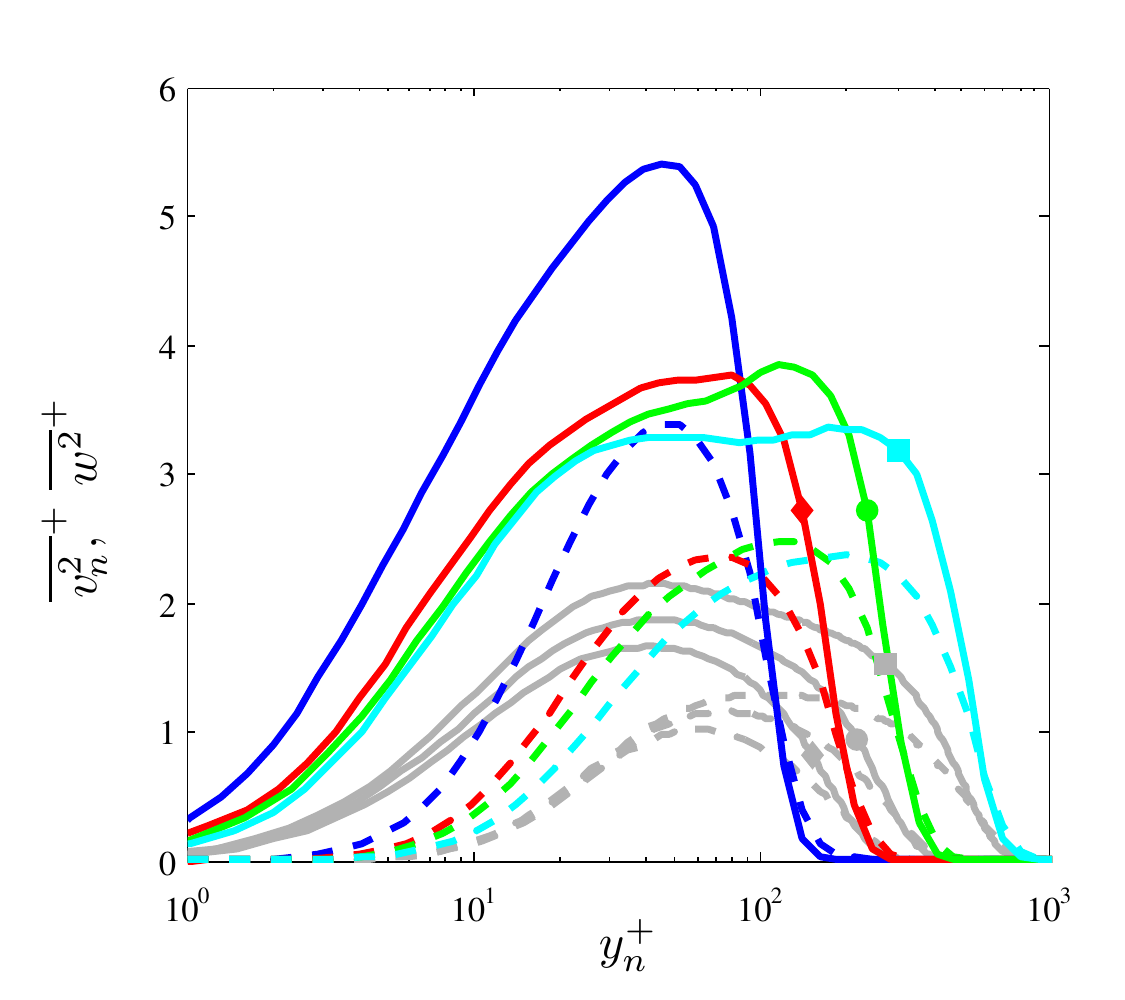}
\includegraphics[width=0.43\textwidth]{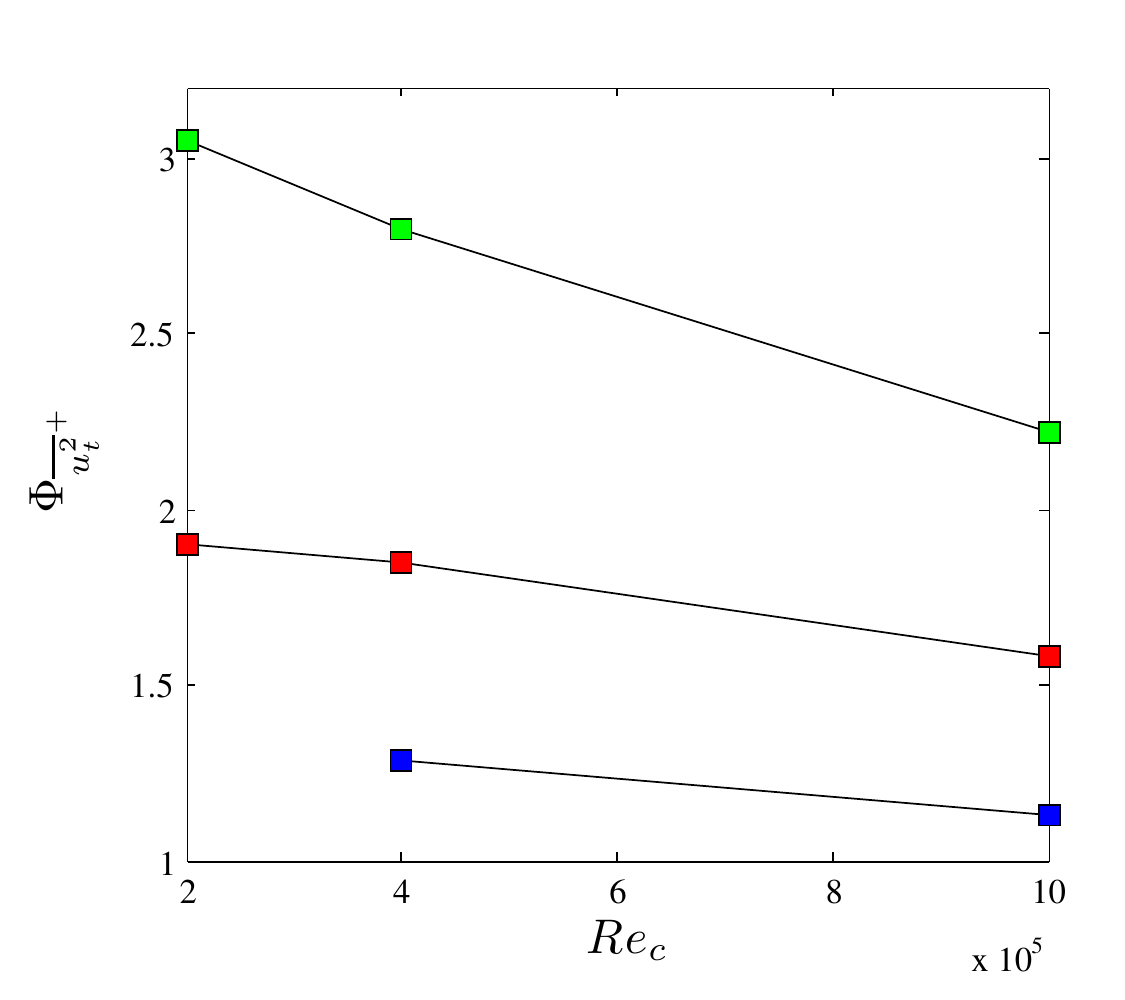}
\includegraphics[width=0.43\textwidth]{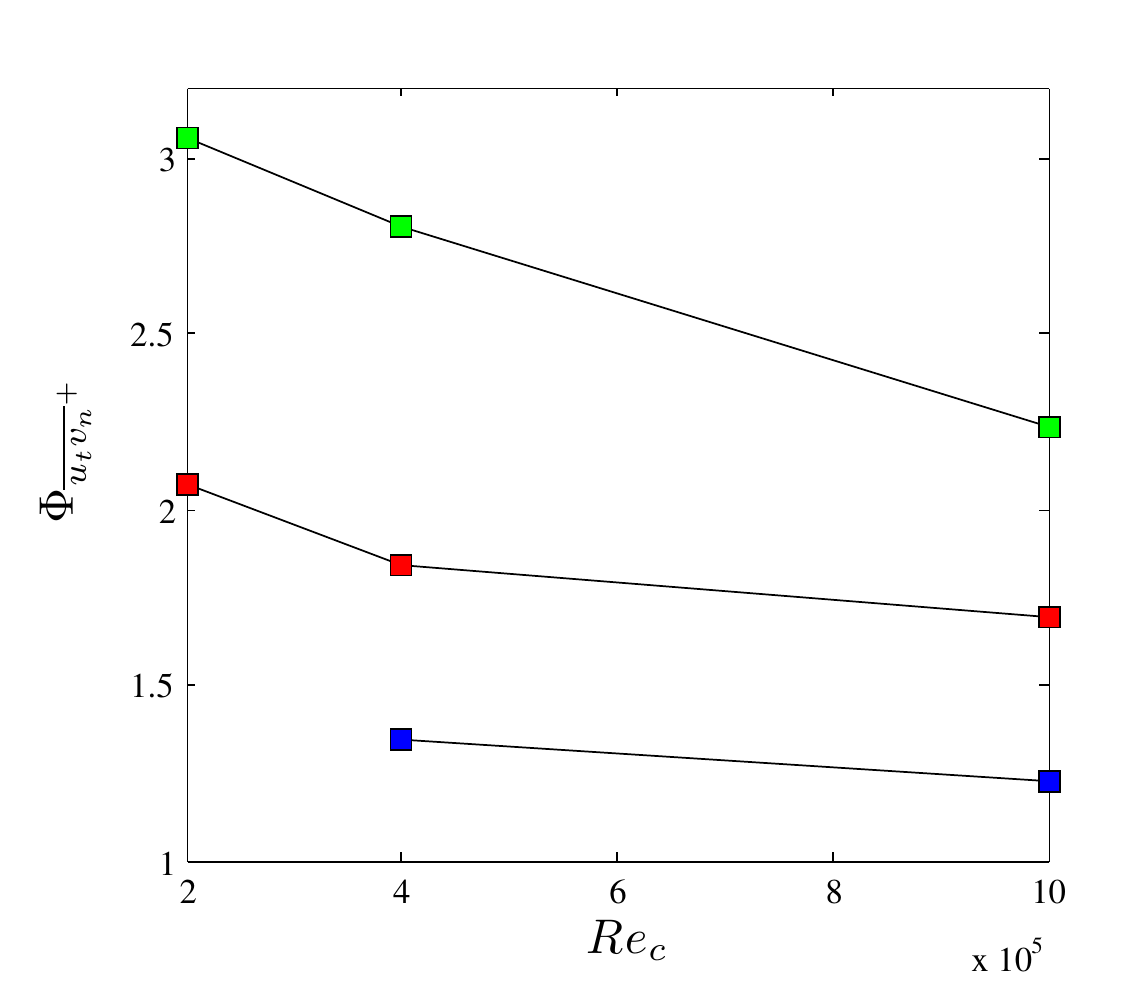}
\caption{Selected inner-scaled components of the Reynolds-stress tensor at (top-left, middle-left) $x_{ss}/c=0.4$ and (top-right, middle-right) $x_{ss}/c=0.7$, for the four wing cases under study, compared with the DNS results of ZPG TBL by \cite{schlatter_orlu10} at approximately matching $Re_{\tau}$ values. Wall-normal profiles of (top panels) tangential velocity fluctuations (solid) and Reynolds-shear stress (dashed), and (middle panels) wall-normal (dashed) and spanwise (solid) velocity fluctuations are shown. Colors from wing cases as in Table \ref{wing_cases}, and {\color{grey}\solid} denotes ZPG TBL data. \textcolor{black}{ The matched $\overline{u^{2}_{t}}^{+}$ and $\overline{u_{t} v_{n}}^{+}$ profiles for W10 and ZPG10 are denoted by $\left ( \blacksquare \right )$, for W4 and ZPG4 by $\left ( \bullet \right )$ and for W2 and ZPG2 by $\left ( \blacklozenge \right )$.} Ratio of (bottom-left) $\overline{u^{2}_{t}}^{+}$ and (bottom-right) $\overline{u_{t}v_{n}}^{+}$ between wing and ZPG at $y_{n}/\delta_{99} \simeq 0.2$. Here (\textcolor{blue}{$\blacksquare$}), (\textcolor{red}{$\blacksquare$}) and (\textcolor{green}{$\blacksquare$}) denote ratios at $x_{ss}/c=0.4$, $0.7$ \textcolor{black}{ and $0.8$}, respectively.}
\label{uup_vs_yp}
\end{figure}

As discussed by \cite{harun_et_al} or \cite{bobke_et_al}, the APG \textcolor{black}{ leads to more energetic turbulent structures in the outer region of the TBL}. This effect is also observed when increasing the Reynolds number in a ZPG TBL, since as the boundary layer develops the outer region exhibits more energetic structures as shown for instance in the experiments by \cite{hutchins_marusic} and the numerical simulations by \cite{eitel_amor_et_al}. However, the mean velocity profiles shown in Figure \ref{Up_vs_yp} suggest that there may be differences in the way that this energizing process takes place, since the the evolution of the mean flow parameters with Reynolds number is not the same in the $\beta=0$ (ZPG) as in the APG cases. In particular, \textcolor{black}{ our results suggest that the effect of the APG is more intense at low Reynolds numbers than at higher $Re$.} This was pointed out in the preliminary work by \cite{vinuesa_negi_lic}, as well as in the experimental study by \cite{apg_piv}. In the latter, the lower-$Re$ numerical data exhibited more pronounced APG features than the measurements. Large-scale energetic motions develop in ZPG TBLs at increasing Reynolds number together with the development of the outer region of the boundary layer. The present results suggest that such development of the outer region takes place in a different way when an APG is present, as illustrated schematically in Figure \ref{sketch_Vp}. In APG TBLs there are two complementing mechanisms responsible for the development of the boundary-layer outer region, namely due to $\beta$ and due to $Re$. In order to further analyze the differences between these mechanisms, several components of the Reynolds-stress tensor are shown for the two wing cases at $x_{ss}/c=0.4$ and $0.7$ in Figure \ref{uup_vs_yp}. Note that we also show the Reynolds-stress profiles from the ZPG DNS by \cite{schlatter_orlu10} at approximately matching $Re_{\tau}$ values (as in Figure \ref{Up_vs_yp}) for comparison. The first important conclusion that can be drawn from Figure \ref{uup_vs_yp} is the fact that all the components of the Reynolds-stress tensor under study exhibit \textcolor{black}{ larger values in the outer region compared to} ZPG TBLs, as discussed for instance by \textcolor{black}{ \cite{kitsios_et_al}} or \cite{bobke_et_al}. As in the case of the mean velocity profiles, the \textcolor{black}{ W1} case exhibits different trends than the ones from the higher-$Re$ cases, which again confirms that \textcolor{black}{ this} boundary layer is not well-behaved. At $x_{ss}/c=0.4$, the near-wall peak of the tangential velocity fluctuation profile is higher than the ones of the cases at higher $Re$, whereas the peaks in the wall-normal and spanwise fluctuation profiles are lower in this case. This can be associated to the influence of the relatively strong APG on a very low-$Re$ TBL, which leads to a different structure of turbulence. A similar behavior was observed in connection with marginally-turbulent flow through hexagonal ducts by \cite{marin_et_al}. On the other hand, at $x_{ss}/c=0.7$ all the components of the Reynolds-stress tensor under study show larger peaks than the higher-$Re$ cases. Focusing on the $\overline{u^{2}_{t}}^{+}$ and $\overline{u_{t} v_{n}}^{+}$ profiles for $Re_{c} \geq 200,000$, it can be noted that the APG TBLs exhibit a much more energetic outer region than the corresponding ZPG cases at the same $Re_{\tau}$. This is further quantified in Figure \ref{uup_vs_yp} (bottom), where the ratios $\Phi_{\overline{u^{2}_{t}}^{+}}$ and $\Phi_{\overline{u_{t} v_{n}}^{+}}$ are shown for the various wing cases, at the two \textcolor{black}{ previous locations and $x_{ss}/c=0.8$.} In this figure, the ratios are defined as the wing values of the tangential velocity fluctuation and the Reynolds shear-stress profiles, at $y_{n} / \delta_{99}=0.2$, divided by the ones from the ZPG profile at approximately matching $Re_{\tau}$. This wall-normal location indicates the end of the overlap region in ZPG TBLs for $Re_{\theta}<6,000$, but evaluating these ratios in other locations within the overlap layer yields qualitatively similar trends. As in the case of the mean flow, these indicators are larger than 1, \textcolor{black}{ where} $\Phi_{\overline{u^{2}_{t}}^{+}}$ ranges from 1.13 to \textcolor{black}{ 3.05}, and $\Phi_{\overline{u_{t} v_{n}}^{+}}$ from 1.23 to \textcolor{black}{ 3.06}. The decreasing trends with $Re$ of these indicators reveal that lower-$Re$ TBLs exhibit a higher energy concentration in the outer region than higher-$Re$ cases. \textcolor{black}{ In fact, the decay with $Re$ becomes steeper at higher $x_{ss}$, {\it i.e.} for stronger $\beta$ conditions.} Moreover, this also reflects the fact that at lower Reynolds numbers the TBLs are more sensitive to APG effects than the higher-$Re$ cases, a conclusion in agreement to what was observed in the mean flow. This is a very relevant result, since it shows not only that the energizing mechanisms of the outer region in the boundary layer are different when they are connected to APG than when they are associated to $Re$, but also that lower-$Re$ TBLs are more sensitive to pressure-gradient effects than high-$Re$ ones. In particular, the tangential velocity fluctuation profiles show larger outer-region values in the lower-$Re$ cases: the values of $\overline{u^{2}_{t}}^{+}$ at $y_{n}/\delta_{99}=0.2$ are, for the wings at $Re_{c}=200,000$, $400,000$ and $1,000,000$, 6.1, 5.7 and 5.0, respectively. This is a manifestation of the more prominent energy accumulation in the large-scale motions at low $Re$ than in the high-$Re$ APG boundary layers.

\begin{figure}
\centering
\includegraphics[width=0.43\textwidth]{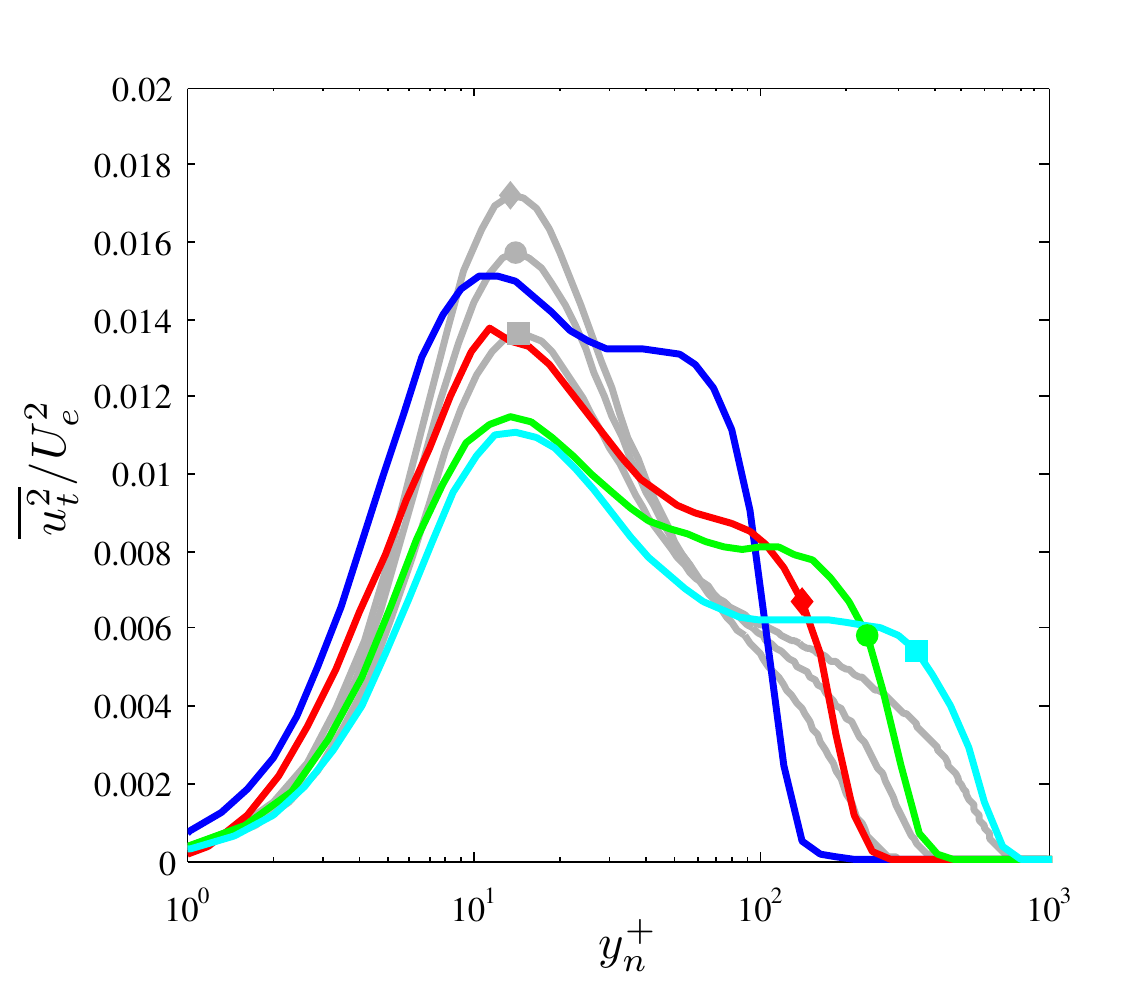}
\includegraphics[width=0.43\textwidth]{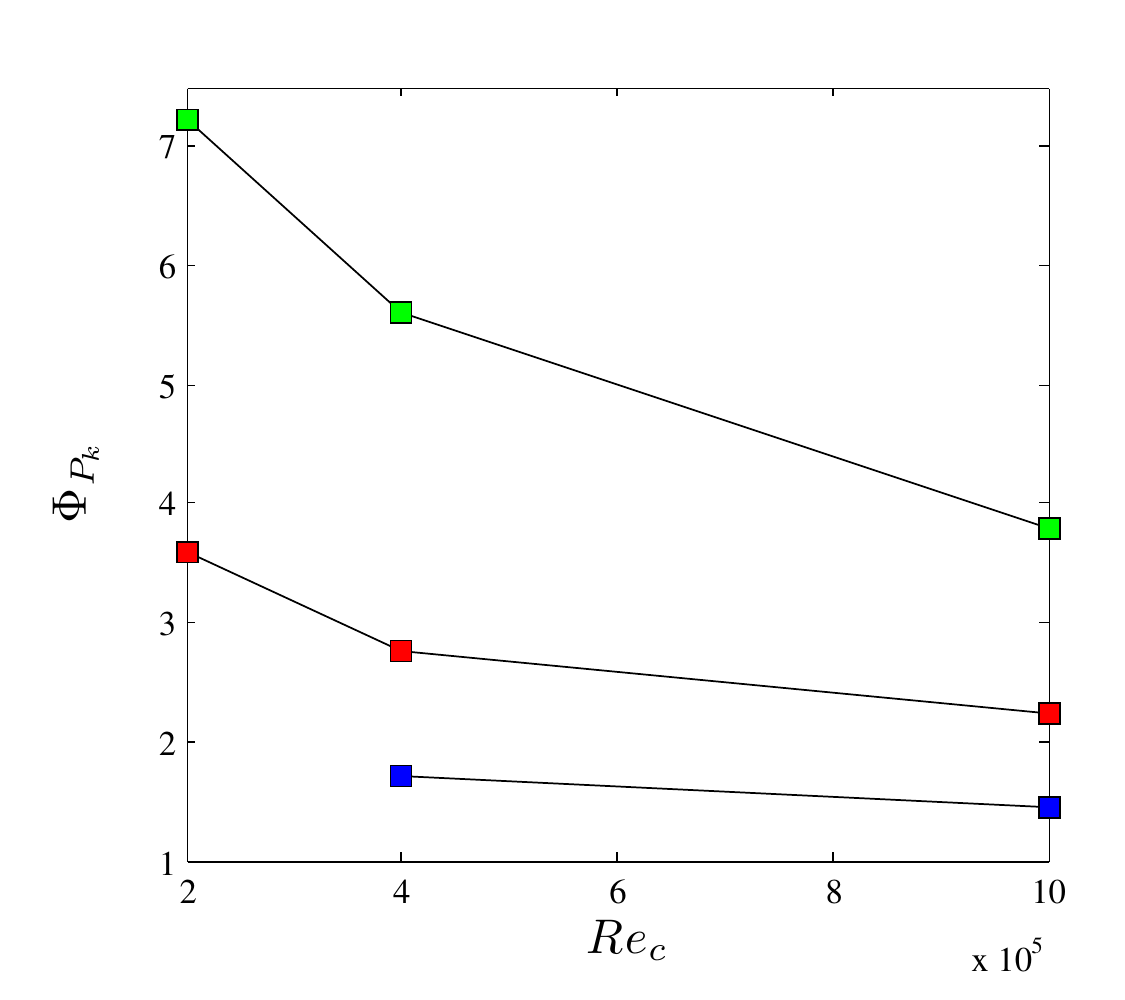}
\caption{\textcolor{black}{ (Left) Outer-scaled tangential velocity fluctuations at $x_{ss}/c=0.7$ for the four wing cases under study, compared with the DNS results of ZPG TBL by \cite{schlatter_orlu10} at approximately matching $Re_{\tau}$ values. Colors from wing cases as in Table~\ref{wing_cases}, and {\color{grey}\solid} denotes ZPG TBL data. The matched $\overline{u^{2}_{t}}/U_{e}^{2}$ profiles for W10 and ZPG10 are denoted by $\left ( \blacksquare \right )$, for W4 and ZPG4 by $\left ( \bullet \right )$ and for W2 and ZPG2 by $\left ( \blacklozenge \right )$. (Right) Ratio of the inner-scaled TKE production between wing and ZPG at $y_{n}/\delta_{99} \simeq 0.2$.} Here (\textcolor{blue}{$\blacksquare$}), (\textcolor{red}{$\blacksquare$}) and (\textcolor{green}{$\blacksquare$}) denote ratios at $x_{ss}/c=0.4$, $0.7$ and $0.8$, respectively.}
\label{uu_Ue_fig}
\end{figure}

\textcolor{black}{ In their experimental study, \cite{harun_et_al} compared three TBLs at approximately the same $Re_{\tau}$, subjected to different values of $\beta$. In addition to inner-scaled statistics, they also compared the streamwise velocity fluctuations in outer scaling, and observed that although the near-wall peak decreased with $\beta$, the outer peak increased (as in inner scaling), thus proving that the energizing effect of the APG was not an artifact of the varying $u_{\tau}$. In Figure~\ref{uu_Ue_fig}~(left) we show the tangential velocity fluctuations at $x_{ss}/c=0.7$ scaled with $U_{e}$, compared with the corresponding outer-scaled ZPG profiles. The ZPG data shows a decrease in the outer-scaled near-wall peak with $Re$, together with an increase of the fluctuations in the outer region (note that \cite{monkewitz_et_al} suggest that at very high $Re$ the outer-scaled fluctuations may reach an asymptotic value in ZPG TBLs). The results from the wing show a trend consistent with that of the $\beta=0$ TBL regarding the near-wall peak, {\it i.e.}, a decrease with Reynolds number. On the other hand, the outer-scaled fluctuations also decrease with $Re$, a result in agreement with the inner-scaled results, which supports the fact that at low $Re$ the APG has a stronger effect (regardless of the change in $u_{\tau}$). Note that, in all the cases, the outer-scaled fluctuations from the wing in the outer region are larger than the corresponding values in ZPGs, but the $Re$ trend is reversed. An additional indicator of the effect of the APG on the TBLs is the TKE production in the outer region. The TKE is defined as $k=1/2 \left (\overline{u^{2}} + \overline{v^{2}} + \overline{w^{2}}  \right )$, and its production as $P_{k}=\overline{u_{i} u_{j}} \partial U_{i} / \partial x_{j}$ (where index notation is used, and inner scaling would be defined in terms of $\nu$ and $u_{\tau}^{4}$). In Figure~\ref{uu_Ue_fig}~(right) we show the ratio $\Phi_{P_{k}}$, defined similarly to $\Phi_{\overline{u^{2}_{t}}^{+}}$ and $\Phi_{\overline{u_{t} v_{n}}^{+}}$, also at $y_{n} / \delta_{99}=0.2$. The more energetic large-scale motions in APGs are associated with higher production in the outer region, as evident from the fact that all the ratios are larger than 1. Note however that the ratios are larger for the production (ranging from 1.46 to 7.24) than for the tangential fluctuations and the Reynolds-shear stress. The larger ratios, together with their steep decay with $Re_{c}$, indicate that the Reynolds-number dependence of the outer-region production is severely affected by the APG, and again low-$Re$ TBLs perceive a bigger effect of the pressure gradient. This effect will be further studied in future work by analyzing the impact of the $\beta(x)$ distribution on the power-spectral densities.}

\begin{figure}
\centering
\includegraphics[width=0.4\textwidth]{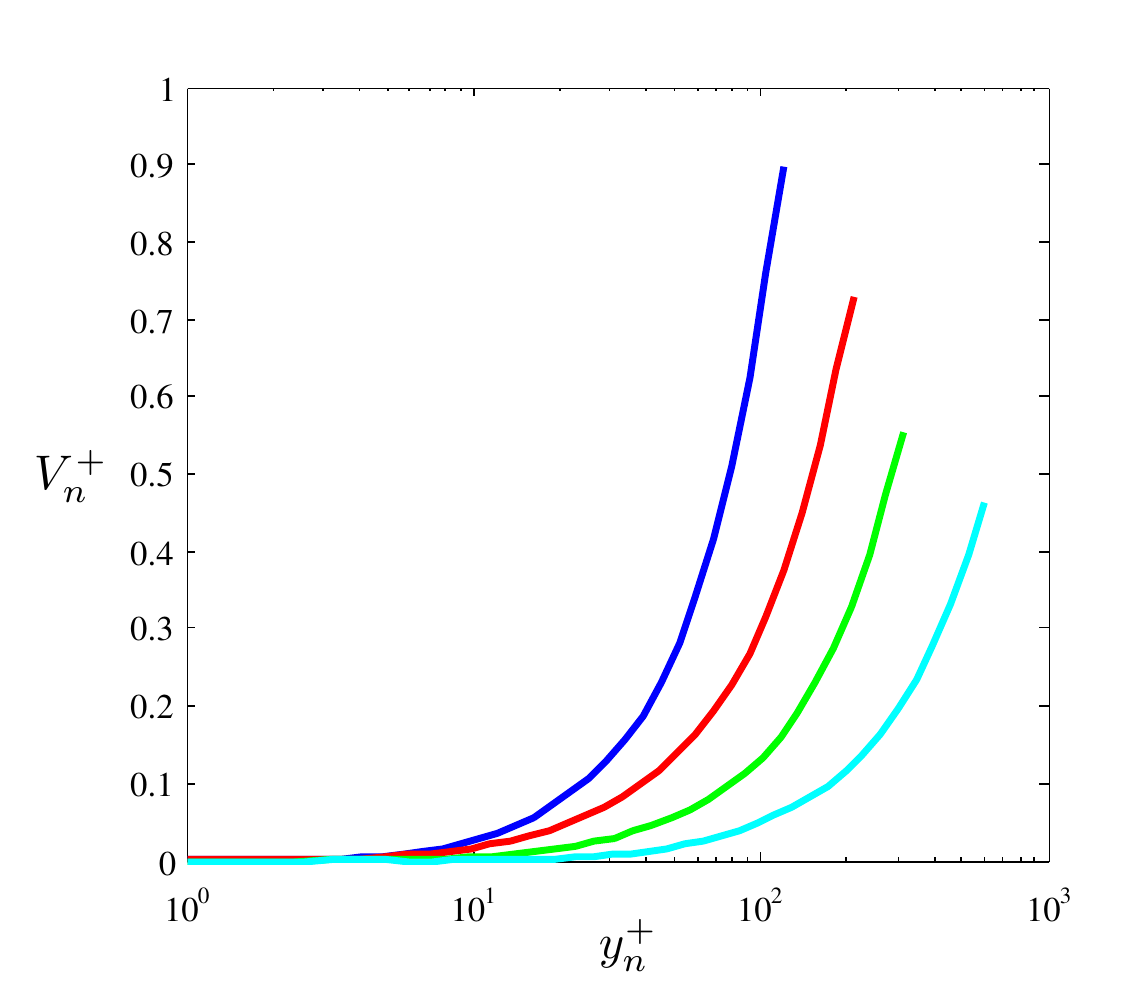}
\includegraphics[width=0.53\textwidth]{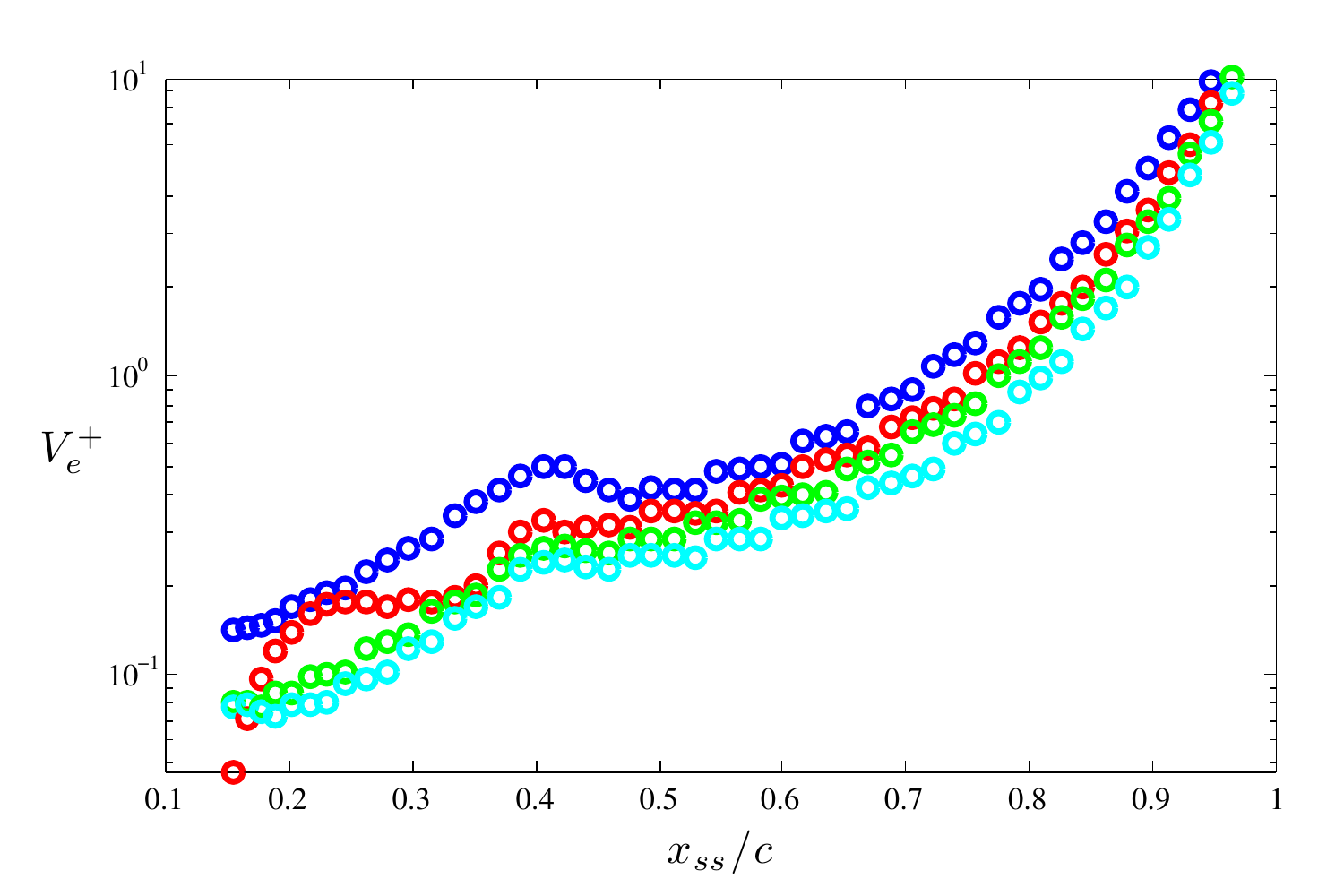}
\caption{(Left) Inner-scaled wall-normal velocity profiles at $x_{ss}/c=0.7$, truncated at the boundary-layer edge and (right) streamwise evolution of the inner-scaled wall-normal edge velocity. Colors from wing cases are as in Table \ref{wing_cases}.}
\label{Vp_figure}
\end{figure}

The \textcolor{black}{ statistics} presented above indicate that, given \textcolor{black}{ approximately} the same streamwise evolution of $\beta$, low-Reynolds-number boundary layers are more significantly affected by the pressure gradient. It is possible to present an additional argument supporting this claim by analyzing the wall-normal velocity distributions since, as illustrated in Figure \ref{sketch_Vp}, an APG increases the wall-normal convection throughout the boundary layer. In Figure \ref{Vp_figure} (left) we compare the inner-scaled wall-normal velocity distributions \textcolor{black}{ from} the various wing cases, all of them at $x_{ss}/c=0.7$. As expected, the lower-$Re$ wings exhibit larger $V^{+}_{n}$ values than the higher-$Re$ cases across the whole boundary layer, a fact that is consistent with the stronger APG effects at low Reynolds numbers. This analysis is extended to the whole streamwise extent of the wing in Figure \ref{Vp_figure} (right), where the wall-normal velocity at the boundary-layer edge $V^{+}_{e}$ is shown for the four cases under study. Note that this figure shows the difficulties in determining the boundary-layer edge in PG TBLs \citep{vinuesa_diagnostic}, which are due to the fact that the tangential velocity is not constant for $y_{n} > \delta_{99}$. In this figure it can also be observed that the \textcolor{black}{ W4} wing exhibits $V^{+}_{e}$ values around $20\%$ larger than those of the \textcolor{black}{ W10} case beyond $x_{ss}/c>0.25$, and in the \textcolor{black}{ W2} case the values are around $40\%$ higher. The curve associated to the \textcolor{black}{ W1} wing also shows the change in trend around $x_{ss}/c \simeq 0.4$ discussed above, which is connected to low-Reynolds-number effects. 

\begin{figure}
\centering
\includegraphics[width=0.65\textwidth]{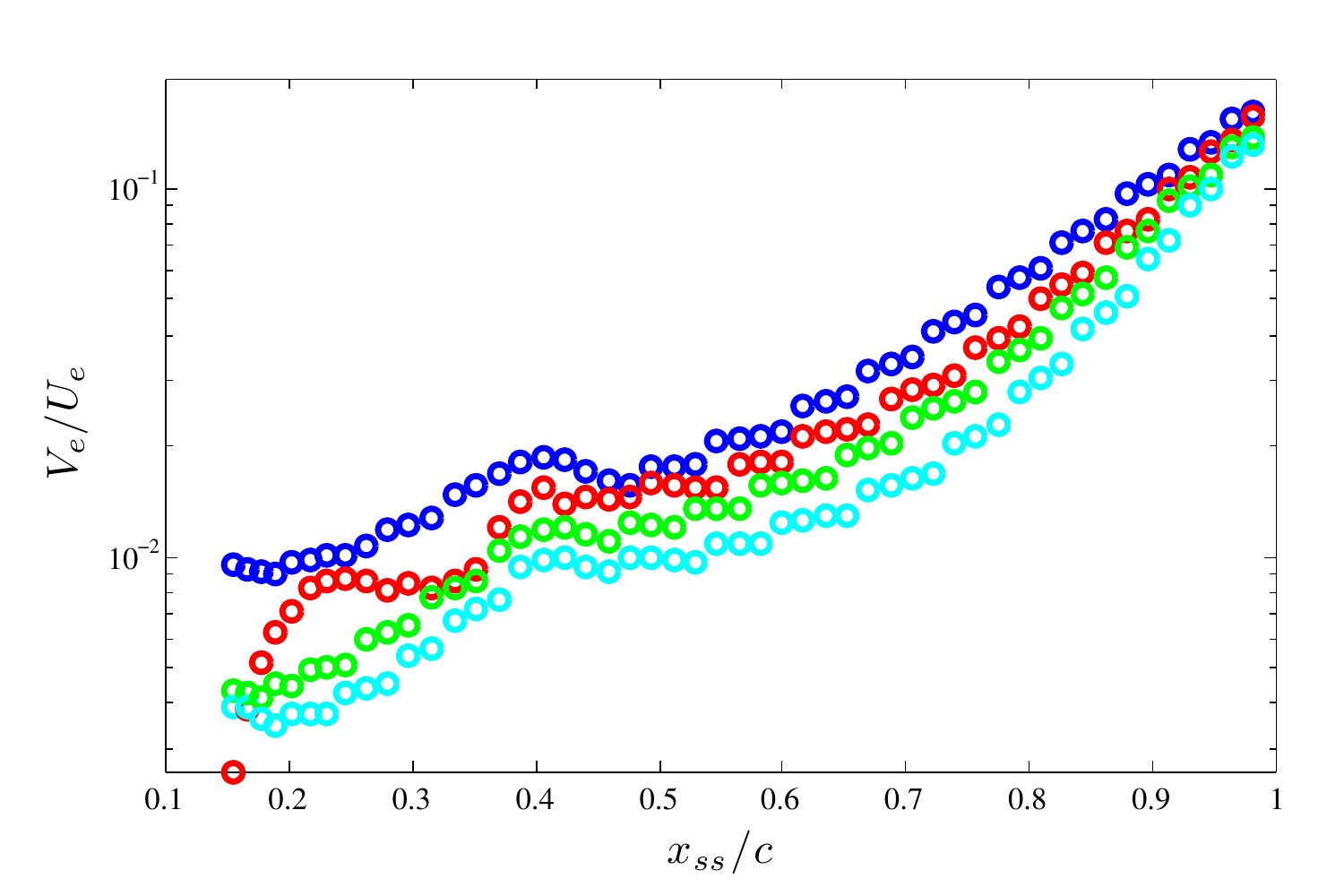}
\caption{\textcolor{black}{ Streamwise evolution of the outer-scaled wall-normal edge velocity. Colors from wing cases are as in Table \ref{wing_cases}.}}
\label{Ve_Ue_fig}
\end{figure}

\textcolor{black}{ The larger wall-normal edge velocities exhibited by the lower-$Re$ wings are also present in outer scaling ({\it i.e.}, in terms of the local tangential edge velocity $U_{e}$) as evident from Figure~\ref{Ve_Ue_fig}. In this case, the W2 and W4 cases have outer-scaled wall-normal edge velocities $55\%$ and $27\%$ larger than the W10 case. This observation, together with the qualitatively similar trends compared to the inner-scaled curves, implies that the larger wall-normal convection experienced at lower $Re$ is not a consequence of the different $u_{\tau}$, but a genuine effect experienced by the TBLs. This is connected to the fact that the outer-scaled APG magnitude $u_{p}/U_{e}$, shown in Figure~\ref{up_Ue_fig}, also exhibits larger values at lower $Re$ despite the fact that the $\beta(x)$ values are approximately the same. The higher wall-normal velocities (and outer-scaled APG magnitude) exhibited by the W2 and W4 cases imply that the development of the outer layer has been more significantly affected by the APG, thus producing more energetic large-scale motions as evident from the turbulence statistics.}


\section{Summary and conclusions} \label{conclusions_section}

The present study is aimed at further understanding the mechanisms responsible for the development of the outer region of TBLs and the energizing of the large-scale motions, as well as their connection with APGs and increasing Reynolds number. To this end, we performed well-resolved LESs of the turbulent boundary layers developing around a NACA4412 wing section at $Re_{c}$ values from $100,000$ to $1,000,000$, all of them with $5^{\circ}$ angle of attack. All the simulations were performed with the spectral-element code Nek5000, using a setup similar to the one employed by \cite{hosseini_et_al} and \cite{wing_ftac} to perform a DNS of the same flow case at $Re_{c}=400,000$. The boundary layers developing on the suction side of the wing sections, for $Re_{c} \geq 200,000$, are subjected to essentially the same streamwise Clauser pressure-gradient distribution $\beta(x)$, a fact that allows to characterize the effect of the Reynolds number in APG TBLs subjected to approximately the same pressure-gradient history. Note that this study complements the one by \cite{bobke_et_al} on flat-plate APG TBLs, in which the effect of different $\beta(x)$ distributions over similar Reynolds-number ranges was assessed.

As a TBL develops, the increasing Reynolds number produces a more energetic outer region, a fact that is manifested in the Reynolds-stress tensor profiles. On the other hand, an APG also produces more energetic large-scale motions in the outer region of the boundary layer due to the increased wall-normal convection associated to it. Our results indicate that the skin-friction curve from the wing at $Re_{c}=1,000,000$ is below the ones at lower Reynolds numbers (up to around $x_{ss}/c \simeq 0.9$), a fact that is consistent with the well-known effect of Reynolds number in ZPG TBLs. Moreover, the shape factor curve in the high-$Re$ wing is also below the ones at lower $Re$, which is associated with another effect of Reynolds number, {\it i.e.}, to reduce $H$. 

We also analyzed, for $Re_{c} \geq 200,000$, the inner-scaled mean velocity profiles at $x_{ss}/c=0.4$ and $0.7$, which are subjected to $\beta$ values of \textcolor{black}{ approximately} $0.6$ and $2$, respectively. For the three wing cases between $Re_{c}=200,000$ and $1,000,000$, we compared the mean profiles with the ones from ZPG TBLs \citep{schlatter_orlu10} at approximately the same $Re_{\tau}$. The ratios $\Phi_{U_{e}^{+}}$ and $\Phi_{H}$ have values larger than 1 (ranging from 1.07 and \textcolor{black}{ 1.48,} and from 1.05 and \textcolor{black}{ 1.27} respectively), which is consistent with the features exhibited by APGs. Interestingly, both ratios decay with $Re_{c}$, which implies that low-$Re$ TBLs are more sensitive to the effect of APGs, when exposed to \textcolor{black}{ approximately} the same $\beta(x)$ flow history. This conclusion is supported by the observations on several components of the Reynolds-stress tensor, in particular in the tangential velocity fluctuation profile and the Reynolds-shear stress. The values of the ratios $\Phi_{\overline{u^{2}_{t}}^{+}}$ and $\Phi_{\overline{u_{t} v_{n}}^{+}}$ (which relate the APG and ZPG profiles with matched $Re_{\tau}$, at $y_{n}/\delta_{99}=0.2$) at $x_{ss}/c=0.4$ and $0.7$ are also larger than 1, and range from 1.13 to \textcolor{black}{ 3.05}, and from 1.23 to \textcolor{black}{ 3.06}, respectively. Also in this case the ratios decrease with $Re_{c}$, which indicates that the outer region of the lower-$Re$ wings is more energetic with respect to the corresponding ZPG than the higher-$Re$; this also implies that the low-$Re$ TBLs are more sensitive to APG effects. In fact, at $x_{ss}/c=0.7$ the value of the $\overline{u^{2}_{t}}^{+}$ profile at $y_{n}/\delta_{99}=0.2$ is largest at $Re_{c}=200,000$, and decays with increasing $Re$. \textcolor{black}{ The decrease in impact of the APG on the TBL with $Re$ is also observed in the outer-region TKE production.}

Our results \textcolor{black}{ suggest} that two complementing mechanisms contribute to the development of the outer region in TBLs and the formation of large-scale energetic structures: one mechanism associated with the increase in Reynolds number, and another one connected to the APG. When \textcolor{black}{ approximately} the same streamwise evolution of the pressure-gradient magnitude is imposed, the low-Reynolds-number boundary layer becomes more severely affected by the APG, as also observed when analyzing the distributions of mean wall-normal velocity. In particular, the $Re_{c}=200,000$ and $400,000$ wings exhibit $V^{+}_{e}$ values around $40\%$ and $20\%$ larger than those of the $1,000,000$, respectively, over a significant portion of the suction side of the wing. \textcolor{black}{ The lower-$Re$ wings also exhibit larger outer-scaled wall-normal velocities and turbulence fluctuations.} As illustrated in Figure \ref{sketch_Vp}, the APG increases wall-normal convection, which thickens the boundary layer allowing a larger outer region and leading to the formation of more energetic large-scale motions. These structures are taller, but shorter in the streamwise direction and more inclined with respect to the wall \citep{maciel_et_al}, due to the increased $V^{+}$. This suggests that a TBL at a higher Reynolds number, with a more ``mature'' outer region, is less affected by the effect of the APG. Further analyses of the current databases, with emphasis on extraction and characterization of coherent structures, will help to elucidate the differences in the mechanisms for outer-region energizing due to APG and Reynolds number. 


\section*{Acknowledgments}

\textcolor{black}{ We acknowledge A. Tanarro for his help running some of the simulations.} The simulations were performed on resources provided by the Swedish National Infrastructure for Computing (SNIC) at the Center for Parallel Computers (PDC), in Stockholm (Sweden), and by the Partnership for Advanced Computing in Europe (PRACE) at the Barcelona Supercomputing Center (BSC) in Barcelona (Spain).  The computer time at the BSC was obtained through the 12h PRACE Project Access Call, Project Number 2015133182. RV and PS acknowledge the funding provided by the Swedish Research Council (VR) and from the Knut and Alice Wallenberg Foundation. This research is also supported by the ERC Grant No. ``2015-AdG-694452, TRANSEP'' to DH \textcolor{black}{ and by the Swedish Foundation for Strategic Research, project ``In-Situ Big Data Analysis for Flow and Climate Simulations'' (ref. number BD15-0082). } 

\section*{References}



\bibliographystyle{elsarticle-harv} 
\bibliography{wing_bib}


%
%
%
\end{document}